%% file: RealMalleable25.tex
\documentclass[runningheads]{llncs}

\usepackage[T1]{fontenc}
\usepackage{microtype}

\usepackage[american]{babel}
\usepackage{listings}
\usepackage{graphicx}
\usepackage{color}
\usepackage{hyperref}
\usepackage{epstopdf}
\usepackage{caption}
\usepackage{subcaption}
\usepackage[table, dvipsnames, xcdraw]{xcolor}
\usepackage{longtable}
\usepackage{booktabs}
\usepackage[group-separator={,}, group-minimum-digits=4]{siunitx}
\usepackage{xspace}
\usepackage{array}
\usepackage[normalem]{ulem}
\useunder{\uline}{\ul}{}
\usepackage{orcidlink}
\usepackage{todonotes}
\usepackage{multirow}
\usepackage{textgreek}
\usepackage{float}
\usepackage{enumitem}
\setlist[itemize]{topsep=5pt, partopsep=5pt}
\usepackage{makecell}
\usepackage{etoolbox}
\patchcmd{\paragraph}{\itshape}{\bfseries\boldmath}{}{}

\usepackage{diagbox}
\usepackage{amsmath}

\usepackage{mathtools, nccmath}
\usepackage{wrapfig}

\newcommand{\haswell}{\textsc{Haswell}\xspace}
\newcommand{\knl}{\textsc{KNL}\xspace}
\renewcommand{\theta}{\textsc{Theta}\xspace}
\newcommand{\eagle}{\textsc{Eagle}\xspace}
\newcommand{\cori}{\textsc{Cori}\xspace}
\newcommand{\easybackfill}{\textsc{EASY-Backfill}\xspace}
\newcommand{\prefkeeper}{\textsc{KeepPref}\xspace}
\newcommand{\pref}{\textsc{Pref}\xspace}
\renewcommand{\min}{\textsc{Min}\xspace}
\newcommand{\avg}{\textsc{Avg}\xspace}

\begin{document}

\title{
    Evaluating Malleable Job Scheduling in HPC Clusters using Real-World Workloads
}

\titlerunning{Evaluating Malleable Job Scheduling in HPC Clusters}
\authorrunning{Patrick Zojer, Jonas Posner, and Taylan Özden}

\author{
    Patrick Zojer\,\inst{1}
    \and
    Jonas Posner\,\inst{1}\,\orcidlink{0000-0002-6491-1626}
    \and
    Taylan Özden\,\inst{2}\,\orcidlink{0000-0002-4540-4717}\,
}

\institute{
    University of Kassel, Kassel, Germany\\
    \email{patrick.zojer@student.uni-kassel.de},
    \email{jonas.posner@uni-kassel.de}
    \and Technical University of Darmstadt, Darmstadt, Germany\\
    \email{taylan.oezden@tu-darmstadt.de}
}

\hypersetup{
    pdftitle={{Evaluating Malleable Job Scheduling in HPC Clusters using Real-World Workloads}},
    pdfsubject={CARLA25},
    pdfauthor={Patrick Zojer, Jonas Posner, Taylan Özden},
    pdfkeywords={{Dynamic Resources, Malleability, Job Scheduling}}
}

\maketitle
\setcounter{footnote}{0}
\interfootnotelinepenalty=10000

\begin{abstract}
    \input{00abstract}
\end{abstract}

\input{01introduction}
\input{02methodology}

\input{03evaluation}

\input{04discussion}
\input{05relatedwork}
\input{06conclusions}

\bibliography{RealMalleable25}
\end{document}

%% file: 00abstract.tex
Optimizing resource utilization in high-performance computing (HPC) clusters is essential for maximizing both system efficiency and user satisfaction.
However, traditional rigid job scheduling often results in underutilized resources and increased job waiting times.

This work evaluates the benefits of resource elasticity, where the job scheduler dynamically adjusts the resource allocation of malleable jobs at runtime.
Using real workload traces from the Cori, Eagle, and Theta supercomputers, we simulate varying proportions (0--100\%) of malleable jobs with the ElastiSim software.

We evaluate five job scheduling strategies, including a novel one that maintains malleable jobs at their preferred resource allocation when possible.
Results show that, compared to fully rigid workloads, malleable jobs yield significant improvements across all key metrics.
Considering the best-performing scheduling strategy for each supercomputer, job turnaround times decrease by 37--67\%, job makespan by 16--65\%, job wait times by 73--99\%, and node utilization improves by 5--52\%.
Although improvements vary, gains remain substantial even at 20\% malleable jobs.

This work highlights important correlations between workload characteristics (e.g., job runtimes and node requirements), malleability proportions, and scheduling strategies.
These findings confirm the potential of malleability to address inefficiencies in current HPC practices and demonstrate that even limited adoption can provide substantial advantages, encouraging its integration into HPC resource management.

%% file: 01introduction.tex
\section{Introduction}\label{sec:01-Introduction}
\textit{High-Performance Computing (HPC)} clusters (\textit{supercomputers}) enable the solution of complex problems in various scientific fields.
Supercomputers comprise multiple interconnected \textit{nodes}, each executing submitted \textit{jobs}.
A job is a program submitted to a \textit{job scheduler} (e.g., Slurm~\cite{Slurm}) with user-defined metadata such as runtime limits and required nodes.
A collection of jobs constitutes a \textit{workload}.

Traditionally, jobs are \textit{rigid} and are allocated a fixed number of nodes for their entire execution.
Schedulers assign jobs to nodes based on availability and fairness.

Maximizing node utilization is crucial, yet often unattainable in practice.
Job size heterogeneity and scheduler constraints inhibit optimal system packing, resulting in idle nodes even when jobs remain queued.
This inefficiency leads to extended job wait times and wasted computational resources.

Supercomputers such as \theta~\cite{ALCF-Theta}, \cori~\cite{NERSC-Cori}, and \eagle~\cite{NREL-Eagle} exhibit average node utilization between 30\% and 80\%~\cite{ALCF-Theta-Workload,elastic-sim,NREL-Eagle-Workload}.
These low values result from both scheduling inefficiencies and factors such as empty queues or system maintenance.

\textit{Resource elasticity} offers a promising approach to address underutilization by allowing jobs to change their number of allocated nodes at runtime.
Elastic jobs include variants~\cite{Feitelson} such as \textit{evolving} jobs that request resource changes from the scheduler.
In contrast, for \textit{malleable} jobs, the scheduler initiates resource changes.
This requires schedulers capable of \textit{expanding} and \textit{shrinking} jobs and applications that support dynamic resource allocation.
While resources may include storage or GPUs, this study focuses solely on CPU nodes.

However, introducing elasticity into real HPC environments poses significant challenges.
It requires not only schedulers with elasticity support but also applications capable of adapting to dynamic node counts.
Many applications cannot efficiently handle such variability, necessitating changes at both the system and application levels.
Therefore, elasticity adoption requires modifications at both system administration and application development levels.
Additionally, the dominant inter-node parallel programming model Message Passing Interface (MPI)~\cite{MPI} offers limited support for elasticity.

Given these constraints, many studies on elasticity rely on simulations, e.g.~\cite{Taylan2024,ScheduleElasticIPDPS22,JonasMallScheduling23}.
This approach is practical given the cost and complexity of real testbeds, but requires accurate modeling of input parameters.

This work evaluates the impact of malleable jobs using real workload traces from \cori, \theta, and \eagle.
For simulations, we use the ElastiSim simulation software~\cite{ElastiSim}.
Using a speedup model and efficiency thresholds, we transform rigid jobs into malleable ones, with malleability proportions varying from 0\% to 100\%.
In addition to the conventional rigid EASY-backfilling~\cite{EasyBackfilling}, we evaluate four malleable strategies.
Three are from prior work~\cite{JonasMallScheduling23}, while the fourth---introduced in this work---aims to preserve each job's preferred number of nodes.
The preferred number of nodes reflects a reasonable speed and efficiency tradeoff~\cite{DowneyModel}.

\enlargethispage{0.1\baselineskip}
We evaluate multiple performance metrics: job turnaround time, job makespan, job wait time, and node utilization.
Results show that, compared to fully rigid workloads, malleability consistently yields significant improvements.
Considering the best-performing scheduling strategy for each supercomputer, job turnaround times decrease by 37--67\%, job makespan by 16--65\%, job wait times by \mbox{73--99\%}, and node utilization improves by 5--52\%.
However, the improvement varies considerably depending on workload characteristics and scheduling strategy.

The remainder of this work is structured as follows.
Section~\ref{sec:03-Methodology} outlines the scheduling strategies, workload preparation, and experimental design.
Section~\ref{sec:04-Evaluation} presents and analyzes the simulation results across different workloads and scheduling configurations, while Section~\ref{sec:05-Discussion} outlines limitations.
Section~\ref{sec:06-Related-Work} situates this study within the context of existing research, and Section~\ref{sec:07-Conclusions} summarizes key insights and concludes this work.

%% file: 02methodology.tex
\section{Methodology}\label{sec:03-Methodology}
This section describes the methodological foundation of our study.
We begin by detailing the job scheduling strategies in Section~\ref{subsec:03-Job-Scheduler-Algorithms}, including both baseline and malleable strategies.
Section~\ref{subsec:03-workload-preparation} explains the preprocessing steps applied to real-world HPC workload traces, such as cleaning, filtering, and transformation into simulation-ready formats.
Section~\ref{subsec:03-Experimental-Design} outlines our experimental setup, including malleability proportions, simulation parameters, and the metrics used to evaluate performance.
Together, these components form a reproducible framework for evaluating the impact of malleable job scheduling in realistic HPC scenarios.

\input{02scheduler}
\input{02figs}
\input{02prep}
\input{02design}

%% file: 02scheduler.tex
\subsection{Job Scheduling Strategies}\label{subsec:03-Job-Scheduler-Algorithms}
We evaluate the following job scheduling strategies in our simulations.

\textbf{Rigid \easybackfill~\cite{EasyBackfilling}:}
In traditional rigid scheduling, each job requests a fixed number of nodes for its entire runtime.
Jobs are placed in a queue and selected for execution according to scheduling policies.

\easybackfill extends the well-known \textit{First-Come, First-Served (FCFS)} strategy, which schedules jobs strictly in submission order.
While fair, pure FCFS often results in poor node utilization when the first job in the queue cannot start due to insufficient available nodes.
\easybackfill addresses this by allowing limited reordering.
If enough nodes are available, the first job starts as usual.
If not, the scheduler scans the queue for smaller jobs that can be started without delaying the first job.
This approach increases node utilization while preserving fairness.
In this work, \easybackfill serves as our rigid baseline.

\textbf{Malleable \avg, \min, and \pref~\cite{JonasMallScheduling23}:}
Based on Iserte~\textsl{et\,al.~}\cite{Iserte2020} and extended in our prior work~\cite{JonasMallScheduling23}, these strategies use malleable jobs that define minimum and maximum numbers of nodes.
Scheduling proceeds in three steps:
\begin{enumerate}[label=Step\,\arabic*:, labelwidth=5em, labelsep=1em, leftmargin=*, align=left]
  \item Waiting jobs are started using \easybackfill.
  The number of nodes for starting a malleable job is determined by \avg, \min, and \pref.

  \item If there are idle nodes but not enough to start jobs, running malleable jobs are shrunk.
  A priority order (described below) determines the order in which jobs are considered for shrinking, starting with the highest-priority job.

  \item Any remaining idle nodes are distributed to expand running malleable jobs.
  Again, a priority order is used, but in this case, the lowest-priority job is considered first.
\end{enumerate}

\vspace{0.5em}
\textbf{\min} prioritizes jobs based on the surplus of allocated over minimum nodes:
\begin{equation}
  \text{priority}_{\text{job}} = \textit{current-nodes}_{\text{job}} - \textit{min-nodes}_{\text{job}}
  \label{eq:02-min}
  \vspace*{-1ex}
\end{equation}
In Step\,1, all jobs are started at their minimum number of nodes.
In Steps\,2 and\,3, resizing is limited to the smallest number of jobs needed to achieve the target allocation.

\vspace{0.5em}
\textbf{\pref} introduces the additional job attribute \textit{preferred number of nodes}, reflecting a reasonable speedup-efficiency trade-off~\cite{DowneyModel}:
\vspace*{-1ex}
\begin{equation}
  \text{priority}_{\text{job}} = \textit{current-nodes}_{\text{job}} - \textit{preferred-nodes}_{\text{job}}
  \label{eq:02-pref}
  \vspace*{-1ex}
\end{equation}
In Step\,1, jobs attempt to start with their preferred nodes; if that is not possible, fewer nodes are assigned.
Resizing behavior in Steps\,2 and\,3 mirrors that of \min.

\vspace{0.5em}
\textbf{\avg} evaluates the relative utilization within the node range:
\vspace*{-1ex}
\begin{equation}
  \text{priority}_{\text{job}} =
  \frac{\textit{current-nodes}_{\text{job}} - \textit{min-nodes}_{\text{job}}}
  {\textit{max-nodes}_{\text{job}} - \textit{min-nodes}_{\text{job}}}
  \label{eq:02-avg}
  \vspace*{-1ex}
\end{equation}
In Step\,1, jobs start with their minimum nodes.
In Steps\,2 and\,3, all malleable jobs are eligible for resizing, enabling a more balanced redistribution of resources.

\textbf{Malleable \prefkeeper:}
Proposed in this work, it is based on Eq.~\ref{eq:02-pref} but explicitly aims to preserve each job's preferred number of nodes.
In Step\,1, jobs are always started with their preferred nodes.
In Step\,2, only jobs currently running with more than their preferred nodes are considered for shrinking.

\vspace{0.5em}
\textbf{Metrics:}
We evaluate scheduling performance using the following metrics:
\begin{itemize}
  \item \textbf{Job Wait Time}:~Time from job submission to job start.
  \item \textbf{Job Makespan}:~Execution time from job start to completion.
  \item \textbf{Job Turnaround Time}:~Total time from submission to completion ($\text{Wait Time} + \text{Makespan}$).
  \item \textbf{Node Utilization}:~Proportion of nodes actively used during the simulation.
\end{itemize}

%% file: 02figs.tex
\begin{figure}
    \centering
    \vspace*{-10pt}

    \begin{minipage}{\linewidth}
        \centering
        \begin{subfigure}{.49\textwidth}
            \centering
            \resizebox{\linewidth}{!}{\input{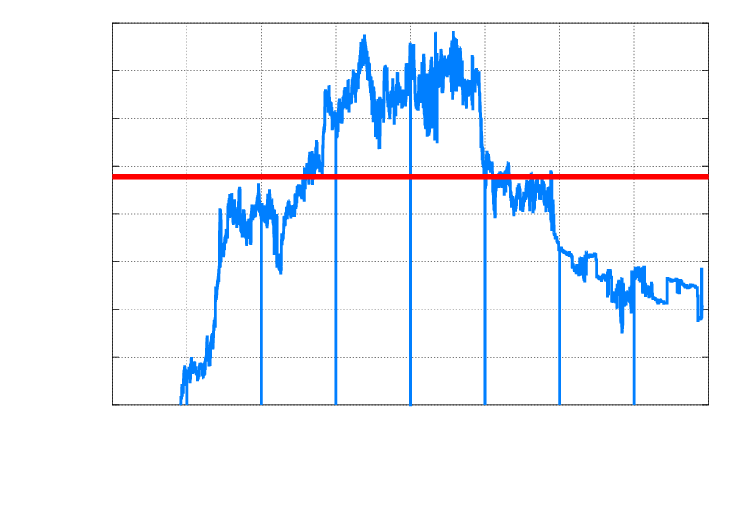}}
            \caption{Raw data}
            \label{fig:cori-raw}
        \end{subfigure}
        \hfill
        \begin{subfigure}{.49\textwidth}
            \centering
            \resizebox{\linewidth}{!}{\input{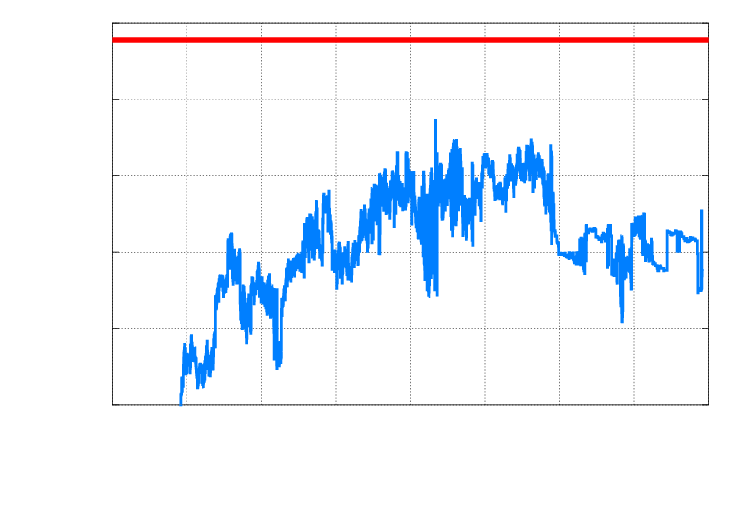}}
            \caption{After cleaning and merging}
            \label{fig:cori-clean}
        \end{subfigure}

        \caption{
            \haswell: Raw data (a) shows artifacts from shared-node jobs and daily splits (vertical blue lines).
            The red line shows the capacity of \num{2388} nodes.
            After cleaning and merging, the data looks realistic (b).
        }
        \label{fig:cori-clean-vs-raw}
    \end{minipage}

    \vspace{1em}

    \begin{minipage}{\linewidth}
        \centering
        \captionof{table}{Raw jobs and cleaned jobs (* = jobs with shared nodes)}
        \label{tab:csv-filter}
        \resizebox{\linewidth}{!}{%
            \begin{tabular}{
                l
                    >{\raggedleft\arraybackslash}m{17mm}
                    >{\raggedleft\arraybackslash}m{24mm}
                    >{\raggedleft\arraybackslash}m{36mm}
            }
                \toprule
                \textbf{Dataset} & \textbf{Raw Jobs} & \textbf{Cleaned Jobs} & \textbf{Runtime Loss} \\
                \midrule
                \mbox{\eagle}   & \num{7390585} & \num{72580}  & \(\sim\)\num{165000} h (\(\sim\)1.275\%)\phantom{*} \\
                \mbox{\knl}     & \num{55587}   & \num{14063}  & \(\sim\)\num{41} h (\(\sim\)0.004\%)* \\
                \mbox{\haswell} & \num{119781}  & \num{91200}  & \(\sim\)\num{33370} h (\(\sim\)13.680\%)* \\
                \bottomrule
            \end{tabular}
        }
    \end{minipage}

    \vspace{1em}

    \begin{minipage}{\linewidth}
        \centering
        \resizebox{0.99\linewidth}{!}{\input{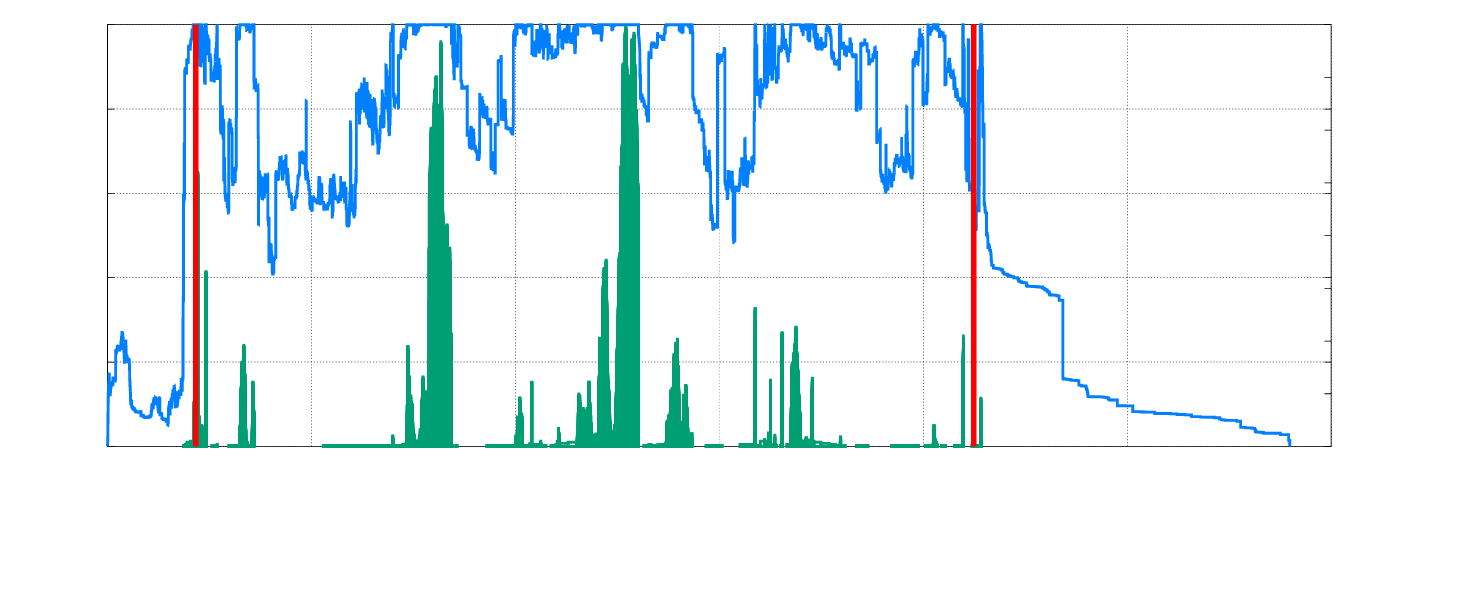}}
        \captionof{figure}{
            \knl: Node utilization (blue line) with 100\% rigid jobs.
            Red lines mark warm-up and final submission; metrics are computed in between.
            Green areas illustrate the number of jobs in the queue.
        }
        \label{fig:cori-knl-rigid-node-util}
    \end{minipage}

    \vspace{1em}

    \begin{minipage}{\linewidth}
        \centering
        \captionof{table}{Simulation configurations}
        \label{tab:sim-configs}
        \resizebox{\linewidth}{!}{%
            \begin{tabular}{
                l
                    >{\centering\arraybackslash}m{18mm}
                    >{\centering\arraybackslash}m{14mm}
                    >{\centering\arraybackslash}m{10mm}
                    >{\centering\arraybackslash}m{12mm}
                    >{\centering\arraybackslash}m{9mm}
            }
                \toprule
                \textbf{Dataset} & \textbf{Start} & \textbf{Duration} & \textbf{Jobs} & \textbf{Tick} & \textbf{Nodes} \\
                \midrule
                \mbox{\theta~\cite{ALCF-Theta}}     & \makecell{2023-08-01} & 28 days                 & \num{2,550}   & 1 s  & \num{4392} \\
                \mbox{\eagle~\cite{NREL-Eagle}}     & \makecell{2022-09-01} & 28 days                 & \num{143829}  & 10 s & \num{2568} \\
                \mbox{\knl~\cite{NERSC-Cori}}       & \makecell{2022-11-06} & 5 days                  & \num{41524}   & 10 s & \num{9688} \\
                \mbox{\haswell~\cite{NERSC-Cori}}   & \makecell{2022-11-07} & 5 days                  & \num{28259}   & 1 s  & \num{2388} \\
                \bottomrule
            \end{tabular}
        }
    \end{minipage}
\end{figure}

%% file: img/HaswellRawUtilization.tex
\begingroup
\Large
  \makeatletter
  \providecommand\color[2][]{%
    \GenericError{(gnuplot) \space\space\space\@spaces}{%
      Package color not loaded in conjunction with
      terminal option `colourtext'%
    }{See the gnuplot documentation for explanation.%
    }{Either use 'blacktext' in gnuplot or load the package
      color.sty in LaTeX.}%
    \renewcommand\color[2][]{}%
  }%
  \providecommand\includegraphics[2][]{%
    \GenericError{(gnuplot) \space\space\space\@spaces}{%
      Package graphicx or graphics not loaded%
    }{See the gnuplot documentation for explanation.%
    }{The gnuplot epslatex terminal needs graphicx.sty or graphics.sty.}%
    \renewcommand\includegraphics[2][]{}%
  }%
  \providecommand\rotatebox[2]{#2}%
  \@ifundefined{ifGPcolor}{%
    \newif\ifGPcolor
    \GPcolorfalse
  }{}%
  \@ifundefined{ifGPblacktext}{%
    \newif\ifGPblacktext
    \GPblacktexttrue
  }{}%
  \let\gplgaddtomacro\g@addto@macro
  \gdef\gplbacktext{}%
  \gdef\gplfronttext{}%
  \makeatother
  \ifGPblacktext
    \def\colorrgb#1{}%
    \def\colorgray#1{}%
  \else
    \ifGPcolor
      \def\colorrgb#1{\color[rgb]{#1}}%
      \def\colorgray#1{\color[gray]{#1}}%
      \expandafter\def\csname LTw\endcsname{\color{white}}%
      \expandafter\def\csname LTb\endcsname{\color{black}}%
      \expandafter\def\csname LTa\endcsname{\color{black}}%
      \expandafter\def\csname LT0\endcsname{\color[rgb]{1,0,0}}%
      \expandafter\def\csname LT1\endcsname{\color[rgb]{0,1,0}}%
      \expandafter\def\csname LT2\endcsname{\color[rgb]{0,0,1}}%
      \expandafter\def\csname LT3\endcsname{\color[rgb]{1,0,1}}%
      \expandafter\def\csname LT4\endcsname{\color[rgb]{0,1,1}}%
      \expandafter\def\csname LT5\endcsname{\color[rgb]{1,1,0}}%
      \expandafter\def\csname LT6\endcsname{\color[rgb]{0,0,0}}%
      \expandafter\def\csname LT7\endcsname{\color[rgb]{1,0.3,0}}%
      \expandafter\def\csname LT8\endcsname{\color[rgb]{0.5,0.5,0.5}}%
    \else
      \def\colorrgb#1{\color{black}}%
      \def\colorgray#1{\color[gray]{#1}}%
      \expandafter\def\csname LTw\endcsname{\color{white}}%
      \expandafter\def\csname LTb\endcsname{\color{black}}%
      \expandafter\def\csname LTa\endcsname{\color{black}}%
      \expandafter\def\csname LT0\endcsname{\color{black}}%
      \expandafter\def\csname LT1\endcsname{\color{black}}%
      \expandafter\def\csname LT2\endcsname{\color{black}}%
      \expandafter\def\csname LT3\endcsname{\color{black}}%
      \expandafter\def\csname LT4\endcsname{\color{black}}%
      \expandafter\def\csname LT5\endcsname{\color{black}}%
      \expandafter\def\csname LT6\endcsname{\color{black}}%
      \expandafter\def\csname LT7\endcsname{\color{black}}%
      \expandafter\def\csname LT8\endcsname{\color{black}}%
    \fi
  \fi
    \setlength{\unitlength}{0.0500bp}%
    \ifx\gptboxheight\undefined%
      \newlength{\gptboxheight}%
      \newlength{\gptboxwidth}%
      \newsavebox{\gptboxtext}%
    \fi%
    \setlength{\fboxrule}{0.5pt}%
    \setlength{\fboxsep}{1pt}%
    \definecolor{tbcol}{rgb}{1,1,1}%
\begin{picture}(7200.00,5040.00)%
    \gplgaddtomacro\gplbacktext{%
      \csname LTb\endcsname
      \put(946,1153){\makebox(0,0)[r]{\strut{}$0$}}%
      \csname LTb\endcsname
      \put(946,1611){\makebox(0,0)[r]{\strut{}$500$}}%
      \csname LTb\endcsname
      \put(946,2070){\makebox(0,0)[r]{\strut{}$1000$}}%
      \csname LTb\endcsname
      \put(946,2528){\makebox(0,0)[r]{\strut{}$1500$}}%
      \csname LTb\endcsname
      \put(946,2986){\makebox(0,0)[r]{\strut{}$2000$}}%
      \csname LTb\endcsname
      \put(946,3444){\makebox(0,0)[r]{\strut{}$2500$}}%
      \csname LTb\endcsname
      \put(946,3903){\makebox(0,0)[r]{\strut{}$3000$}}%
      \csname LTb\endcsname
      \put(946,4361){\makebox(0,0)[r]{\strut{}$3500$}}%
      \csname LTb\endcsname
      \put(946,4819){\makebox(0,0)[r]{\strut{}$4000$}}%
      \csname LTb\endcsname
      \put(1078,1021){\rotatebox{-45.00}{\makebox(0,0)[l]{\strut{}2022-11-06}}}%
      \csname LTb\endcsname
      \put(1794,1021){\rotatebox{-45.00}{\makebox(0,0)[l]{\strut{}2022-11-07}}}%
      \csname LTb\endcsname
      \put(2509,1021){\rotatebox{-45.00}{\makebox(0,0)[l]{\strut{}2022-11-08}}}%
      \csname LTb\endcsname
      \put(3225,1021){\rotatebox{-45.00}{\makebox(0,0)[l]{\strut{}2022-11-09}}}%
      \csname LTb\endcsname
      \put(3941,1021){\rotatebox{-45.00}{\makebox(0,0)[l]{\strut{}2022-11-10}}}%
      \csname LTb\endcsname
      \put(4656,1021){\rotatebox{-45.00}{\makebox(0,0)[l]{\strut{}2022-11-11}}}%
      \csname LTb\endcsname
      \put(5372,1021){\rotatebox{-45.00}{\makebox(0,0)[l]{\strut{}2022-11-12}}}%
      \csname LTb\endcsname
      \put(6087,1021){\rotatebox{-45.00}{\makebox(0,0)[l]{\strut{}2022-11-13}}}%
      \csname LTb\endcsname
      \put(6803,1021){\rotatebox{-45.00}{\makebox(0,0)[l]{\strut{}2022-11-14}}}%
    }%
    \gplgaddtomacro\gplfronttext{%
      \csname LTb\endcsname
      \put(209,2986){\rotatebox{-270.00}{\makebox(0,0){\strut{}Nodes}}}%
    }%
    \gplbacktext
    \put(0,0){\includegraphics[width={360.00bp},height={252.00bp}]{img//HaswellRawUtilization}}%
    \gplfronttext
  \end{picture}%
\endgroup

%% file: img/HaswellCleanUtilization.tex
\begingroup
\Large
  \makeatletter
  \providecommand\color[2][]{%
    \GenericError{(gnuplot) \space\space\space\@spaces}{%
      Package color not loaded in conjunction with
      terminal option `colourtext'%
    }{See the gnuplot documentation for explanation.%
    }{Either use 'blacktext' in gnuplot or load the package
      color.sty in LaTeX.}%
    \renewcommand\color[2][]{}%
  }%
  \providecommand\includegraphics[2][]{%
    \GenericError{(gnuplot) \space\space\space\@spaces}{%
      Package graphicx or graphics not loaded%
    }{See the gnuplot documentation for explanation.%
    }{The gnuplot epslatex terminal needs graphicx.sty or graphics.sty.}%
    \renewcommand\includegraphics[2][]{}%
  }%
  \providecommand\rotatebox[2]{#2}%
  \@ifundefined{ifGPcolor}{%
    \newif\ifGPcolor
    \GPcolorfalse
  }{}%
  \@ifundefined{ifGPblacktext}{%
    \newif\ifGPblacktext
    \GPblacktexttrue
  }{}%
  \let\gplgaddtomacro\g@addto@macro
  \gdef\gplbacktext{}%
  \gdef\gplfronttext{}%
  \makeatother
  \ifGPblacktext
    \def\colorrgb#1{}%
    \def\colorgray#1{}%
  \else
    \ifGPcolor
      \def\colorrgb#1{\color[rgb]{#1}}%
      \def\colorgray#1{\color[gray]{#1}}%
      \expandafter\def\csname LTw\endcsname{\color{white}}%
      \expandafter\def\csname LTb\endcsname{\color{black}}%
      \expandafter\def\csname LTa\endcsname{\color{black}}%
      \expandafter\def\csname LT0\endcsname{\color[rgb]{1,0,0}}%
      \expandafter\def\csname LT1\endcsname{\color[rgb]{0,1,0}}%
      \expandafter\def\csname LT2\endcsname{\color[rgb]{0,0,1}}%
      \expandafter\def\csname LT3\endcsname{\color[rgb]{1,0,1}}%
      \expandafter\def\csname LT4\endcsname{\color[rgb]{0,1,1}}%
      \expandafter\def\csname LT5\endcsname{\color[rgb]{1,1,0}}%
      \expandafter\def\csname LT6\endcsname{\color[rgb]{0,0,0}}%
      \expandafter\def\csname LT7\endcsname{\color[rgb]{1,0.3,0}}%
      \expandafter\def\csname LT8\endcsname{\color[rgb]{0.5,0.5,0.5}}%
    \else
      \def\colorrgb#1{\color{black}}%
      \def\colorgray#1{\color[gray]{#1}}%
      \expandafter\def\csname LTw\endcsname{\color{white}}%
      \expandafter\def\csname LTb\endcsname{\color{black}}%
      \expandafter\def\csname LTa\endcsname{\color{black}}%
      \expandafter\def\csname LT0\endcsname{\color{black}}%
      \expandafter\def\csname LT1\endcsname{\color{black}}%
      \expandafter\def\csname LT2\endcsname{\color{black}}%
      \expandafter\def\csname LT3\endcsname{\color{black}}%
      \expandafter\def\csname LT4\endcsname{\color{black}}%
      \expandafter\def\csname LT5\endcsname{\color{black}}%
      \expandafter\def\csname LT6\endcsname{\color{black}}%
      \expandafter\def\csname LT7\endcsname{\color{black}}%
      \expandafter\def\csname LT8\endcsname{\color{black}}%
    \fi
  \fi
    \setlength{\unitlength}{0.0500bp}%
    \ifx\gptboxheight\undefined%
      \newlength{\gptboxheight}%
      \newlength{\gptboxwidth}%
      \newsavebox{\gptboxtext}%
    \fi%
    \setlength{\fboxrule}{0.5pt}%
    \setlength{\fboxsep}{1pt}%
    \definecolor{tbcol}{rgb}{1,1,1}%
\begin{picture}(7200.00,5040.00)%
    \gplgaddtomacro\gplbacktext{%
      \csname LTb\endcsname
      \put(946,1153){\makebox(0,0)[r]{\strut{}$0$}}%
      \csname LTb\endcsname
      \put(946,1886){\makebox(0,0)[r]{\strut{}$500$}}%
      \csname LTb\endcsname
      \put(946,2619){\makebox(0,0)[r]{\strut{}$1000$}}%
      \csname LTb\endcsname
      \put(946,3353){\makebox(0,0)[r]{\strut{}$1500$}}%
      \csname LTb\endcsname
      \put(946,4086){\makebox(0,0)[r]{\strut{}$2000$}}%
      \csname LTb\endcsname
      \put(946,4819){\makebox(0,0)[r]{\strut{}$2500$}}%
      \csname LTb\endcsname
      \put(1078,1021){\rotatebox{-45.00}{\makebox(0,0)[l]{\strut{}2022-11-06}}}%
      \csname LTb\endcsname
      \put(1794,1021){\rotatebox{-45.00}{\makebox(0,0)[l]{\strut{}2022-11-07}}}%
      \csname LTb\endcsname
      \put(2509,1021){\rotatebox{-45.00}{\makebox(0,0)[l]{\strut{}2022-11-08}}}%
      \csname LTb\endcsname
      \put(3225,1021){\rotatebox{-45.00}{\makebox(0,0)[l]{\strut{}2022-11-09}}}%
      \csname LTb\endcsname
      \put(3941,1021){\rotatebox{-45.00}{\makebox(0,0)[l]{\strut{}2022-11-10}}}%
      \csname LTb\endcsname
      \put(4656,1021){\rotatebox{-45.00}{\makebox(0,0)[l]{\strut{}2022-11-11}}}%
      \csname LTb\endcsname
      \put(5372,1021){\rotatebox{-45.00}{\makebox(0,0)[l]{\strut{}2022-11-12}}}%
      \csname LTb\endcsname
      \put(6087,1021){\rotatebox{-45.00}{\makebox(0,0)[l]{\strut{}2022-11-13}}}%
      \csname LTb\endcsname
      \put(6803,1021){\rotatebox{-45.00}{\makebox(0,0)[l]{\strut{}2022-11-14}}}%
    }%
    \gplgaddtomacro\gplfronttext{%
      \csname LTb\endcsname
      \put(209,2986){\rotatebox{-270.00}{\makebox(0,0){\strut{}Nodes}}}%
    }%
    \gplbacktext
    \put(0,0){\includegraphics[width={360.00bp},height={252.00bp}]{img//HaswellCleanUtilization}}%
    \gplfronttext
  \end{picture}%
\endgroup

%% file: img/KNLRigidEasyBackfillUtilization.tex
\begingroup
\LARGE
  \makeatletter
  \providecommand\color[2][]{%
    \GenericError{(gnuplot) \space\space\space\@spaces}{%
      Package color not loaded in conjunction with
      terminal option `colourtext'%
    }{See the gnuplot documentation for explanation.%
    }{Either use 'blacktext' in gnuplot or load the package
      color.sty in LaTeX.}%
    \renewcommand\color[2][]{}%
  }%
  \providecommand\includegraphics[2][]{%
    \GenericError{(gnuplot) \space\space\space\@spaces}{%
      Package graphicx or graphics not loaded%
    }{See the gnuplot documentation for explanation.%
    }{The gnuplot epslatex terminal needs graphicx.sty or graphics.sty.}%
    \renewcommand\includegraphics[2][]{}%
  }%
  \providecommand\rotatebox[2]{#2}%
  \@ifundefined{ifGPcolor}{%
    \newif\ifGPcolor
    \GPcolortrue
  }{}%
  \@ifundefined{ifGPblacktext}{%
    \newif\ifGPblacktext
    \GPblacktextfalse
  }{}%
  \let\gplgaddtomacro\g@addto@macro
  \gdef\gplbacktext{}%
  \gdef\gplfronttext{}%
  \makeatother
  \ifGPblacktext
    \def\colorrgb#1{}%
    \def\colorgray#1{}%
  \else
    \ifGPcolor
      \def\colorrgb#1{\color[rgb]{#1}}%
      \def\colorgray#1{\color[gray]{#1}}%
      \expandafter\def\csname LTw\endcsname{\color{white}}%
      \expandafter\def\csname LTb\endcsname{\color{black}}%
      \expandafter\def\csname LTa\endcsname{\color{black}}%
      \expandafter\def\csname LT0\endcsname{\color[rgb]{1,0,0}}%
      \expandafter\def\csname LT1\endcsname{\color[rgb]{0,1,0}}%
      \expandafter\def\csname LT2\endcsname{\color[rgb]{0,0,1}}%
      \expandafter\def\csname LT3\endcsname{\color[rgb]{1,0,1}}%
      \expandafter\def\csname LT4\endcsname{\color[rgb]{0,1,1}}%
      \expandafter\def\csname LT5\endcsname{\color[rgb]{1,1,0}}%
      \expandafter\def\csname LT6\endcsname{\color[rgb]{0,0,0}}%
      \expandafter\def\csname LT7\endcsname{\color[rgb]{1,0.3,0}}%
      \expandafter\def\csname LT8\endcsname{\color[rgb]{0.5,0.5,0.5}}%
    \else
      \def\colorrgb#1{\color{black}}%
      \def\colorgray#1{\color[gray]{#1}}%
      \expandafter\def\csname LTw\endcsname{\color{white}}%
      \expandafter\def\csname LTb\endcsname{\color{black}}%
      \expandafter\def\csname LTa\endcsname{\color{black}}%
      \expandafter\def\csname LT0\endcsname{\color{black}}%
      \expandafter\def\csname LT1\endcsname{\color{black}}%
      \expandafter\def\csname LT2\endcsname{\color{black}}%
      \expandafter\def\csname LT3\endcsname{\color{black}}%
      \expandafter\def\csname LT4\endcsname{\color{black}}%
      \expandafter\def\csname LT5\endcsname{\color{black}}%
      \expandafter\def\csname LT6\endcsname{\color{black}}%
      \expandafter\def\csname LT7\endcsname{\color{black}}%
      \expandafter\def\csname LT8\endcsname{\color{black}}%
    \fi
  \fi
    \setlength{\unitlength}{0.0500bp}%
    \ifx\gptboxheight\undefined%
      \newlength{\gptboxheight}%
      \newlength{\gptboxwidth}%
      \newsavebox{\gptboxtext}%
    \fi%
    \setlength{\fboxrule}{0.5pt}%
    \setlength{\fboxsep}{1pt}%
    \definecolor{tbcol}{rgb}{1,1,1}%
\begin{picture}(14172.00,5668.00)%
    \gplgaddtomacro\gplbacktext{%
      \csname LTb\endcsname
      \put(888,1378){\makebox(0,0)[r]{\strut{}$0$}}%
      \csname LTb\endcsname
      \put(888,2188){\makebox(0,0)[r]{\strut{}$20$}}%
      \csname LTb\endcsname
      \put(888,2998){\makebox(0,0)[r]{\strut{}$40$}}%
      \csname LTb\endcsname
      \put(888,3807){\makebox(0,0)[r]{\strut{}$60$}}%
      \csname LTb\endcsname
      \put(888,4617){\makebox(0,0)[r]{\strut{}$80$}}%
      \csname LTb\endcsname
      \put(888,5427){\makebox(0,0)[r]{\strut{}$100$}}%
      \csname LTb\endcsname
      \put(1032,1234){\rotatebox{-45.00}{\makebox(0,0)[l]{\strut{}$0$}}}%
      \csname LTb\endcsname
      \put(2990,1234){\rotatebox{-45.00}{\makebox(0,0)[l]{\strut{}$100000$}}}%
      \csname LTb\endcsname
      \put(4948,1234){\rotatebox{-45.00}{\makebox(0,0)[l]{\strut{}$200000$}}}%
      \csname LTb\endcsname
      \put(6906,1234){\rotatebox{-45.00}{\makebox(0,0)[l]{\strut{}$300000$}}}%
      \csname LTb\endcsname
      \put(8863,1234){\rotatebox{-45.00}{\makebox(0,0)[l]{\strut{}$400000$}}}%
      \csname LTb\endcsname
      \put(10821,1234){\rotatebox{-45.00}{\makebox(0,0)[l]{\strut{}$500000$}}}%
      \csname LTb\endcsname
      \put(12779,1234){\rotatebox{-45.00}{\makebox(0,0)[l]{\strut{}$600000$}}}%
      \put(12923,1378){\makebox(0,0)[l]{\strut{}$0$}}%
      \put(12923,1884){\makebox(0,0)[l]{\strut{}$50$}}%
      \put(12923,2390){\makebox(0,0)[l]{\strut{}$100$}}%
      \put(12923,2896){\makebox(0,0)[l]{\strut{}$150$}}%
      \put(12923,3403){\makebox(0,0)[l]{\strut{}$200$}}%
      \put(12923,3909){\makebox(0,0)[l]{\strut{}$250$}}%
      \put(12923,4415){\makebox(0,0)[l]{\strut{}$300$}}%
      \put(12923,4921){\makebox(0,0)[l]{\strut{}$350$}}%
      \put(12923,5427){\makebox(0,0)[l]{\strut{}$400$}}%
    }%
    \gplgaddtomacro\gplfronttext{%
      \csname LTb\endcsname
      \put(228,3402){\rotatebox{-270.00}{\makebox(0,0){\strut{}Nodeutilization (\%)}}}%
      \put(13619,3402){\rotatebox{-270.00}{\makebox(0,0){\strut{}Jobs in Queue (\#)}}}%
      \put(6905,168){\makebox(0,0){\strut{}Time (s)}}%
    }%
    \gplbacktext
    \put(0,0){\includegraphics[width={708.60bp},height={283.40bp}]{img//KNLRigidEasyBackfillUtilization}}%
    \gplfronttext
  \end{picture}%
\endgroup

%% file: 02prep.tex
\subsection{Workload Preparation}\label{subsec:03-workload-preparation}
We utilize real-world workload traces from the following supercomputers:
\begin{itemize}
    \item \textbf{Theta~\cite{ALCF-Theta}:} Operated by ALCF (2017--2023) with \num{4392} Xeon Phi 7230 nodes, reaching 11.7 PetaFLOPS (Top500 rank 16, 06/2017).
    24~GPU nodes excluded.

    \item \textbf{Eagle~\cite{NREL-Eagle}:} Operated by NREL (2018--2024) with \num{2618} Skylake 6154 nodes, delivering 8 PetaFLOPS (Top500 rank 35, 11/2018).
    50~GPU nodes excluded.

    \item \textbf{Cori~\cite{NERSC-Cori}:} Operated by NERSC (2015--2023) with two partitions: \haswell (\num{2388} Xeon E5-2698 nodes, 2.81 PetaFLOPS) and \knl (\num{9688} Xeon Phi 7250 nodes, 29.5 PetaFLOPS).
\end{itemize}

The \cori datasets required substantial preprocessing due to structural inconsistencies.
Jobs were recorded in daily segments, artificially inflating node utilization due to split entries and shared-node (oversubscribed) execution.
As shown in Figure~\ref{fig:cori-raw}, raw \haswell data showed utilization exceeding \num{4000} nodes, well above the actual capacity of \num{2388}.
To correct this, we merged split entries and removed all shared-node jobs, resulting in the utilization shown in Figure~\ref{fig:cori-clean}.

Table~\ref{tab:csv-filter} summarizes our job filtering results.
\haswell experienced the largest reduction, primarily due to the exclusion of shared-node jobs.
Although this significantly reduced the job count, the removed jobs typically consumed only partial node resources while contributing disproportionately to total runtime, and their removal provides a more accurate picture of actual system utilization.
In contrast, \knl and \eagle required minimal filtering, and \theta remained unmodified, as it contained no shared-node jobs, GPUs, or structural anomalies.

For simulations, we used ElastiSim~\cite{ElastiSim}, an open-source simulator for malleable job scheduling
ElastiSim takes a set of jobs as input and simulates their execution on a supercomputer using a given scheduling algorithm.
To integrate with ElastiSim, job data were converted to JSON with metadata (job ID, submission time, nodes, runtime, etc.).
Missing time limits were set to 125\% of runtime based on empirical averages.
Based on previous work, malleable jobs received min/max/preferred node counts using a speedup model with efficiency thresholds to ensure realistic scaling behavior~\cite{JonasMall21}.

For \theta and \eagle, we selected representative time windows to balance data quality and simulation feasibility.
Anomalous periods, including scheduled maintenance, end-of-year test jobs, or irregular submission spikes, were excluded.
Segment selection was guided by statistical analysis and visual inspection of job and utilization distributions.
The final time windows are detailed in Section~\ref{subsec:03-Experimental-Design}, with workload characteristics discussed in Section~\ref{sec:04-Evaluation}.

%% file: 02design.tex
\subsection{Experimental Design}\label{subsec:03-Experimental-Design}
Each workload is simulated with increasing proportions of malleable jobs: 0\%, 20\%, \dots, 100\%.
For each proportion, we perform 10 simulation runs using different pseudo-random seeds to determine which jobs are converted to malleable variants.
All metrics reported in Section~\ref{sec:04-Evaluation} are averaged over these 10 runs.
We report interquartile ranges (IQR) as error bars, as real workload metrics often exhibit non-normal distributions, making IQR preferable to standard deviation.

To ensure fair comparisons, we exclude the first 12 hours of each simulation as a warm-up phase, which builds a representative job queue but behaves atypically.
We also exclude the time after the last job submission to avoid skewing metrics due to drain-down.
Figure~\ref{fig:cori-knl-rigid-node-util} illustrates this approach for \knl, where only data between the red lines (after warm-up, before drain-down) are used for analysis.

Table~\ref{tab:sim-configs} summarizes simulation parameters: start time, duration, job count, node count, and simulator tick rate.
The tick rate defines the interval between scheduling points and directly impacts simulation runtime.
For larger workloads like \eagle, a \num{10}\,s tick is used.
Smaller workloads such as \haswell are simulated at \num{1}\,s for finer granularity.
These values have a negligible impact on results but enable practical completion times.
Reconfiguration costs are not explicitly modeled, but the tick rate introduces idle times that approximate overheads based on prior work~\cite{DPPvsAPGASWAMTA25} (i.e., 2--4\,s for add/remove 8~nodes).

%% file: 03evaluation.tex
\section{Evaluation}\label{sec:04-Evaluation}
This section presents simulation results, analyzing how real HPC workloads respond to varying malleability proportions under different scheduling strategies.
Sections~\ref{subsec:04-cori-haswell}--\ref{subsec:04-theta} examine each workload individually.
Section~\ref{subsec:04-cross} discusses trends across all workloads and schedulers.

\input{03cori-haswell-figs-1}
\input{03cori-haswell}

\begin{center}
    \begin{minipage}{\linewidth}
        \centering
        \resizebox{0.99\linewidth}{!}{\input{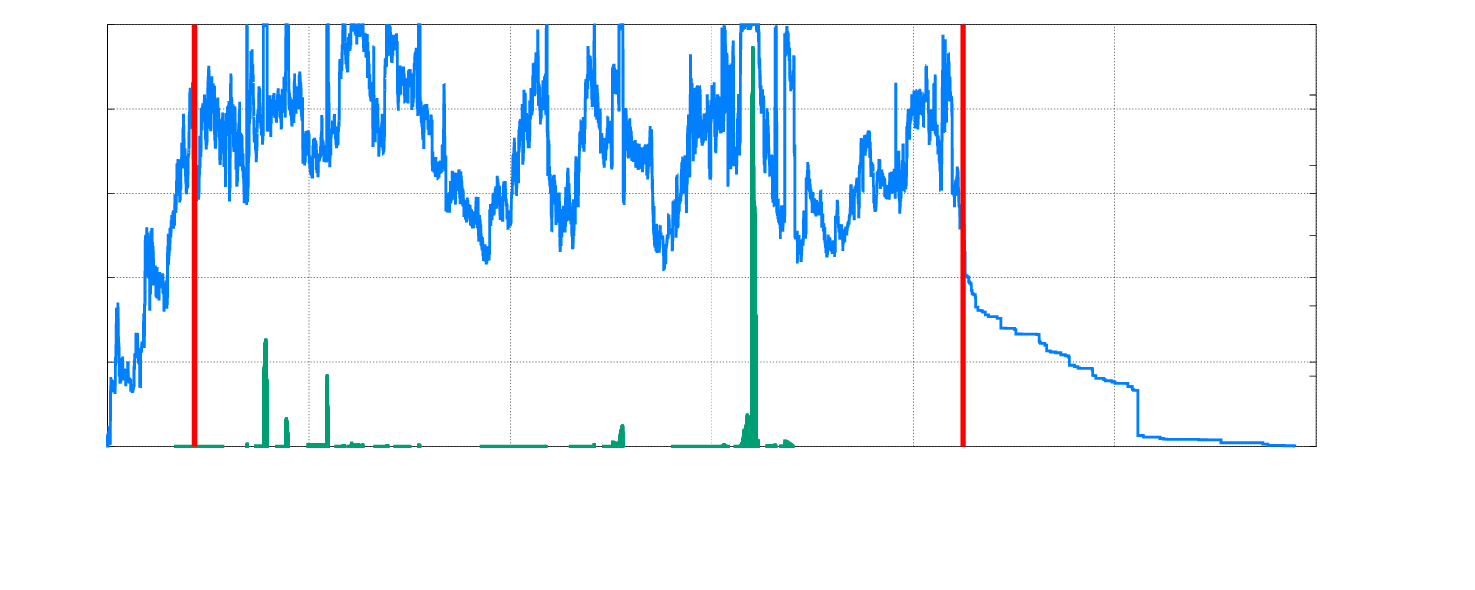}}

        \vspace*{0pt}
        \captionof{figure}{
            \haswell: Node utilization with 100\% rigid jobs.
            Red lines mark warm-up and final submission; metrics are computed in between.
        }
        \label{fig:cori-haswell-rigid-node-util}
    \end{minipage}
\end{center}

\input{03cori-knl}
\input{03eagle}
\input{03theta}
\input{03cross}
\input{03jobs-figs}
\input{03cori-both-figs-2}
\input{03eagle-theta-figs-2}

%% file: 03cori-haswell-figs-1.tex
\begin{center}
    \begin{figure}
        \centering

        \begin{subfigure}{.49\linewidth}
            \centering
            \resizebox{\linewidth}{!}{\input{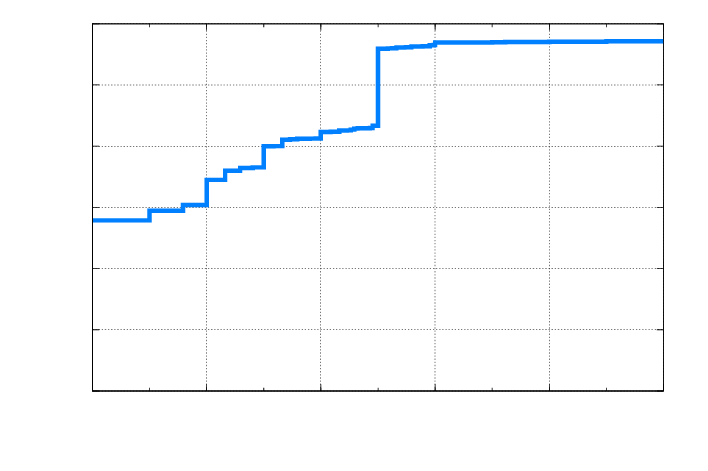}}
            \caption{Number of nodes required by jobs}
            \label{fig:haswell-nodes}
        \end{subfigure}
        \hfill
        \begin{subfigure}{.49\linewidth}
            \centering
            \resizebox{\linewidth}{!}{\input{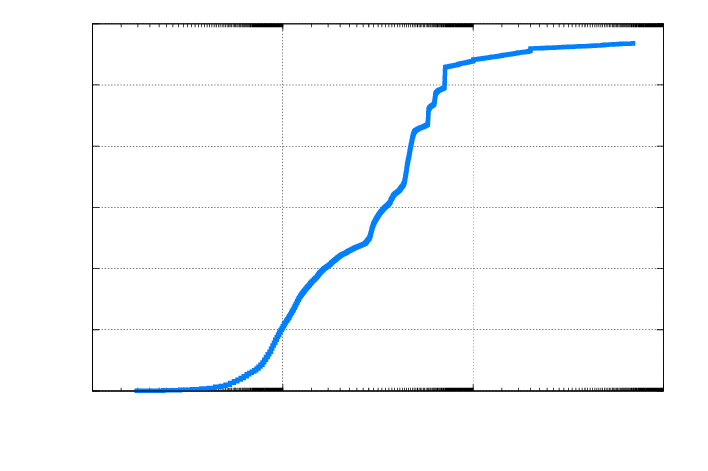}}
            \subcaption{Runtime of jobs}
            \label{fig:haswell-runtimes}
        \end{subfigure}

        \vspace*{1.2\baselineskip}

        \caption{
            \haswell: Node counts and runtimes---most jobs are small and short.
        }
        \label{fig:haswell-stats}
    \end{figure}
\end{center}
\vspace*{-2em}

%% file: img/HaswellNodesAllocated.tex
\begingroup
\LARGE
  \makeatletter
  \providecommand\color[2][]{%
    \GenericError{(gnuplot) \space\space\space\@spaces}{%
      Package color not loaded in conjunction with
      terminal option `colourtext'%
    }{See the gnuplot documentation for explanation.%
    }{Either use 'blacktext' in gnuplot or load the package
      color.sty in LaTeX.}%
    \renewcommand\color[2][]{}%
  }%
  \providecommand\includegraphics[2][]{%
    \GenericError{(gnuplot) \space\space\space\@spaces}{%
      Package graphicx or graphics not loaded%
    }{See the gnuplot documentation for explanation.%
    }{The gnuplot epslatex terminal needs graphicx.sty or graphics.sty.}%
    \renewcommand\includegraphics[2][]{}%
  }%
  \providecommand\rotatebox[2]{#2}%
  \@ifundefined{ifGPcolor}{%
    \newif\ifGPcolor
    \GPcolortrue
  }{}%
  \@ifundefined{ifGPblacktext}{%
    \newif\ifGPblacktext
    \GPblacktextfalse
  }{}%
  \let\gplgaddtomacro\g@addto@macro
  \gdef\gplbacktext{}%
  \gdef\gplfronttext{}%
  \makeatother
  \ifGPblacktext
    \def\colorrgb#1{}%
    \def\colorgray#1{}%
  \else
    \ifGPcolor
      \def\colorrgb#1{\color[rgb]{#1}}%
      \def\colorgray#1{\color[gray]{#1}}%
      \expandafter\def\csname LTw\endcsname{\color{white}}%
      \expandafter\def\csname LTb\endcsname{\color{black}}%
      \expandafter\def\csname LTa\endcsname{\color{black}}%
      \expandafter\def\csname LT0\endcsname{\color[rgb]{1,0,0}}%
      \expandafter\def\csname LT1\endcsname{\color[rgb]{0,1,0}}%
      \expandafter\def\csname LT2\endcsname{\color[rgb]{0,0,1}}%
      \expandafter\def\csname LT3\endcsname{\color[rgb]{1,0,1}}%
      \expandafter\def\csname LT4\endcsname{\color[rgb]{0,1,1}}%
      \expandafter\def\csname LT5\endcsname{\color[rgb]{1,1,0}}%
      \expandafter\def\csname LT6\endcsname{\color[rgb]{0,0,0}}%
      \expandafter\def\csname LT7\endcsname{\color[rgb]{1,0.3,0}}%
      \expandafter\def\csname LT8\endcsname{\color[rgb]{0.5,0.5,0.5}}%
    \else
      \def\colorrgb#1{\color{black}}%
      \def\colorgray#1{\color[gray]{#1}}%
      \expandafter\def\csname LTw\endcsname{\color{white}}%
      \expandafter\def\csname LTb\endcsname{\color{black}}%
      \expandafter\def\csname LTa\endcsname{\color{black}}%
      \expandafter\def\csname LT0\endcsname{\color{black}}%
      \expandafter\def\csname LT1\endcsname{\color{black}}%
      \expandafter\def\csname LT2\endcsname{\color{black}}%
      \expandafter\def\csname LT3\endcsname{\color{black}}%
      \expandafter\def\csname LT4\endcsname{\color{black}}%
      \expandafter\def\csname LT5\endcsname{\color{black}}%
      \expandafter\def\csname LT6\endcsname{\color{black}}%
      \expandafter\def\csname LT7\endcsname{\color{black}}%
      \expandafter\def\csname LT8\endcsname{\color{black}}%
    \fi
  \fi
    \setlength{\unitlength}{0.0500bp}%
    \ifx\gptboxheight\undefined%
      \newlength{\gptboxheight}%
      \newlength{\gptboxwidth}%
      \newsavebox{\gptboxtext}%
    \fi%
    \setlength{\fboxrule}{0.5pt}%
    \setlength{\fboxsep}{1pt}%
    \definecolor{tbcol}{rgb}{1,1,1}%
\begin{picture}(6802.00,4534.00)%
    \gplgaddtomacro\gplbacktext{%
      \csname LTb\endcsname
      \put(744,768){\makebox(0,0)[r]{\strut{}$0$}}%
      \csname LTb\endcsname
      \put(744,1356){\makebox(0,0)[r]{\strut{}$5$}}%
      \csname LTb\endcsname
      \put(744,1943){\makebox(0,0)[r]{\strut{}$10$}}%
      \csname LTb\endcsname
      \put(744,2531){\makebox(0,0)[r]{\strut{}$15$}}%
      \csname LTb\endcsname
      \put(744,3118){\makebox(0,0)[r]{\strut{}$20$}}%
      \csname LTb\endcsname
      \put(744,3706){\makebox(0,0)[r]{\strut{}$25$}}%
      \csname LTb\endcsname
      \put(744,4293){\makebox(0,0)[r]{\strut{}$30$}}%
      \csname LTb\endcsname
      \put(888,528){\makebox(0,0){\strut{}$1$}}%
      \csname LTb\endcsname
      \put(1984,528){\makebox(0,0){\strut{}$4$}}%
      \csname LTb\endcsname
      \put(3080,528){\makebox(0,0){\strut{}$16$}}%
      \csname LTb\endcsname
      \put(4177,528){\makebox(0,0){\strut{}$64$}}%
      \csname LTb\endcsname
      \put(5273,528){\makebox(0,0){\strut{}$256$}}%
      \csname LTb\endcsname
      \put(6369,528){\makebox(0,0){\strut{}$1024$}}%
    }%
    \gplgaddtomacro\gplfronttext{%
      \csname LTb\endcsname
      \put(228,2530){\rotatebox{-270.00}{\makebox(0,0){\strut{}Jobs [×1000] cumulative}}}%
      \put(3628,168){\makebox(0,0){\strut{}Required nodes}}%
    }%
    \gplbacktext
    \put(0,0){\includegraphics[width={340.10bp},height={226.70bp}]{img//HaswellNodesAllocated}}%
    \gplfronttext
  \end{picture}%
\endgroup

%% file: img/HaswellRuntimeSeconds.tex
\begingroup
\LARGE
  \makeatletter
  \providecommand\color[2][]{%
    \GenericError{(gnuplot) \space\space\space\@spaces}{%
      Package color not loaded in conjunction with
      terminal option `colourtext'%
    }{See the gnuplot documentation for explanation.%
    }{Either use 'blacktext' in gnuplot or load the package
      color.sty in LaTeX.}%
    \renewcommand\color[2][]{}%
  }%
  \providecommand\includegraphics[2][]{%
    \GenericError{(gnuplot) \space\space\space\@spaces}{%
      Package graphicx or graphics not loaded%
    }{See the gnuplot documentation for explanation.%
    }{The gnuplot epslatex terminal needs graphicx.sty or graphics.sty.}%
    \renewcommand\includegraphics[2][]{}%
  }%
  \providecommand\rotatebox[2]{#2}%
  \@ifundefined{ifGPcolor}{%
    \newif\ifGPcolor
    \GPcolortrue
  }{}%
  \@ifundefined{ifGPblacktext}{%
    \newif\ifGPblacktext
    \GPblacktextfalse
  }{}%
  \let\gplgaddtomacro\g@addto@macro
  \gdef\gplbacktext{}%
  \gdef\gplfronttext{}%
  \makeatother
  \ifGPblacktext
    \def\colorrgb#1{}%
    \def\colorgray#1{}%
  \else
    \ifGPcolor
      \def\colorrgb#1{\color[rgb]{#1}}%
      \def\colorgray#1{\color[gray]{#1}}%
      \expandafter\def\csname LTw\endcsname{\color{white}}%
      \expandafter\def\csname LTb\endcsname{\color{black}}%
      \expandafter\def\csname LTa\endcsname{\color{black}}%
      \expandafter\def\csname LT0\endcsname{\color[rgb]{1,0,0}}%
      \expandafter\def\csname LT1\endcsname{\color[rgb]{0,1,0}}%
      \expandafter\def\csname LT2\endcsname{\color[rgb]{0,0,1}}%
      \expandafter\def\csname LT3\endcsname{\color[rgb]{1,0,1}}%
      \expandafter\def\csname LT4\endcsname{\color[rgb]{0,1,1}}%
      \expandafter\def\csname LT5\endcsname{\color[rgb]{1,1,0}}%
      \expandafter\def\csname LT6\endcsname{\color[rgb]{0,0,0}}%
      \expandafter\def\csname LT7\endcsname{\color[rgb]{1,0.3,0}}%
      \expandafter\def\csname LT8\endcsname{\color[rgb]{0.5,0.5,0.5}}%
    \else
      \def\colorrgb#1{\color{black}}%
      \def\colorgray#1{\color[gray]{#1}}%
      \expandafter\def\csname LTw\endcsname{\color{white}}%
      \expandafter\def\csname LTb\endcsname{\color{black}}%
      \expandafter\def\csname LTa\endcsname{\color{black}}%
      \expandafter\def\csname LT0\endcsname{\color{black}}%
      \expandafter\def\csname LT1\endcsname{\color{black}}%
      \expandafter\def\csname LT2\endcsname{\color{black}}%
      \expandafter\def\csname LT3\endcsname{\color{black}}%
      \expandafter\def\csname LT4\endcsname{\color{black}}%
      \expandafter\def\csname LT5\endcsname{\color{black}}%
      \expandafter\def\csname LT6\endcsname{\color{black}}%
      \expandafter\def\csname LT7\endcsname{\color{black}}%
      \expandafter\def\csname LT8\endcsname{\color{black}}%
    \fi
  \fi
    \setlength{\unitlength}{0.0500bp}%
    \ifx\gptboxheight\undefined%
      \newlength{\gptboxheight}%
      \newlength{\gptboxwidth}%
      \newsavebox{\gptboxtext}%
    \fi%
    \setlength{\fboxrule}{0.5pt}%
    \setlength{\fboxsep}{1pt}%
    \definecolor{tbcol}{rgb}{1,1,1}%
\begin{picture}(6802.00,4534.00)%
    \gplgaddtomacro\gplbacktext{%
      \csname LTb\endcsname
      \put(744,768){\makebox(0,0)[r]{\strut{}$0$}}%
      \csname LTb\endcsname
      \put(744,1356){\makebox(0,0)[r]{\strut{}$5$}}%
      \csname LTb\endcsname
      \put(744,1943){\makebox(0,0)[r]{\strut{}$10$}}%
      \csname LTb\endcsname
      \put(744,2531){\makebox(0,0)[r]{\strut{}$15$}}%
      \csname LTb\endcsname
      \put(744,3118){\makebox(0,0)[r]{\strut{}$20$}}%
      \csname LTb\endcsname
      \put(744,3706){\makebox(0,0)[r]{\strut{}$25$}}%
      \csname LTb\endcsname
      \put(744,4293){\makebox(0,0)[r]{\strut{}$30$}}%
      \csname LTb\endcsname
      \put(888,528){\makebox(0,0){\strut{}$0.0001$}}%
      \csname LTb\endcsname
      \put(2715,528){\makebox(0,0){\strut{}$0.01$}}%
      \csname LTb\endcsname
      \put(4542,528){\makebox(0,0){\strut{}$1$}}%
      \csname LTb\endcsname
      \put(6369,528){\makebox(0,0){\strut{}$100$}}%
    }%
    \gplgaddtomacro\gplfronttext{%
      \csname LTb\endcsname
      \put(228,2530){\rotatebox{-270.00}{\makebox(0,0){\strut{}Jobs [×1000] cumulative}}}%
      \put(3628,168){\makebox(0,0){\strut{}Runtime [hours]}}%
    }%
    \gplbacktext
    \put(0,0){\includegraphics[width={340.10bp},height={226.70bp}]{img//HaswellRuntimeSeconds}}%
    \gplfronttext
  \end{picture}%
\endgroup

%% file: 03cori-haswell.tex
\subsection{Cori Haswell}\label{subsec:04-cori-haswell}
Figure~\ref{fig:haswell-stats} illustrates the characteristics of the \haswell workload.
The distribution is dominated by small-scale jobs: 97.8\% request $\leq$\,32 nodes, with 50\% requesting a single node (Fig.\,\ref{fig:haswell-nodes}).
In terms of runtime, 75\% of jobs complete within \num{1000}\,s (Fig.\,\ref{fig:haswell-runtimes}), consistent with the high throughput of \num{28259} jobs over the five-day simulation period.

Figure~\ref{fig:cori-haswell-rigid-node-util} shows node utilization under the rigid baseline.
A job burst at \num{300000}\,s leads to elevated queue pressure and corresponding increases in wait times.

Figure~\ref{fig:cori-haswell-all-metrics} presents the impact of increasing malleability.
Job turnaround time (Fig.\,\ref{fig:cori-haswell-statistics-turnaround}) decreases consistently with increasing malleability, dropping from \num{2391}\,s to \num{792}\,s---a 66.87\% improvement.
Job makespan (Fig.\,\ref{fig:cori-haswell-statistics-makespan}) follows a similar trend, improving by up to 65.32\%.

Job wait time (Fig.\,\ref{fig:cori-haswell-statistics-waittime}) improves by up to 99.71\% and stabilizes below \num{10}\,s once malleability reaches 40\%.
While \prefkeeper shows slightly higher average wait times, the IQR reveals that most jobs wait less than one second even at modest 20\% malleability.
The previously identified job submission burst (Fig.\,\ref{fig:cori-haswell-rigid-node-util}) contributes to occasional high-value outliers.

Node utilization (Fig.,\ref{fig:cori-haswell-statistics-nodeutil}) rises from 72.33\% to 98.73\% with increasing malleability.
Variability across seeds decreases, indicating consistent resource usage.

Figures~\ref{fig:cori-haswell-expands} and~\ref{fig:cori-haswell-shrinks} show expand and shrink operations per job, respectively.
Expand operations are most frequent with \prefkeeper, which averages over 11~expand operations per job.
In contrast, \avg remains steady at $\approx$\,5 expand operations.
Shrink operations start at $\approx$\,25 per job across all schedulers at low malleability.
Only \avg and \pref show clear declines as malleability increases, while \min and \prefkeeper maintain higher shrink rates, indicating a more aggressive approach.

%% file: img/HaswellRigidEasyBackfillUtilization.tex
\begingroup
\LARGE
  \makeatletter
  \providecommand\color[2][]{%
    \GenericError{(gnuplot) \space\space\space\@spaces}{%
      Package color not loaded in conjunction with
      terminal option `colourtext'%
    }{See the gnuplot documentation for explanation.%
    }{Either use 'blacktext' in gnuplot or load the package
      color.sty in LaTeX.}%
    \renewcommand\color[2][]{}%
  }%
  \providecommand\includegraphics[2][]{%
    \GenericError{(gnuplot) \space\space\space\@spaces}{%
      Package graphicx or graphics not loaded%
    }{See the gnuplot documentation for explanation.%
    }{The gnuplot epslatex terminal needs graphicx.sty or graphics.sty.}%
    \renewcommand\includegraphics[2][]{}%
  }%
  \providecommand\rotatebox[2]{#2}%
  \@ifundefined{ifGPcolor}{%
    \newif\ifGPcolor
    \GPcolortrue
  }{}%
  \@ifundefined{ifGPblacktext}{%
    \newif\ifGPblacktext
    \GPblacktextfalse
  }{}%
  \let\gplgaddtomacro\g@addto@macro
  \gdef\gplbacktext{}%
  \gdef\gplfronttext{}%
  \makeatother
  \ifGPblacktext
    \def\colorrgb#1{}%
    \def\colorgray#1{}%
  \else
    \ifGPcolor
      \def\colorrgb#1{\color[rgb]{#1}}%
      \def\colorgray#1{\color[gray]{#1}}%
      \expandafter\def\csname LTw\endcsname{\color{white}}%
      \expandafter\def\csname LTb\endcsname{\color{black}}%
      \expandafter\def\csname LTa\endcsname{\color{black}}%
      \expandafter\def\csname LT0\endcsname{\color[rgb]{1,0,0}}%
      \expandafter\def\csname LT1\endcsname{\color[rgb]{0,1,0}}%
      \expandafter\def\csname LT2\endcsname{\color[rgb]{0,0,1}}%
      \expandafter\def\csname LT3\endcsname{\color[rgb]{1,0,1}}%
      \expandafter\def\csname LT4\endcsname{\color[rgb]{0,1,1}}%
      \expandafter\def\csname LT5\endcsname{\color[rgb]{1,1,0}}%
      \expandafter\def\csname LT6\endcsname{\color[rgb]{0,0,0}}%
      \expandafter\def\csname LT7\endcsname{\color[rgb]{1,0.3,0}}%
      \expandafter\def\csname LT8\endcsname{\color[rgb]{0.5,0.5,0.5}}%
    \else
      \def\colorrgb#1{\color{black}}%
      \def\colorgray#1{\color[gray]{#1}}%
      \expandafter\def\csname LTw\endcsname{\color{white}}%
      \expandafter\def\csname LTb\endcsname{\color{black}}%
      \expandafter\def\csname LTa\endcsname{\color{black}}%
      \expandafter\def\csname LT0\endcsname{\color{black}}%
      \expandafter\def\csname LT1\endcsname{\color{black}}%
      \expandafter\def\csname LT2\endcsname{\color{black}}%
      \expandafter\def\csname LT3\endcsname{\color{black}}%
      \expandafter\def\csname LT4\endcsname{\color{black}}%
      \expandafter\def\csname LT5\endcsname{\color{black}}%
      \expandafter\def\csname LT6\endcsname{\color{black}}%
      \expandafter\def\csname LT7\endcsname{\color{black}}%
      \expandafter\def\csname LT8\endcsname{\color{black}}%
    \fi
  \fi
    \setlength{\unitlength}{0.0500bp}%
    \ifx\gptboxheight\undefined%
      \newlength{\gptboxheight}%
      \newlength{\gptboxwidth}%
      \newsavebox{\gptboxtext}%
    \fi%
    \setlength{\fboxrule}{0.5pt}%
    \setlength{\fboxsep}{1pt}%
    \definecolor{tbcol}{rgb}{1,1,1}%
\begin{picture}(14172.00,5668.00)%
    \gplgaddtomacro\gplbacktext{%
      \csname LTb\endcsname
      \put(888,1378){\makebox(0,0)[r]{\strut{}$0$}}%
      \csname LTb\endcsname
      \put(888,2188){\makebox(0,0)[r]{\strut{}$20$}}%
      \csname LTb\endcsname
      \put(888,2998){\makebox(0,0)[r]{\strut{}$40$}}%
      \csname LTb\endcsname
      \put(888,3807){\makebox(0,0)[r]{\strut{}$60$}}%
      \csname LTb\endcsname
      \put(888,4617){\makebox(0,0)[r]{\strut{}$80$}}%
      \csname LTb\endcsname
      \put(888,5427){\makebox(0,0)[r]{\strut{}$100$}}%
      \csname LTb\endcsname
      \put(1032,1234){\rotatebox{-45.00}{\makebox(0,0)[l]{\strut{}$0$}}}%
      \csname LTb\endcsname
      \put(2966,1234){\rotatebox{-45.00}{\makebox(0,0)[l]{\strut{}$100000$}}}%
      \csname LTb\endcsname
      \put(4900,1234){\rotatebox{-45.00}{\makebox(0,0)[l]{\strut{}$200000$}}}%
      \csname LTb\endcsname
      \put(6834,1234){\rotatebox{-45.00}{\makebox(0,0)[l]{\strut{}$300000$}}}%
      \csname LTb\endcsname
      \put(8767,1234){\rotatebox{-45.00}{\makebox(0,0)[l]{\strut{}$400000$}}}%
      \csname LTb\endcsname
      \put(10701,1234){\rotatebox{-45.00}{\makebox(0,0)[l]{\strut{}$500000$}}}%
      \csname LTb\endcsname
      \put(12635,1234){\rotatebox{-45.00}{\makebox(0,0)[l]{\strut{}$600000$}}}%
      \put(12779,1378){\makebox(0,0)[l]{\strut{}$0$}}%
      \put(12779,2053){\makebox(0,0)[l]{\strut{}$200$}}%
      \put(12779,2728){\makebox(0,0)[l]{\strut{}$400$}}%
      \put(12779,3403){\makebox(0,0)[l]{\strut{}$600$}}%
      \put(12779,4077){\makebox(0,0)[l]{\strut{}$800$}}%
      \put(12779,4752){\makebox(0,0)[l]{\strut{}$1000$}}%
      \put(12779,5427){\makebox(0,0)[l]{\strut{}$1200$}}%
    }%
    \gplgaddtomacro\gplfronttext{%
      \csname LTb\endcsname
      \put(228,3402){\rotatebox{-270.00}{\makebox(0,0){\strut{}Nodeutilization (\%)}}}%
      \put(13619,3402){\rotatebox{-270.00}{\makebox(0,0){\strut{}Jobs in Queue (\#)}}}%
      \put(6833,168){\makebox(0,0){\strut{}Time (s)}}%
    }%
    \gplbacktext
    \put(0,0){\includegraphics[width={708.60bp},height={283.40bp}]{img//HaswellRigidEasyBackfillUtilization}}%
    \gplfronttext
  \end{picture}%
\endgroup

%% file: 03cori-knl.tex
\subsection{Cori KNL}\label{subsec:04-cori-knl}
Like \haswell, \knl is dominated by small jobs: 94.4\% request $\leq$\,32 nodes, with 63\% requesting exactly four nodes (Fig.\,\ref{fig:knl-nodes}).
Job runtimes are similarly compact, with 80\% of jobs completing within \num{1000}\,s and a visible cluster between \num{600}--\num{800}\,s (Fig.\,\ref{fig:knl-runtimes}).
The simulation window includes \num{41524} jobs submitted over five days.

Figure~\ref{fig:cori-knl-all-metrics} presents the impact of increasing malleability.
Job turnaround time (Fig.\,\ref{fig:cori-knl-statistics-turnaround}) improves with increasing malleability, dropping from \num{4551}\,s to \num{1696}\,s---a 62.74\% improvement.
Job makespan (Fig.\,\ref{fig:cori-knl-statistics-makespan}) follows a similar trend, improving by up to 60.87\%.

Job wait time (Fig.\,\ref{fig:cori-knl-statistics-waittime}) improves by up to 98.81\% and falls below \num{20}\,s at $\geq$\,40\% malleability.
Notably, \prefkeeper shows higher averages and wider IQRs, particularly at 40\% and 60\% malleability.
This behavior correlates with job submission spikes observed in node utilization (see Fig.\,\ref{fig:cori-knl-rigid-node-util} in Section~\ref{subsec:03-Experimental-Design}).

Node utilization (Fig.\,\ref{fig:cori-knl-statistics-nodeutil}) increases from 85.5\% to $\approx$\,95\% at 60\% malleability, with minimal additional gains thereafter.
The progressively narrowing IQR indicates more consistent system usage as malleability increases.

Figures~\ref{fig:cori-knl-expands} and~\ref{fig:cori-knl-shrinks} show expand and shrink operations per job, respectively.
Expand operations decrease steadily as malleability increases, with \prefkeeper and \avg exhibiting more than twice the frequency of other strategies.
Shrink operations remain fairly stable at $\approx$\,20 per job, except for \min, which increases slightly---indicating more aggressive downscaling as malleability increases.

%% file: 03eagle.tex
\subsection{Eagle}\label{subsec:04-eagle}
Figures~\ref{fig:eagle-nodes} and~\ref{fig:eagle-runtimes} show that the \eagle workload is homogeneous: 96.6\% of jobs request exactly one node, and 86.8\% complete within \num{10000}\,s.

Job turnaround time (Fig.\,\ref{fig:eagle-statistics-turnaround}) improves by 65.55\%, dropping from \num{6138}\,s to \num{2114}\,s.
Although the bars appear similar due to axis scaling, the large IQR in the rigid case (\num{29377}\,s) highlights a wide spread in job turnaround times.
Job makespan (Fig.\,\ref{fig:eagle-statistics-makespan}) follows a similar pattern, decreasing by up to 65.43\%.

Job wait time (Fig.\,\ref{fig:eagle-statistics-waittime}) decreases from \num{330}\,s to \num{89}\,s---a 73.17\% improvement.
The reduced IQR indicates that this benefit extends beyond the average case.

Node utilization (Fig.\,\ref{fig:eagle-statistics-nodeutil}) increases from 28.71\% to 43.61\%.
The low initial value suggests that underutilization is primarily due to workload structure rather than scheduling inefficiencies.

Figures~\ref{fig:eagle-expands} and~\ref{fig:eagle-shrinks} show expand and shrink operations per job, respectively.
Expand operations decrease moderately with increasing malleability: \prefkeeper and \avg drop from 8--9 expands per job to 6--7, while \pref and \min drop from 3 to 1--2.
Shrink operations decline slightly across all strategies, with the most notable reduction occurring between 20\% and 40\% malleability.

%% file: 03theta.tex
\subsection{Theta}\label{subsec:04-theta}
For \theta, node requests concentrate at 1 (34.8\%), 8 (20.3\%), and 256 nodes (12.6\%) (Fig.\,\ref{fig:theta-nodes}).
A total of 84.7\% of jobs complete within \num{10000}\,s (Fig.\,\ref{fig:theta-runtimes}).

Job turnaround time (Fig.\,\ref{fig:theta-statistics-turnaround}) improves by up to 36.91\% for three of the four strategies, while \pref shows a slight increase for $\geq$\,60\% malleability---an anomaly unique to this workload.
Job makespan (Fig.\,\ref{fig:theta-statistics-makespan}) follows a similar pattern: only \prefkeeper achieves consistent improvements (up to 16.3\%), while the remaining strategies \textit{degrade} performance by up to 60.85\%.

Job wait time (Fig.\,\ref{fig:theta-statistics-waittime}) decreases by up 99.98\% for all strategies except \prefkeeper, which achieves only a 16.53\% reduction.
This suggests limited adaptability of \prefkeeper to the workload's heterogeneity.
Node utilization (Fig.\,\ref{fig:theta-statistics-nodeutil}) further confirms this divergence.
\prefkeeper slightly increases utilization (72.67\% to 76.26\%), whereas other strategies exhibit minor drops (up to 3.82\%).

Figures~\ref{fig:theta-expands} and~\ref{fig:theta-shrinks} show expand and shrink operations per job, respectively.
Expand operations remain stable across all malleability proportions, with \prefkeeper maintaining 6 expand operations per job.
Shrink operations also vary little, but \prefkeeper consistently triggers about 50\% fewer shrink operations than the other strategies, reflecting its more conservative reallocation policy.

%% file: 03cross.tex
\subsection{Cross-Workload Summary}\label{subsec:04-cross}
To enable meaningful comparisons across workloads with differing timeframes, Table~\ref{tab:submission-rate} reports job submission rates in \textit{jobs per hour}.
Among the four supercomputers, the \theta workload exhibits the lowest job submission rate, reflecting a workload dominated by fewer, longer-running jobs with higher node requirements.
In contrast, the \eagle workload consists of many short, single-node jobs, yet having a moderate job submission rate, which explains its persistently low node utilization.
The \haswell and \knl workloads show both high job submission rates and the most consistent improvements across all evaluation metrics.

Although this work primarily aims to evaluate the impact of malleability on real workloads, we also propose \prefkeeper, a novel strategy that preserves each job's number of preferred nodes where feasible.
The differences observed among strategies yield valuable insights into malleable scheduling behavior.

\easybackfill serves as the rigid baseline and, as expected, performs worst in malleable settings.
The three malleable strategies (\avg, \min, and \pref) differ in how they redistribute nodes: either based on average, minimum, or preferred number of nodes.
In contrast, \prefkeeper applies a more conservative approach, shrinking only when necessary and prioritizing stability in resource allocation.

Across all workloads except \theta, malleability results in substantial gains in job turnaround time, job wait time, job makespan, and node utilization.
While \prefkeeper occasionally shows slightly higher job wait times, these remain within acceptable limits for production systems.
Notably, \prefkeeper performs best on \theta, which features a diverse job mix, while \pref consistently triggers the fewest shrink and expand operations.

In summary, three key findings emerge:
(1)~even modest malleability proportions (e.g., 20\%) deliver notable improvements;
(2)~workload characteristics strongly impact the effectiveness of the job scheduling strategy;
and (3)~the consistent gains across varied configurations demonstrate benefits, given the applicability of malleability.

%% file: 03jobs-figs.tex
\begin{figure}
    \centering
    \vspace*{-10pt}

    \begin{subfigure}{.49\linewidth}
        \centering
        \resizebox{\linewidth}{!}{\input{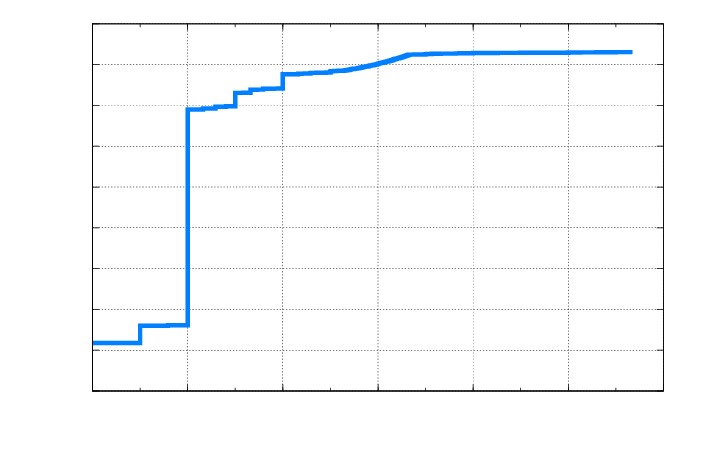}}
        \caption{\knl: Required nodes by jobs}
        \label{fig:knl-nodes}
    \end{subfigure}
    \hfill
    \begin{subfigure}{.49\linewidth}
        \centering
        \resizebox{\linewidth}{!}{\input{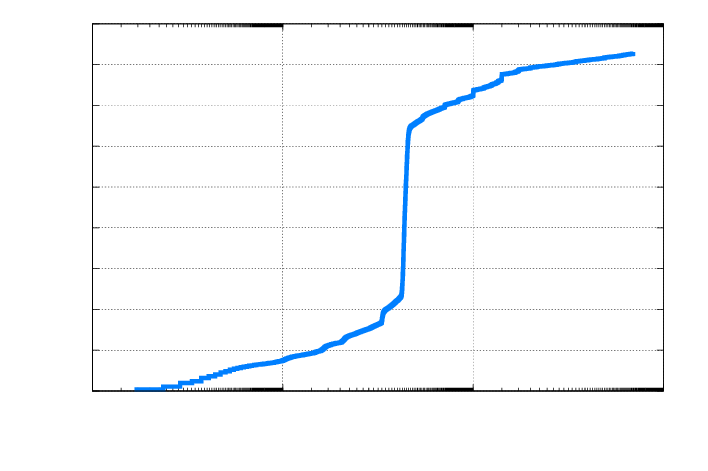}}
        \caption{\knl: Runtime of jobs}
        \label{fig:knl-runtimes}
    \end{subfigure}

    \vspace*{5pt}

    \begin{subfigure}{.49\linewidth}
        \centering
        \resizebox{\linewidth}{!}{\input{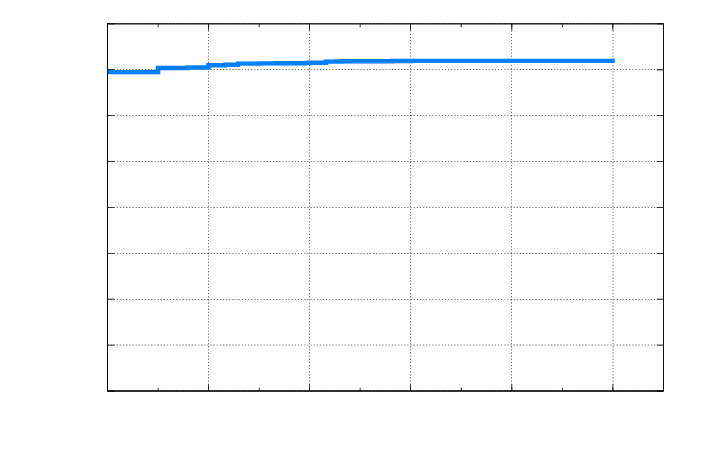}}
        \caption{\eagle: Required nodes by jobs}
        \label{fig:eagle-nodes}
    \end{subfigure}
    \hfill
    \begin{subfigure}{.49\linewidth}
        \centering
        \resizebox{\linewidth}{!}{\input{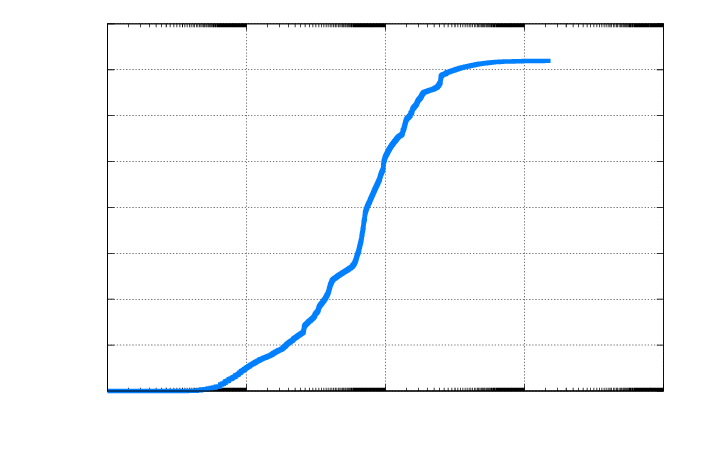}}
        \caption{\eagle: Runtime of jobs}
        \label{fig:eagle-runtimes}
    \end{subfigure}

    \vspace*{5pt}

    \begin{subfigure}{.49\linewidth}
        \centering
        \resizebox{\linewidth}{!}{\input{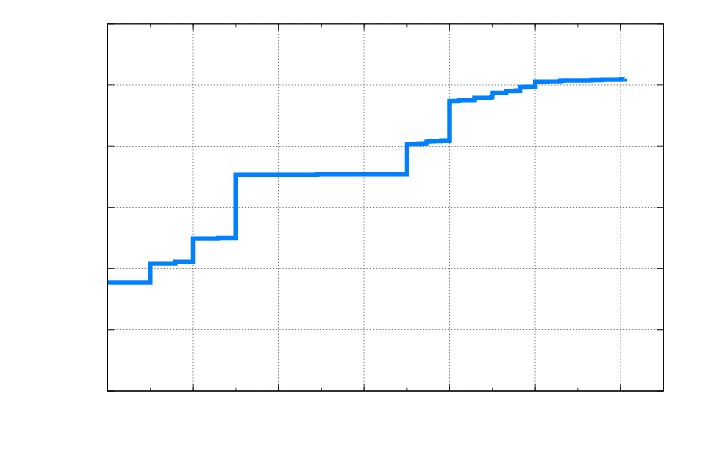}}
        \caption{\theta: Required nodes by jobs}
        \label{fig:theta-nodes}
    \end{subfigure}
    \hfill
    \begin{subfigure}{.49\linewidth}
        \centering
        \resizebox{\linewidth}{!}{\input{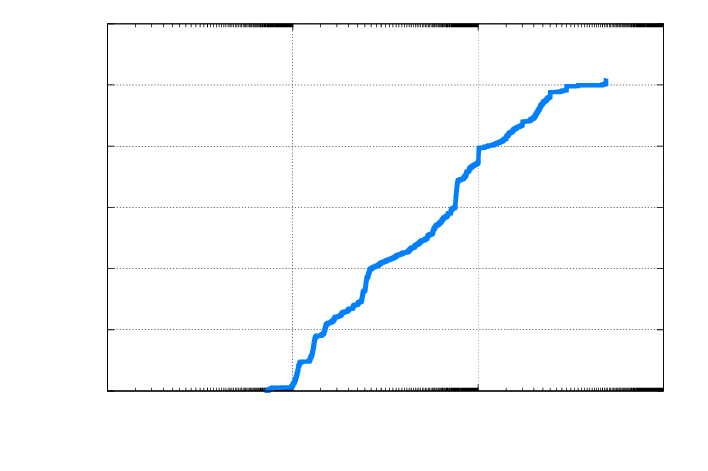}}
        \caption{\theta: Runtime of jobs}
        \label{fig:theta-runtimes}
    \end{subfigure}

    \vspace*{5pt}
    \caption{
        Distribution of job sizes and runtimes.
        \knl: Most jobs use 4 or fewer nodes and run under \num{1000}\,s.
        \eagle: Jobs are nearly all single-node and short-lived.
        \theta: More even distribution, with peaks at 1, 8, and 256 nodes.
    }
    \label{fig:job-distribution-comparison}
    \vspace*{5pt}

    \captionof{table}{Job submission rates in \textit{jobs per hour}}
    \label{tab:submission-rate}
    \begin{tabular}{l|cccc}
        \toprule
        \textbf{Dataset} & \textbf{\haswell} & \textbf{\knl} & \textbf{\eagle} & \textbf{\theta}   \\
        \midrule
        \makecell{Job submission rate\\[0.3ex][jobs/hour]}
        & \num{235.49} & \num{340.36} & \num{214.03} & \num{3.79} \\
        \bottomrule
    \end{tabular}

\end{figure}

%% file: img/KNLNodesAllocated.tex
\begingroup
\LARGE
  \makeatletter
  \providecommand\color[2][]{%
    \GenericError{(gnuplot) \space\space\space\@spaces}{%
      Package color not loaded in conjunction with
      terminal option `colourtext'%
    }{See the gnuplot documentation for explanation.%
    }{Either use 'blacktext' in gnuplot or load the package
      color.sty in LaTeX.}%
    \renewcommand\color[2][]{}%
  }%
  \providecommand\includegraphics[2][]{%
    \GenericError{(gnuplot) \space\space\space\@spaces}{%
      Package graphicx or graphics not loaded%
    }{See the gnuplot documentation for explanation.%
    }{The gnuplot epslatex terminal needs graphicx.sty or graphics.sty.}%
    \renewcommand\includegraphics[2][]{}%
  }%
  \providecommand\rotatebox[2]{#2}%
  \@ifundefined{ifGPcolor}{%
    \newif\ifGPcolor
    \GPcolortrue
  }{}%
  \@ifundefined{ifGPblacktext}{%
    \newif\ifGPblacktext
    \GPblacktextfalse
  }{}%
  \let\gplgaddtomacro\g@addto@macro
  \gdef\gplbacktext{}%
  \gdef\gplfronttext{}%
  \makeatother
  \ifGPblacktext
    \def\colorrgb#1{}%
    \def\colorgray#1{}%
  \else
    \ifGPcolor
      \def\colorrgb#1{\color[rgb]{#1}}%
      \def\colorgray#1{\color[gray]{#1}}%
      \expandafter\def\csname LTw\endcsname{\color{white}}%
      \expandafter\def\csname LTb\endcsname{\color{black}}%
      \expandafter\def\csname LTa\endcsname{\color{black}}%
      \expandafter\def\csname LT0\endcsname{\color[rgb]{1,0,0}}%
      \expandafter\def\csname LT1\endcsname{\color[rgb]{0,1,0}}%
      \expandafter\def\csname LT2\endcsname{\color[rgb]{0,0,1}}%
      \expandafter\def\csname LT3\endcsname{\color[rgb]{1,0,1}}%
      \expandafter\def\csname LT4\endcsname{\color[rgb]{0,1,1}}%
      \expandafter\def\csname LT5\endcsname{\color[rgb]{1,1,0}}%
      \expandafter\def\csname LT6\endcsname{\color[rgb]{0,0,0}}%
      \expandafter\def\csname LT7\endcsname{\color[rgb]{1,0.3,0}}%
      \expandafter\def\csname LT8\endcsname{\color[rgb]{0.5,0.5,0.5}}%
    \else
      \def\colorrgb#1{\color{black}}%
      \def\colorgray#1{\color[gray]{#1}}%
      \expandafter\def\csname LTw\endcsname{\color{white}}%
      \expandafter\def\csname LTb\endcsname{\color{black}}%
      \expandafter\def\csname LTa\endcsname{\color{black}}%
      \expandafter\def\csname LT0\endcsname{\color{black}}%
      \expandafter\def\csname LT1\endcsname{\color{black}}%
      \expandafter\def\csname LT2\endcsname{\color{black}}%
      \expandafter\def\csname LT3\endcsname{\color{black}}%
      \expandafter\def\csname LT4\endcsname{\color{black}}%
      \expandafter\def\csname LT5\endcsname{\color{black}}%
      \expandafter\def\csname LT6\endcsname{\color{black}}%
      \expandafter\def\csname LT7\endcsname{\color{black}}%
      \expandafter\def\csname LT8\endcsname{\color{black}}%
    \fi
  \fi
    \setlength{\unitlength}{0.0500bp}%
    \ifx\gptboxheight\undefined%
      \newlength{\gptboxheight}%
      \newlength{\gptboxwidth}%
      \newsavebox{\gptboxtext}%
    \fi%
    \setlength{\fboxrule}{0.5pt}%
    \setlength{\fboxsep}{1pt}%
    \definecolor{tbcol}{rgb}{1,1,1}%
\begin{picture}(6802.00,4534.00)%
    \gplgaddtomacro\gplbacktext{%
      \csname LTb\endcsname
      \put(744,768){\makebox(0,0)[r]{\strut{}$0$}}%
      \csname LTb\endcsname
      \put(744,1160){\makebox(0,0)[r]{\strut{}$5$}}%
      \csname LTb\endcsname
      \put(744,1551){\makebox(0,0)[r]{\strut{}$10$}}%
      \csname LTb\endcsname
      \put(744,1943){\makebox(0,0)[r]{\strut{}$15$}}%
      \csname LTb\endcsname
      \put(744,2335){\makebox(0,0)[r]{\strut{}$20$}}%
      \csname LTb\endcsname
      \put(744,2726){\makebox(0,0)[r]{\strut{}$25$}}%
      \csname LTb\endcsname
      \put(744,3118){\makebox(0,0)[r]{\strut{}$30$}}%
      \csname LTb\endcsname
      \put(744,3510){\makebox(0,0)[r]{\strut{}$35$}}%
      \csname LTb\endcsname
      \put(744,3901){\makebox(0,0)[r]{\strut{}$40$}}%
      \csname LTb\endcsname
      \put(744,4293){\makebox(0,0)[r]{\strut{}$45$}}%
      \csname LTb\endcsname
      \put(888,528){\makebox(0,0){\strut{}$1$}}%
      \csname LTb\endcsname
      \put(1802,528){\makebox(0,0){\strut{}$4$}}%
      \csname LTb\endcsname
      \put(2715,528){\makebox(0,0){\strut{}$16$}}%
      \csname LTb\endcsname
      \put(3629,528){\makebox(0,0){\strut{}$64$}}%
      \csname LTb\endcsname
      \put(4542,528){\makebox(0,0){\strut{}$256$}}%
      \csname LTb\endcsname
      \put(5456,528){\makebox(0,0){\strut{}$1024$}}%
      \csname LTb\endcsname
      \put(6369,528){\makebox(0,0){\strut{}$4096$}}%
    }%
    \gplgaddtomacro\gplfronttext{%
      \csname LTb\endcsname
      \put(228,2530){\rotatebox{-270.00}{\makebox(0,0){\strut{}Jobs [×1000] cumulative}}}%
      \put(3628,168){\makebox(0,0){\strut{}Required nodes}}%
    }%
    \gplbacktext
    \put(0,0){\includegraphics[width={340.10bp},height={226.70bp}]{img//KNLNodesAllocated}}%
    \gplfronttext
  \end{picture}%
\endgroup

%% file: img/KNLRuntimeSeconds.tex
\begingroup
\LARGE
  \makeatletter
  \providecommand\color[2][]{%
    \GenericError{(gnuplot) \space\space\space\@spaces}{%
      Package color not loaded in conjunction with
      terminal option `colourtext'%
    }{See the gnuplot documentation for explanation.%
    }{Either use 'blacktext' in gnuplot or load the package
      color.sty in LaTeX.}%
    \renewcommand\color[2][]{}%
  }%
  \providecommand\includegraphics[2][]{%
    \GenericError{(gnuplot) \space\space\space\@spaces}{%
      Package graphicx or graphics not loaded%
    }{See the gnuplot documentation for explanation.%
    }{The gnuplot epslatex terminal needs graphicx.sty or graphics.sty.}%
    \renewcommand\includegraphics[2][]{}%
  }%
  \providecommand\rotatebox[2]{#2}%
  \@ifundefined{ifGPcolor}{%
    \newif\ifGPcolor
    \GPcolortrue
  }{}%
  \@ifundefined{ifGPblacktext}{%
    \newif\ifGPblacktext
    \GPblacktextfalse
  }{}%
  \let\gplgaddtomacro\g@addto@macro
  \gdef\gplbacktext{}%
  \gdef\gplfronttext{}%
  \makeatother
  \ifGPblacktext
    \def\colorrgb#1{}%
    \def\colorgray#1{}%
  \else
    \ifGPcolor
      \def\colorrgb#1{\color[rgb]{#1}}%
      \def\colorgray#1{\color[gray]{#1}}%
      \expandafter\def\csname LTw\endcsname{\color{white}}%
      \expandafter\def\csname LTb\endcsname{\color{black}}%
      \expandafter\def\csname LTa\endcsname{\color{black}}%
      \expandafter\def\csname LT0\endcsname{\color[rgb]{1,0,0}}%
      \expandafter\def\csname LT1\endcsname{\color[rgb]{0,1,0}}%
      \expandafter\def\csname LT2\endcsname{\color[rgb]{0,0,1}}%
      \expandafter\def\csname LT3\endcsname{\color[rgb]{1,0,1}}%
      \expandafter\def\csname LT4\endcsname{\color[rgb]{0,1,1}}%
      \expandafter\def\csname LT5\endcsname{\color[rgb]{1,1,0}}%
      \expandafter\def\csname LT6\endcsname{\color[rgb]{0,0,0}}%
      \expandafter\def\csname LT7\endcsname{\color[rgb]{1,0.3,0}}%
      \expandafter\def\csname LT8\endcsname{\color[rgb]{0.5,0.5,0.5}}%
    \else
      \def\colorrgb#1{\color{black}}%
      \def\colorgray#1{\color[gray]{#1}}%
      \expandafter\def\csname LTw\endcsname{\color{white}}%
      \expandafter\def\csname LTb\endcsname{\color{black}}%
      \expandafter\def\csname LTa\endcsname{\color{black}}%
      \expandafter\def\csname LT0\endcsname{\color{black}}%
      \expandafter\def\csname LT1\endcsname{\color{black}}%
      \expandafter\def\csname LT2\endcsname{\color{black}}%
      \expandafter\def\csname LT3\endcsname{\color{black}}%
      \expandafter\def\csname LT4\endcsname{\color{black}}%
      \expandafter\def\csname LT5\endcsname{\color{black}}%
      \expandafter\def\csname LT6\endcsname{\color{black}}%
      \expandafter\def\csname LT7\endcsname{\color{black}}%
      \expandafter\def\csname LT8\endcsname{\color{black}}%
    \fi
  \fi
    \setlength{\unitlength}{0.0500bp}%
    \ifx\gptboxheight\undefined%
      \newlength{\gptboxheight}%
      \newlength{\gptboxwidth}%
      \newsavebox{\gptboxtext}%
    \fi%
    \setlength{\fboxrule}{0.5pt}%
    \setlength{\fboxsep}{1pt}%
    \definecolor{tbcol}{rgb}{1,1,1}%
\begin{picture}(6802.00,4534.00)%
    \gplgaddtomacro\gplbacktext{%
      \csname LTb\endcsname
      \put(744,768){\makebox(0,0)[r]{\strut{}$0$}}%
      \csname LTb\endcsname
      \put(744,1160){\makebox(0,0)[r]{\strut{}$5$}}%
      \csname LTb\endcsname
      \put(744,1551){\makebox(0,0)[r]{\strut{}$10$}}%
      \csname LTb\endcsname
      \put(744,1943){\makebox(0,0)[r]{\strut{}$15$}}%
      \csname LTb\endcsname
      \put(744,2335){\makebox(0,0)[r]{\strut{}$20$}}%
      \csname LTb\endcsname
      \put(744,2726){\makebox(0,0)[r]{\strut{}$25$}}%
      \csname LTb\endcsname
      \put(744,3118){\makebox(0,0)[r]{\strut{}$30$}}%
      \csname LTb\endcsname
      \put(744,3510){\makebox(0,0)[r]{\strut{}$35$}}%
      \csname LTb\endcsname
      \put(744,3901){\makebox(0,0)[r]{\strut{}$40$}}%
      \csname LTb\endcsname
      \put(744,4293){\makebox(0,0)[r]{\strut{}$45$}}%
      \csname LTb\endcsname
      \put(888,528){\makebox(0,0){\strut{}$0.0001$}}%
      \csname LTb\endcsname
      \put(2715,528){\makebox(0,0){\strut{}$0.01$}}%
      \csname LTb\endcsname
      \put(4542,528){\makebox(0,0){\strut{}$1$}}%
      \csname LTb\endcsname
      \put(6369,528){\makebox(0,0){\strut{}$100$}}%
    }%
    \gplgaddtomacro\gplfronttext{%
      \csname LTb\endcsname
      \put(228,2530){\rotatebox{-270.00}{\makebox(0,0){\strut{}Jobs [×1000] cumulative}}}%
      \put(3628,168){\makebox(0,0){\strut{}Runtime [hours]}}%
    }%
    \gplbacktext
    \put(0,0){\includegraphics[width={340.10bp},height={226.70bp}]{img//KNLRuntimeSeconds}}%
    \gplfronttext
  \end{picture}%
\endgroup

%% file: img/EagleNodesAllocated.tex
\begingroup
\LARGE
  \makeatletter
  \providecommand\color[2][]{%
    \GenericError{(gnuplot) \space\space\space\@spaces}{%
      Package color not loaded in conjunction with
      terminal option `colourtext'%
    }{See the gnuplot documentation for explanation.%
    }{Either use 'blacktext' in gnuplot or load the package
      color.sty in LaTeX.}%
    \renewcommand\color[2][]{}%
  }%
  \providecommand\includegraphics[2][]{%
    \GenericError{(gnuplot) \space\space\space\@spaces}{%
      Package graphicx or graphics not loaded%
    }{See the gnuplot documentation for explanation.%
    }{The gnuplot epslatex terminal needs graphicx.sty or graphics.sty.}%
    \renewcommand\includegraphics[2][]{}%
  }%
  \providecommand\rotatebox[2]{#2}%
  \@ifundefined{ifGPcolor}{%
    \newif\ifGPcolor
    \GPcolortrue
  }{}%
  \@ifundefined{ifGPblacktext}{%
    \newif\ifGPblacktext
    \GPblacktextfalse
  }{}%
  \let\gplgaddtomacro\g@addto@macro
  \gdef\gplbacktext{}%
  \gdef\gplfronttext{}%
  \makeatother
  \ifGPblacktext
    \def\colorrgb#1{}%
    \def\colorgray#1{}%
  \else
    \ifGPcolor
      \def\colorrgb#1{\color[rgb]{#1}}%
      \def\colorgray#1{\color[gray]{#1}}%
      \expandafter\def\csname LTw\endcsname{\color{white}}%
      \expandafter\def\csname LTb\endcsname{\color{black}}%
      \expandafter\def\csname LTa\endcsname{\color{black}}%
      \expandafter\def\csname LT0\endcsname{\color[rgb]{1,0,0}}%
      \expandafter\def\csname LT1\endcsname{\color[rgb]{0,1,0}}%
      \expandafter\def\csname LT2\endcsname{\color[rgb]{0,0,1}}%
      \expandafter\def\csname LT3\endcsname{\color[rgb]{1,0,1}}%
      \expandafter\def\csname LT4\endcsname{\color[rgb]{0,1,1}}%
      \expandafter\def\csname LT5\endcsname{\color[rgb]{1,1,0}}%
      \expandafter\def\csname LT6\endcsname{\color[rgb]{0,0,0}}%
      \expandafter\def\csname LT7\endcsname{\color[rgb]{1,0.3,0}}%
      \expandafter\def\csname LT8\endcsname{\color[rgb]{0.5,0.5,0.5}}%
    \else
      \def\colorrgb#1{\color{black}}%
      \def\colorgray#1{\color[gray]{#1}}%
      \expandafter\def\csname LTw\endcsname{\color{white}}%
      \expandafter\def\csname LTb\endcsname{\color{black}}%
      \expandafter\def\csname LTa\endcsname{\color{black}}%
      \expandafter\def\csname LT0\endcsname{\color{black}}%
      \expandafter\def\csname LT1\endcsname{\color{black}}%
      \expandafter\def\csname LT2\endcsname{\color{black}}%
      \expandafter\def\csname LT3\endcsname{\color{black}}%
      \expandafter\def\csname LT4\endcsname{\color{black}}%
      \expandafter\def\csname LT5\endcsname{\color{black}}%
      \expandafter\def\csname LT6\endcsname{\color{black}}%
      \expandafter\def\csname LT7\endcsname{\color{black}}%
      \expandafter\def\csname LT8\endcsname{\color{black}}%
    \fi
  \fi
    \setlength{\unitlength}{0.0500bp}%
    \ifx\gptboxheight\undefined%
      \newlength{\gptboxheight}%
      \newlength{\gptboxwidth}%
      \newsavebox{\gptboxtext}%
    \fi%
    \setlength{\fboxrule}{0.5pt}%
    \setlength{\fboxsep}{1pt}%
    \definecolor{tbcol}{rgb}{1,1,1}%
\begin{picture}(6802.00,4534.00)%
    \gplgaddtomacro\gplbacktext{%
      \csname LTb\endcsname
      \put(888,768){\makebox(0,0)[r]{\strut{}$0$}}%
      \csname LTb\endcsname
      \put(888,1209){\makebox(0,0)[r]{\strut{}$20$}}%
      \csname LTb\endcsname
      \put(888,1649){\makebox(0,0)[r]{\strut{}$40$}}%
      \csname LTb\endcsname
      \put(888,2090){\makebox(0,0)[r]{\strut{}$60$}}%
      \csname LTb\endcsname
      \put(888,2531){\makebox(0,0)[r]{\strut{}$80$}}%
      \csname LTb\endcsname
      \put(888,2971){\makebox(0,0)[r]{\strut{}$100$}}%
      \csname LTb\endcsname
      \put(888,3412){\makebox(0,0)[r]{\strut{}$120$}}%
      \csname LTb\endcsname
      \put(888,3852){\makebox(0,0)[r]{\strut{}$140$}}%
      \csname LTb\endcsname
      \put(888,4293){\makebox(0,0)[r]{\strut{}$160$}}%
      \csname LTb\endcsname
      \put(1032,528){\makebox(0,0){\strut{}$1$}}%
      \csname LTb\endcsname
      \put(2002,528){\makebox(0,0){\strut{}$4$}}%
      \csname LTb\endcsname
      \put(2973,528){\makebox(0,0){\strut{}$16$}}%
      \csname LTb\endcsname
      \put(3943,528){\makebox(0,0){\strut{}$64$}}%
      \csname LTb\endcsname
      \put(4913,528){\makebox(0,0){\strut{}$256$}}%
      \csname LTb\endcsname
      \put(5884,528){\makebox(0,0){\strut{}$1024$}}%
    }%
    \gplgaddtomacro\gplfronttext{%
      \csname LTb\endcsname
      \put(228,2530){\rotatebox{-270.00}{\makebox(0,0){\strut{}Jobs [×1000] cumulative}}}%
      \put(3700,168){\makebox(0,0){\strut{}Required nodes}}%
    }%
    \gplbacktext
    \put(0,0){\includegraphics[width={340.10bp},height={226.70bp}]{img//EagleNodesAllocated}}%
    \gplfronttext
  \end{picture}%
\endgroup

%% file: img/EagleRuntimeSeconds.tex
\begingroup
\LARGE
  \makeatletter
  \providecommand\color[2][]{%
    \GenericError{(gnuplot) \space\space\space\@spaces}{%
      Package color not loaded in conjunction with
      terminal option `colourtext'%
    }{See the gnuplot documentation for explanation.%
    }{Either use 'blacktext' in gnuplot or load the package
      color.sty in LaTeX.}%
    \renewcommand\color[2][]{}%
  }%
  \providecommand\includegraphics[2][]{%
    \GenericError{(gnuplot) \space\space\space\@spaces}{%
      Package graphicx or graphics not loaded%
    }{See the gnuplot documentation for explanation.%
    }{The gnuplot epslatex terminal needs graphicx.sty or graphics.sty.}%
    \renewcommand\includegraphics[2][]{}%
  }%
  \providecommand\rotatebox[2]{#2}%
  \@ifundefined{ifGPcolor}{%
    \newif\ifGPcolor
    \GPcolortrue
  }{}%
  \@ifundefined{ifGPblacktext}{%
    \newif\ifGPblacktext
    \GPblacktextfalse
  }{}%
  \let\gplgaddtomacro\g@addto@macro
  \gdef\gplbacktext{}%
  \gdef\gplfronttext{}%
  \makeatother
  \ifGPblacktext
    \def\colorrgb#1{}%
    \def\colorgray#1{}%
  \else
    \ifGPcolor
      \def\colorrgb#1{\color[rgb]{#1}}%
      \def\colorgray#1{\color[gray]{#1}}%
      \expandafter\def\csname LTw\endcsname{\color{white}}%
      \expandafter\def\csname LTb\endcsname{\color{black}}%
      \expandafter\def\csname LTa\endcsname{\color{black}}%
      \expandafter\def\csname LT0\endcsname{\color[rgb]{1,0,0}}%
      \expandafter\def\csname LT1\endcsname{\color[rgb]{0,1,0}}%
      \expandafter\def\csname LT2\endcsname{\color[rgb]{0,0,1}}%
      \expandafter\def\csname LT3\endcsname{\color[rgb]{1,0,1}}%
      \expandafter\def\csname LT4\endcsname{\color[rgb]{0,1,1}}%
      \expandafter\def\csname LT5\endcsname{\color[rgb]{1,1,0}}%
      \expandafter\def\csname LT6\endcsname{\color[rgb]{0,0,0}}%
      \expandafter\def\csname LT7\endcsname{\color[rgb]{1,0.3,0}}%
      \expandafter\def\csname LT8\endcsname{\color[rgb]{0.5,0.5,0.5}}%
    \else
      \def\colorrgb#1{\color{black}}%
      \def\colorgray#1{\color[gray]{#1}}%
      \expandafter\def\csname LTw\endcsname{\color{white}}%
      \expandafter\def\csname LTb\endcsname{\color{black}}%
      \expandafter\def\csname LTa\endcsname{\color{black}}%
      \expandafter\def\csname LT0\endcsname{\color{black}}%
      \expandafter\def\csname LT1\endcsname{\color{black}}%
      \expandafter\def\csname LT2\endcsname{\color{black}}%
      \expandafter\def\csname LT3\endcsname{\color{black}}%
      \expandafter\def\csname LT4\endcsname{\color{black}}%
      \expandafter\def\csname LT5\endcsname{\color{black}}%
      \expandafter\def\csname LT6\endcsname{\color{black}}%
      \expandafter\def\csname LT7\endcsname{\color{black}}%
      \expandafter\def\csname LT8\endcsname{\color{black}}%
    \fi
  \fi
    \setlength{\unitlength}{0.0500bp}%
    \ifx\gptboxheight\undefined%
      \newlength{\gptboxheight}%
      \newlength{\gptboxwidth}%
      \newsavebox{\gptboxtext}%
    \fi%
    \setlength{\fboxrule}{0.5pt}%
    \setlength{\fboxsep}{1pt}%
    \definecolor{tbcol}{rgb}{1,1,1}%
\begin{picture}(6802.00,4534.00)%
    \gplgaddtomacro\gplbacktext{%
      \csname LTb\endcsname
      \put(888,768){\makebox(0,0)[r]{\strut{}$0$}}%
      \csname LTb\endcsname
      \put(888,1209){\makebox(0,0)[r]{\strut{}$20$}}%
      \csname LTb\endcsname
      \put(888,1649){\makebox(0,0)[r]{\strut{}$40$}}%
      \csname LTb\endcsname
      \put(888,2090){\makebox(0,0)[r]{\strut{}$60$}}%
      \csname LTb\endcsname
      \put(888,2531){\makebox(0,0)[r]{\strut{}$80$}}%
      \csname LTb\endcsname
      \put(888,2971){\makebox(0,0)[r]{\strut{}$100$}}%
      \csname LTb\endcsname
      \put(888,3412){\makebox(0,0)[r]{\strut{}$120$}}%
      \csname LTb\endcsname
      \put(888,3852){\makebox(0,0)[r]{\strut{}$140$}}%
      \csname LTb\endcsname
      \put(888,4293){\makebox(0,0)[r]{\strut{}$160$}}%
      \csname LTb\endcsname
      \put(1032,528){\makebox(0,0){\strut{}$0.0001$}}%
      \csname LTb\endcsname
      \put(2366,528){\makebox(0,0){\strut{}$0.01$}}%
      \csname LTb\endcsname
      \put(3701,528){\makebox(0,0){\strut{}$1$}}%
      \csname LTb\endcsname
      \put(5035,528){\makebox(0,0){\strut{}$100$}}%
      \csname LTb\endcsname
      \put(6369,528){\makebox(0,0){\strut{}$10000$}}%
    }%
    \gplgaddtomacro\gplfronttext{%
      \csname LTb\endcsname
      \put(228,2530){\rotatebox{-270.00}{\makebox(0,0){\strut{}Jobs [×1000] cumulative}}}%
      \put(3700,168){\makebox(0,0){\strut{}Runtime [hours]}}%
    }%
    \gplbacktext
    \put(0,0){\includegraphics[width={340.10bp},height={226.70bp}]{img//EagleRuntimeSeconds}}%
    \gplfronttext
  \end{picture}%
\endgroup

%% file: img/ThetaNodesAllocated.tex
\begingroup
\LARGE
  \makeatletter
  \providecommand\color[2][]{%
    \GenericError{(gnuplot) \space\space\space\@spaces}{%
      Package color not loaded in conjunction with
      terminal option `colourtext'%
    }{See the gnuplot documentation for explanation.%
    }{Either use 'blacktext' in gnuplot or load the package
      color.sty in LaTeX.}%
    \renewcommand\color[2][]{}%
  }%
  \providecommand\includegraphics[2][]{%
    \GenericError{(gnuplot) \space\space\space\@spaces}{%
      Package graphicx or graphics not loaded%
    }{See the gnuplot documentation for explanation.%
    }{The gnuplot epslatex terminal needs graphicx.sty or graphics.sty.}%
    \renewcommand\includegraphics[2][]{}%
  }%
  \providecommand\rotatebox[2]{#2}%
  \@ifundefined{ifGPcolor}{%
    \newif\ifGPcolor
    \GPcolortrue
  }{}%
  \@ifundefined{ifGPblacktext}{%
    \newif\ifGPblacktext
    \GPblacktextfalse
  }{}%
  \let\gplgaddtomacro\g@addto@macro
  \gdef\gplbacktext{}%
  \gdef\gplfronttext{}%
  \makeatother
  \ifGPblacktext
    \def\colorrgb#1{}%
    \def\colorgray#1{}%
  \else
    \ifGPcolor
      \def\colorrgb#1{\color[rgb]{#1}}%
      \def\colorgray#1{\color[gray]{#1}}%
      \expandafter\def\csname LTw\endcsname{\color{white}}%
      \expandafter\def\csname LTb\endcsname{\color{black}}%
      \expandafter\def\csname LTa\endcsname{\color{black}}%
      \expandafter\def\csname LT0\endcsname{\color[rgb]{1,0,0}}%
      \expandafter\def\csname LT1\endcsname{\color[rgb]{0,1,0}}%
      \expandafter\def\csname LT2\endcsname{\color[rgb]{0,0,1}}%
      \expandafter\def\csname LT3\endcsname{\color[rgb]{1,0,1}}%
      \expandafter\def\csname LT4\endcsname{\color[rgb]{0,1,1}}%
      \expandafter\def\csname LT5\endcsname{\color[rgb]{1,1,0}}%
      \expandafter\def\csname LT6\endcsname{\color[rgb]{0,0,0}}%
      \expandafter\def\csname LT7\endcsname{\color[rgb]{1,0.3,0}}%
      \expandafter\def\csname LT8\endcsname{\color[rgb]{0.5,0.5,0.5}}%
    \else
      \def\colorrgb#1{\color{black}}%
      \def\colorgray#1{\color[gray]{#1}}%
      \expandafter\def\csname LTw\endcsname{\color{white}}%
      \expandafter\def\csname LTb\endcsname{\color{black}}%
      \expandafter\def\csname LTa\endcsname{\color{black}}%
      \expandafter\def\csname LT0\endcsname{\color{black}}%
      \expandafter\def\csname LT1\endcsname{\color{black}}%
      \expandafter\def\csname LT2\endcsname{\color{black}}%
      \expandafter\def\csname LT3\endcsname{\color{black}}%
      \expandafter\def\csname LT4\endcsname{\color{black}}%
      \expandafter\def\csname LT5\endcsname{\color{black}}%
      \expandafter\def\csname LT6\endcsname{\color{black}}%
      \expandafter\def\csname LT7\endcsname{\color{black}}%
      \expandafter\def\csname LT8\endcsname{\color{black}}%
    \fi
  \fi
    \setlength{\unitlength}{0.0500bp}%
    \ifx\gptboxheight\undefined%
      \newlength{\gptboxheight}%
      \newlength{\gptboxwidth}%
      \newsavebox{\gptboxtext}%
    \fi%
    \setlength{\fboxrule}{0.5pt}%
    \setlength{\fboxsep}{1pt}%
    \definecolor{tbcol}{rgb}{1,1,1}%
\begin{picture}(6802.00,4534.00)%
    \gplgaddtomacro\gplbacktext{%
      \csname LTb\endcsname
      \put(888,768){\makebox(0,0)[r]{\strut{}$0$}}%
      \csname LTb\endcsname
      \put(888,1356){\makebox(0,0)[r]{\strut{}$0.5$}}%
      \csname LTb\endcsname
      \put(888,1943){\makebox(0,0)[r]{\strut{}$1$}}%
      \csname LTb\endcsname
      \put(888,2531){\makebox(0,0)[r]{\strut{}$1.5$}}%
      \csname LTb\endcsname
      \put(888,3118){\makebox(0,0)[r]{\strut{}$2$}}%
      \csname LTb\endcsname
      \put(888,3706){\makebox(0,0)[r]{\strut{}$2.5$}}%
      \csname LTb\endcsname
      \put(888,4293){\makebox(0,0)[r]{\strut{}$3$}}%
      \csname LTb\endcsname
      \put(1032,528){\makebox(0,0){\strut{}$1$}}%
      \csname LTb\endcsname
      \put(1853,528){\makebox(0,0){\strut{}$4$}}%
      \csname LTb\endcsname
      \put(2674,528){\makebox(0,0){\strut{}$16$}}%
      \csname LTb\endcsname
      \put(3495,528){\makebox(0,0){\strut{}$64$}}%
      \csname LTb\endcsname
      \put(4316,528){\makebox(0,0){\strut{}$256$}}%
      \csname LTb\endcsname
      \put(5137,528){\makebox(0,0){\strut{}$1024$}}%
      \csname LTb\endcsname
      \put(5958,528){\makebox(0,0){\strut{}$4096$}}%
    }%
    \gplgaddtomacro\gplfronttext{%
      \csname LTb\endcsname
      \put(228,2530){\rotatebox{-270.00}{\makebox(0,0){\strut{}Jobs [×1000] cumulative}}}%
      \put(3700,168){\makebox(0,0){\strut{}Required nodes}}%
    }%
    \gplbacktext
    \put(0,0){\includegraphics[width={340.10bp},height={226.70bp}]{img//ThetaNodesAllocated}}%
    \gplfronttext
  \end{picture}%
\endgroup

%% file: img/ThetaRuntimeSeconds.tex
\begingroup
\LARGE
  \makeatletter
  \providecommand\color[2][]{%
    \GenericError{(gnuplot) \space\space\space\@spaces}{%
      Package color not loaded in conjunction with
      terminal option `colourtext'%
    }{See the gnuplot documentation for explanation.%
    }{Either use 'blacktext' in gnuplot or load the package
      color.sty in LaTeX.}%
    \renewcommand\color[2][]{}%
  }%
  \providecommand\includegraphics[2][]{%
    \GenericError{(gnuplot) \space\space\space\@spaces}{%
      Package graphicx or graphics not loaded%
    }{See the gnuplot documentation for explanation.%
    }{The gnuplot epslatex terminal needs graphicx.sty or graphics.sty.}%
    \renewcommand\includegraphics[2][]{}%
  }%
  \providecommand\rotatebox[2]{#2}%
  \@ifundefined{ifGPcolor}{%
    \newif\ifGPcolor
    \GPcolortrue
  }{}%
  \@ifundefined{ifGPblacktext}{%
    \newif\ifGPblacktext
    \GPblacktextfalse
  }{}%
  \let\gplgaddtomacro\g@addto@macro
  \gdef\gplbacktext{}%
  \gdef\gplfronttext{}%
  \makeatother
  \ifGPblacktext
    \def\colorrgb#1{}%
    \def\colorgray#1{}%
  \else
    \ifGPcolor
      \def\colorrgb#1{\color[rgb]{#1}}%
      \def\colorgray#1{\color[gray]{#1}}%
      \expandafter\def\csname LTw\endcsname{\color{white}}%
      \expandafter\def\csname LTb\endcsname{\color{black}}%
      \expandafter\def\csname LTa\endcsname{\color{black}}%
      \expandafter\def\csname LT0\endcsname{\color[rgb]{1,0,0}}%
      \expandafter\def\csname LT1\endcsname{\color[rgb]{0,1,0}}%
      \expandafter\def\csname LT2\endcsname{\color[rgb]{0,0,1}}%
      \expandafter\def\csname LT3\endcsname{\color[rgb]{1,0,1}}%
      \expandafter\def\csname LT4\endcsname{\color[rgb]{0,1,1}}%
      \expandafter\def\csname LT5\endcsname{\color[rgb]{1,1,0}}%
      \expandafter\def\csname LT6\endcsname{\color[rgb]{0,0,0}}%
      \expandafter\def\csname LT7\endcsname{\color[rgb]{1,0.3,0}}%
      \expandafter\def\csname LT8\endcsname{\color[rgb]{0.5,0.5,0.5}}%
    \else
      \def\colorrgb#1{\color{black}}%
      \def\colorgray#1{\color[gray]{#1}}%
      \expandafter\def\csname LTw\endcsname{\color{white}}%
      \expandafter\def\csname LTb\endcsname{\color{black}}%
      \expandafter\def\csname LTa\endcsname{\color{black}}%
      \expandafter\def\csname LT0\endcsname{\color{black}}%
      \expandafter\def\csname LT1\endcsname{\color{black}}%
      \expandafter\def\csname LT2\endcsname{\color{black}}%
      \expandafter\def\csname LT3\endcsname{\color{black}}%
      \expandafter\def\csname LT4\endcsname{\color{black}}%
      \expandafter\def\csname LT5\endcsname{\color{black}}%
      \expandafter\def\csname LT6\endcsname{\color{black}}%
      \expandafter\def\csname LT7\endcsname{\color{black}}%
      \expandafter\def\csname LT8\endcsname{\color{black}}%
    \fi
  \fi
    \setlength{\unitlength}{0.0500bp}%
    \ifx\gptboxheight\undefined%
      \newlength{\gptboxheight}%
      \newlength{\gptboxwidth}%
      \newsavebox{\gptboxtext}%
    \fi%
    \setlength{\fboxrule}{0.5pt}%
    \setlength{\fboxsep}{1pt}%
    \definecolor{tbcol}{rgb}{1,1,1}%
\begin{picture}(6802.00,4534.00)%
    \gplgaddtomacro\gplbacktext{%
      \csname LTb\endcsname
      \put(888,768){\makebox(0,0)[r]{\strut{}$0$}}%
      \csname LTb\endcsname
      \put(888,1356){\makebox(0,0)[r]{\strut{}$0.5$}}%
      \csname LTb\endcsname
      \put(888,1943){\makebox(0,0)[r]{\strut{}$1$}}%
      \csname LTb\endcsname
      \put(888,2531){\makebox(0,0)[r]{\strut{}$1.5$}}%
      \csname LTb\endcsname
      \put(888,3118){\makebox(0,0)[r]{\strut{}$2$}}%
      \csname LTb\endcsname
      \put(888,3706){\makebox(0,0)[r]{\strut{}$2.5$}}%
      \csname LTb\endcsname
      \put(888,4293){\makebox(0,0)[r]{\strut{}$3$}}%
      \csname LTb\endcsname
      \put(1032,528){\makebox(0,0){\strut{}$0.0001$}}%
      \csname LTb\endcsname
      \put(2811,528){\makebox(0,0){\strut{}$0.01$}}%
      \csname LTb\endcsname
      \put(4590,528){\makebox(0,0){\strut{}$1$}}%
      \csname LTb\endcsname
      \put(6369,528){\makebox(0,0){\strut{}$100$}}%
    }%
    \gplgaddtomacro\gplfronttext{%
      \csname LTb\endcsname
      \put(228,2530){\rotatebox{-270.00}{\makebox(0,0){\strut{}Jobs [×1000] cumulative}}}%
      \put(3700,168){\makebox(0,0){\strut{}Runtime [hours]}}%
    }%
    \gplbacktext
    \put(0,0){\includegraphics[width={340.10bp},height={226.70bp}]{img//ThetaRuntimeSeconds}}%
    \gplfronttext
  \end{picture}%
\endgroup

%% file: 03cori-both-figs-2.tex
\begin{figure}
    \centering
    \vspace*{-10pt}

    \begin{minipage}{\linewidth}
        \centering

        \begin{subfigure}{.32\linewidth}
            \centering
            \resizebox{\linewidth}{!}{\input{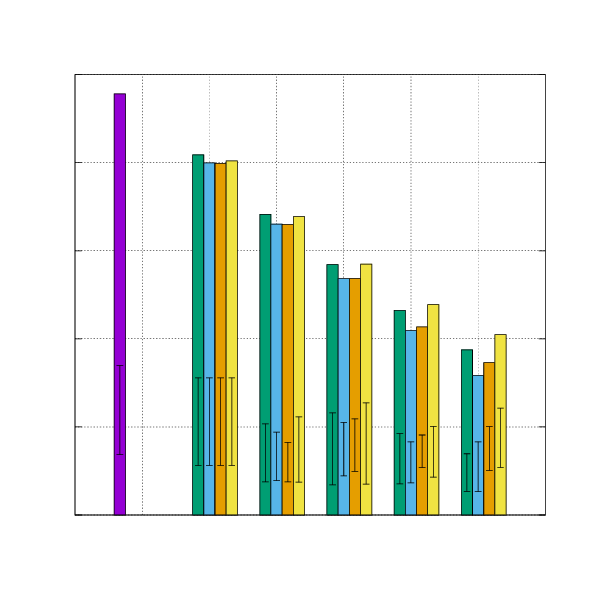}}
            \caption{Turnaround Time}
            \label{fig:cori-haswell-statistics-turnaround}
        \end{subfigure}
        \hfill
        \begin{subfigure}{.32\linewidth}
            \centering
            \resizebox{\linewidth}{!}{\input{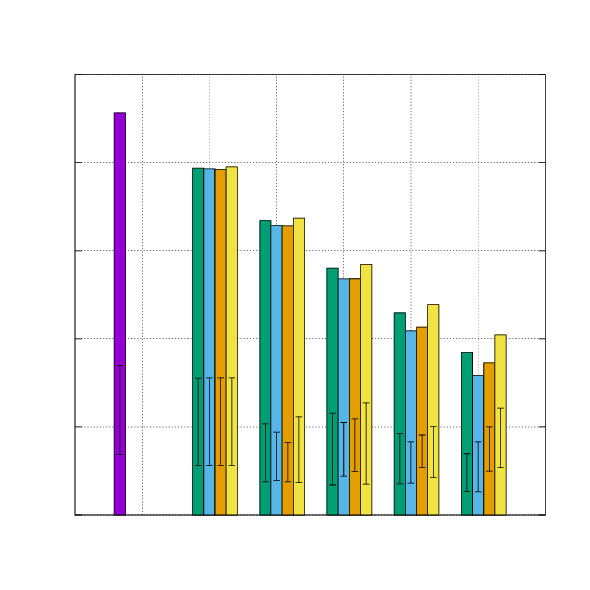}}
            \caption{Makespan}
            \label{fig:cori-haswell-statistics-makespan}
        \end{subfigure}
        \hfill
        \begin{subfigure}{.32\linewidth}
            \centering
            \resizebox{\linewidth}{!}{\input{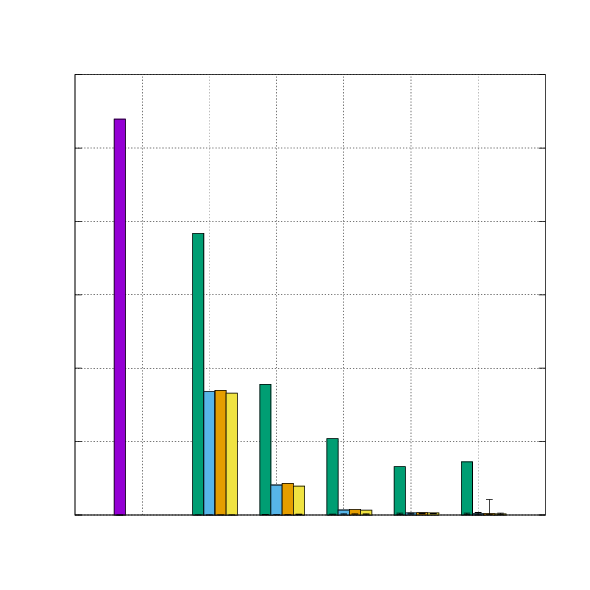}}
            \caption{Wait Time}
            \label{fig:cori-haswell-statistics-waittime}
        \end{subfigure}

        \vspace{1pt}
        \begin{subfigure}{.32\linewidth}
            \centering
            \resizebox{\linewidth}{!}{\input{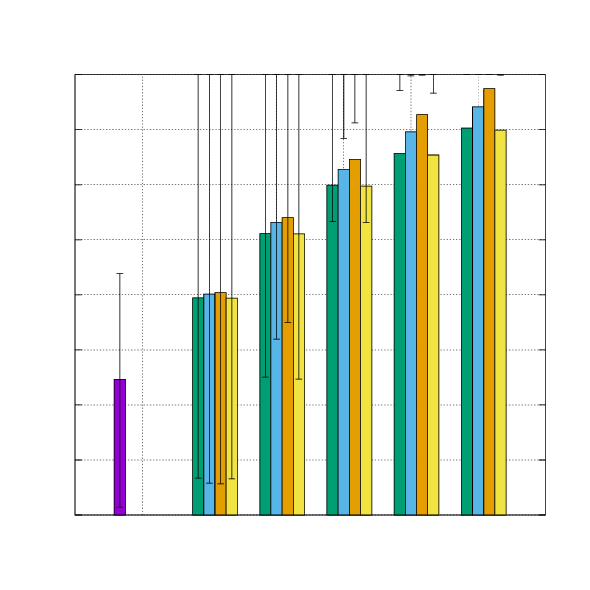}}
            \caption{Node Utilization}
            \label{fig:cori-haswell-statistics-nodeutil}
        \end{subfigure}
        \hfill
        \begin{subfigure}{.32\linewidth}
            \centering
            \resizebox{\linewidth}{!}{\input{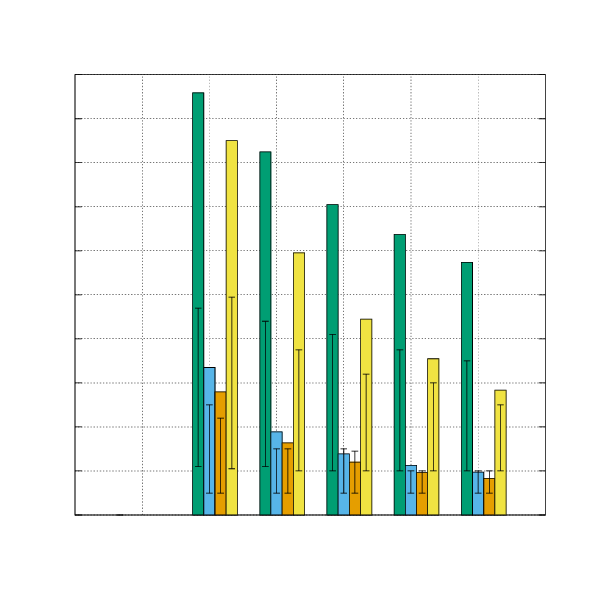}}
            \caption{Expands per Job}
            \label{fig:cori-haswell-expands}
        \end{subfigure}
        \hfill
        \begin{subfigure}{.32\linewidth}
            \centering
            \resizebox{\linewidth}{!}{\input{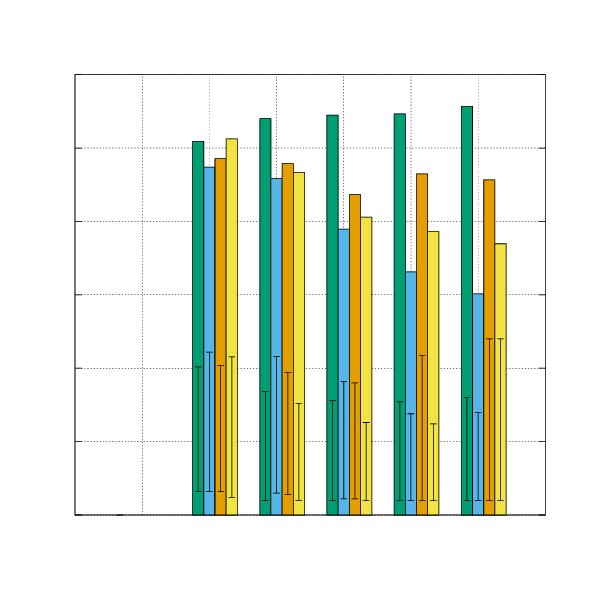}}
            \caption{Shrinks per Job}
            \label{fig:cori-haswell-shrinks}
        \end{subfigure}
        \vspace{0pt}
        \resizebox{0.99\linewidth}{!}{\input{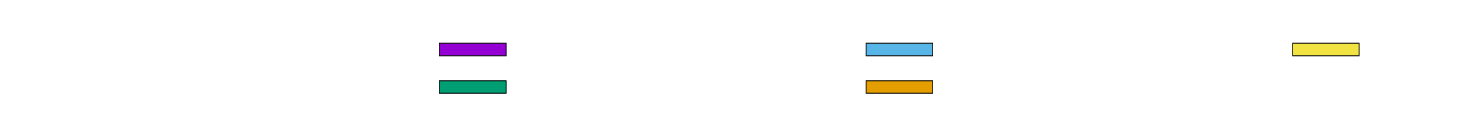}}
        \vspace*{-6pt}
        \caption{\haswell: Simulation results with IQR error bars}
        \label{fig:cori-haswell-all-metrics}
    \end{minipage}

    \vspace{0em}

    \begin{minipage}{\linewidth}
        \centering

        \begin{subfigure}[t]{.32\linewidth}
            \centering
            \resizebox{\linewidth}{!}{\input{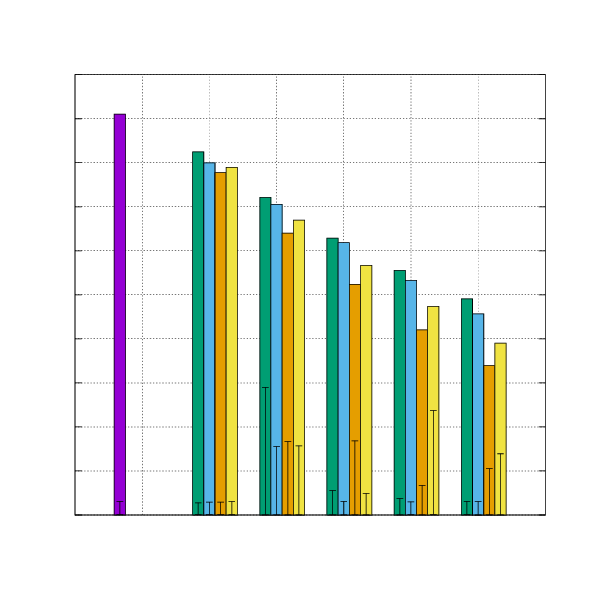}}
            \caption{Turnaround Time}
            \label{fig:cori-knl-statistics-turnaround}
        \end{subfigure}
        \hfill
        \begin{subfigure}[t]{.32\linewidth}
            \centering
            \resizebox{\linewidth}{!}{\input{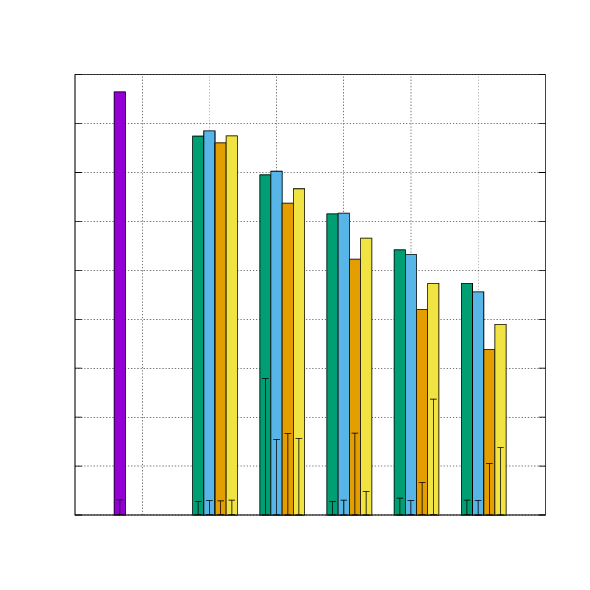}}
            \caption{Makespan}
            \label{fig:cori-knl-statistics-makespan}
        \end{subfigure}
        \hfill
        \begin{subfigure}[t]{.32\linewidth}
            \centering
            \resizebox{\linewidth}{!}{\input{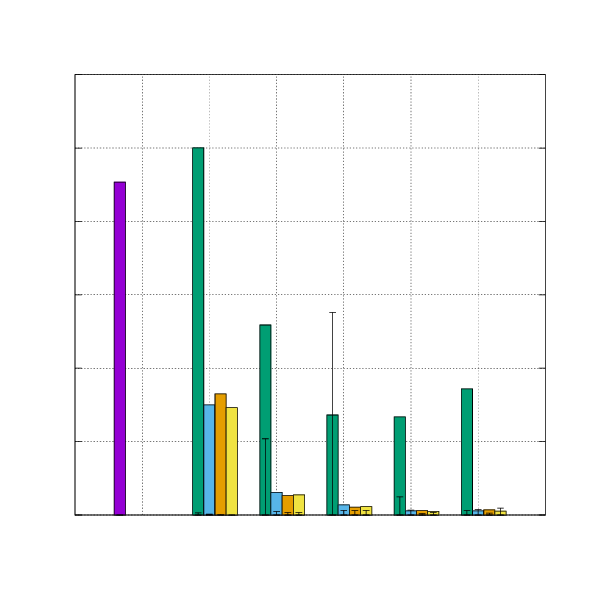}}
            \caption{Wait Time}
            \label{fig:cori-knl-statistics-waittime}
        \end{subfigure}

        \vspace{-10pt}
        \begin{subfigure}{.32\linewidth}
            \centering
            \resizebox{\linewidth}{!}{\input{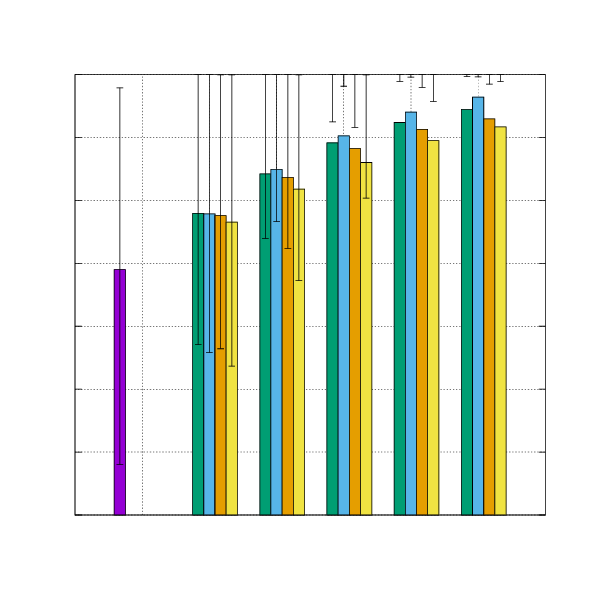}}
            \caption{Node Utilization}
            \label{fig:cori-knl-statistics-nodeutil}
        \end{subfigure}
        \hfill
        \begin{subfigure}{.32\linewidth}
            \centering
            \resizebox{\linewidth}{!}{\input{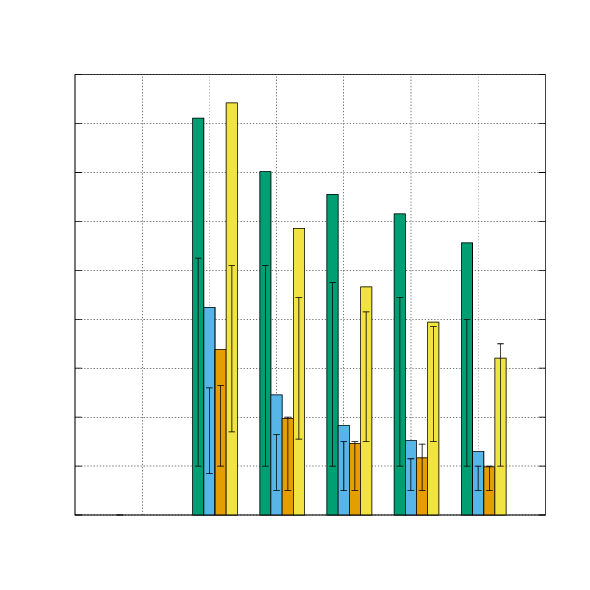}}
            \caption{Expands per Job}
            \label{fig:cori-knl-expands}
        \end{subfigure}
        \hfill
        \begin{subfigure}{.32\linewidth}
            \centering
            \resizebox{\linewidth}{!}{\input{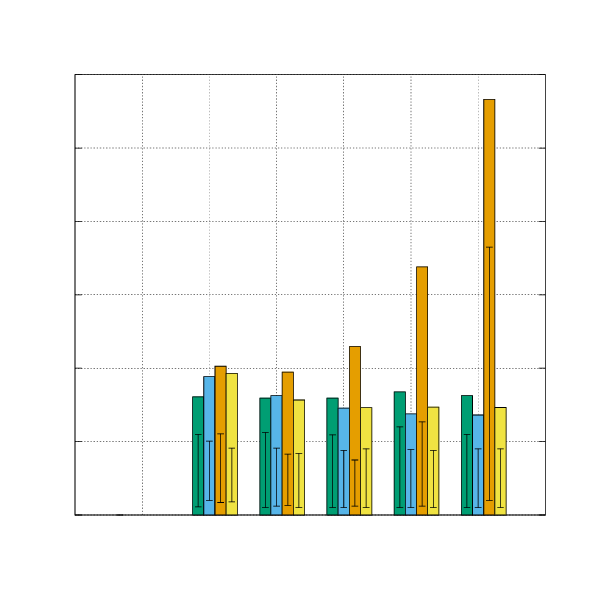}}
            \caption{Shrinks per Job}
            \label{fig:cori-knl-shrinks}
        \end{subfigure}
        \vspace{0pt}
        \resizebox{0.99\linewidth}{!}{\input{img/Legend}}
        \vspace*{-6pt}
        \caption{\knl: Simulation results with IQR error bars}
        \label{fig:cori-knl-all-metrics}
    \end{minipage}
\end{figure}

%% file: img/HaswellTurnaroundMeanPartialStatistic.tex
\begingroup
\LARGE
  \makeatletter
  \providecommand\color[2][]{%
    \GenericError{(gnuplot) \space\space\space\@spaces}{%
      Package color not loaded in conjunction with
      terminal option `colourtext'%
    }{See the gnuplot documentation for explanation.%
    }{Either use 'blacktext' in gnuplot or load the package
      color.sty in LaTeX.}%
    \renewcommand\color[2][]{}%
  }%
  \providecommand\includegraphics[2][]{%
    \GenericError{(gnuplot) \space\space\space\@spaces}{%
      Package graphicx or graphics not loaded%
    }{See the gnuplot documentation for explanation.%
    }{The gnuplot epslatex terminal needs graphicx.sty or graphics.sty.}%
    \renewcommand\includegraphics[2][]{}%
  }%
  \providecommand\rotatebox[2]{#2}%
  \@ifundefined{ifGPcolor}{%
    \newif\ifGPcolor
    \GPcolortrue
  }{}%
  \@ifundefined{ifGPblacktext}{%
    \newif\ifGPblacktext
    \GPblacktextfalse
  }{}%
  \let\gplgaddtomacro\g@addto@macro
  \gdef\gplbacktext{}%
  \gdef\gplfronttext{}%
  \makeatother
  \ifGPblacktext
    \def\colorrgb#1{}%
    \def\colorgray#1{}%
  \else
    \ifGPcolor
      \def\colorrgb#1{\color[rgb]{#1}}%
      \def\colorgray#1{\color[gray]{#1}}%
      \expandafter\def\csname LTw\endcsname{\color{white}}%
      \expandafter\def\csname LTb\endcsname{\color{black}}%
      \expandafter\def\csname LTa\endcsname{\color{black}}%
      \expandafter\def\csname LT0\endcsname{\color[rgb]{1,0,0}}%
      \expandafter\def\csname LT1\endcsname{\color[rgb]{0,1,0}}%
      \expandafter\def\csname LT2\endcsname{\color[rgb]{0,0,1}}%
      \expandafter\def\csname LT3\endcsname{\color[rgb]{1,0,1}}%
      \expandafter\def\csname LT4\endcsname{\color[rgb]{0,1,1}}%
      \expandafter\def\csname LT5\endcsname{\color[rgb]{1,1,0}}%
      \expandafter\def\csname LT6\endcsname{\color[rgb]{0,0,0}}%
      \expandafter\def\csname LT7\endcsname{\color[rgb]{1,0.3,0}}%
      \expandafter\def\csname LT8\endcsname{\color[rgb]{0.5,0.5,0.5}}%
    \else
      \def\colorrgb#1{\color{black}}%
      \def\colorgray#1{\color[gray]{#1}}%
      \expandafter\def\csname LTw\endcsname{\color{white}}%
      \expandafter\def\csname LTb\endcsname{\color{black}}%
      \expandafter\def\csname LTa\endcsname{\color{black}}%
      \expandafter\def\csname LT0\endcsname{\color{black}}%
      \expandafter\def\csname LT1\endcsname{\color{black}}%
      \expandafter\def\csname LT2\endcsname{\color{black}}%
      \expandafter\def\csname LT3\endcsname{\color{black}}%
      \expandafter\def\csname LT4\endcsname{\color{black}}%
      \expandafter\def\csname LT5\endcsname{\color{black}}%
      \expandafter\def\csname LT6\endcsname{\color{black}}%
      \expandafter\def\csname LT7\endcsname{\color{black}}%
      \expandafter\def\csname LT8\endcsname{\color{black}}%
    \fi
  \fi
    \setlength{\unitlength}{0.0500bp}%
    \ifx\gptboxheight\undefined%
      \newlength{\gptboxheight}%
      \newlength{\gptboxwidth}%
      \newsavebox{\gptboxtext}%
    \fi%
    \setlength{\fboxrule}{0.5pt}%
    \setlength{\fboxsep}{1pt}%
    \definecolor{tbcol}{rgb}{1,1,1}%
\begin{picture}(5668.00,5668.00)%
    \gplgaddtomacro\gplbacktext{%
      \csname LTb\endcsname
      \put(576,720){\makebox(0,0)[r]{\strut{}$0$}}%
      \csname LTb\endcsname
      \put(576,1565){\makebox(0,0)[r]{\strut{}$500$}}%
      \csname LTb\endcsname
      \put(576,2411){\makebox(0,0)[r]{\strut{}$1000$}}%
      \csname LTb\endcsname
      \put(576,3256){\makebox(0,0)[r]{\strut{}$1500$}}%
      \csname LTb\endcsname
      \put(576,4102){\makebox(0,0)[r]{\strut{}$2000$}}%
      \csname LTb\endcsname
      \put(576,4947){\makebox(0,0)[r]{\strut{}$2500$}}%
      \csname LTb\endcsname
      \put(1365,576){\rotatebox{-45.00}{\makebox(0,0)[l]{\strut{}0}}}%
      \csname LTb\endcsname
      \put(2010,576){\rotatebox{-45.00}{\makebox(0,0)[l]{\strut{}20}}}%
      \csname LTb\endcsname
      \put(2655,576){\rotatebox{-45.00}{\makebox(0,0)[l]{\strut{}40}}}%
      \csname LTb\endcsname
      \put(3300,576){\rotatebox{-45.00}{\makebox(0,0)[l]{\strut{}60}}}%
      \csname LTb\endcsname
      \put(3945,576){\rotatebox{-45.00}{\makebox(0,0)[l]{\strut{}80}}}%
      \csname LTb\endcsname
      \put(4590,576){\rotatebox{-45.00}{\makebox(0,0)[l]{\strut{}100}}}%
    }%
    \gplgaddtomacro\gplfronttext{%
      \csname LTb\endcsname
      \put(-228,2833){\rotatebox{-270.00}{\makebox(0,0){\strut{}Time (s)}}}%
      \put(2977,55){\makebox(0,0){\strut{}Malleability (\%)}}%
    }%
    \gplbacktext
    \put(0,0){\includegraphics[width={283.40bp},height={283.40bp}]{img//HaswellTurnaroundMeanPartialStatistic}}%
    \gplfronttext
  \end{picture}%
\endgroup

%% file: img/HaswellMakespanMeanPartialStatistic.tex
\begingroup
\LARGE
  \makeatletter
  \providecommand\color[2][]{%
    \GenericError{(gnuplot) \space\space\space\@spaces}{%
      Package color not loaded in conjunction with
      terminal option `colourtext'%
    }{See the gnuplot documentation for explanation.%
    }{Either use 'blacktext' in gnuplot or load the package
      color.sty in LaTeX.}%
    \renewcommand\color[2][]{}%
  }%
  \providecommand\includegraphics[2][]{%
    \GenericError{(gnuplot) \space\space\space\@spaces}{%
      Package graphicx or graphics not loaded%
    }{See the gnuplot documentation for explanation.%
    }{The gnuplot epslatex terminal needs graphicx.sty or graphics.sty.}%
    \renewcommand\includegraphics[2][]{}%
  }%
  \providecommand\rotatebox[2]{#2}%
  \@ifundefined{ifGPcolor}{%
    \newif\ifGPcolor
    \GPcolortrue
  }{}%
  \@ifundefined{ifGPblacktext}{%
    \newif\ifGPblacktext
    \GPblacktextfalse
  }{}%
  \let\gplgaddtomacro\g@addto@macro
  \gdef\gplbacktext{}%
  \gdef\gplfronttext{}%
  \makeatother
  \ifGPblacktext
    \def\colorrgb#1{}%
    \def\colorgray#1{}%
  \else
    \ifGPcolor
      \def\colorrgb#1{\color[rgb]{#1}}%
      \def\colorgray#1{\color[gray]{#1}}%
      \expandafter\def\csname LTw\endcsname{\color{white}}%
      \expandafter\def\csname LTb\endcsname{\color{black}}%
      \expandafter\def\csname LTa\endcsname{\color{black}}%
      \expandafter\def\csname LT0\endcsname{\color[rgb]{1,0,0}}%
      \expandafter\def\csname LT1\endcsname{\color[rgb]{0,1,0}}%
      \expandafter\def\csname LT2\endcsname{\color[rgb]{0,0,1}}%
      \expandafter\def\csname LT3\endcsname{\color[rgb]{1,0,1}}%
      \expandafter\def\csname LT4\endcsname{\color[rgb]{0,1,1}}%
      \expandafter\def\csname LT5\endcsname{\color[rgb]{1,1,0}}%
      \expandafter\def\csname LT6\endcsname{\color[rgb]{0,0,0}}%
      \expandafter\def\csname LT7\endcsname{\color[rgb]{1,0.3,0}}%
      \expandafter\def\csname LT8\endcsname{\color[rgb]{0.5,0.5,0.5}}%
    \else
      \def\colorrgb#1{\color{black}}%
      \def\colorgray#1{\color[gray]{#1}}%
      \expandafter\def\csname LTw\endcsname{\color{white}}%
      \expandafter\def\csname LTb\endcsname{\color{black}}%
      \expandafter\def\csname LTa\endcsname{\color{black}}%
      \expandafter\def\csname LT0\endcsname{\color{black}}%
      \expandafter\def\csname LT1\endcsname{\color{black}}%
      \expandafter\def\csname LT2\endcsname{\color{black}}%
      \expandafter\def\csname LT3\endcsname{\color{black}}%
      \expandafter\def\csname LT4\endcsname{\color{black}}%
      \expandafter\def\csname LT5\endcsname{\color{black}}%
      \expandafter\def\csname LT6\endcsname{\color{black}}%
      \expandafter\def\csname LT7\endcsname{\color{black}}%
      \expandafter\def\csname LT8\endcsname{\color{black}}%
    \fi
  \fi
    \setlength{\unitlength}{0.0500bp}%
    \ifx\gptboxheight\undefined%
      \newlength{\gptboxheight}%
      \newlength{\gptboxwidth}%
      \newsavebox{\gptboxtext}%
    \fi%
    \setlength{\fboxrule}{0.5pt}%
    \setlength{\fboxsep}{1pt}%
    \definecolor{tbcol}{rgb}{1,1,1}%
\begin{picture}(5668.00,5668.00)%
    \gplgaddtomacro\gplbacktext{%
      \csname LTb\endcsname
      \put(576,720){\makebox(0,0)[r]{\strut{}$0$}}%
      \csname LTb\endcsname
      \put(576,1565){\makebox(0,0)[r]{\strut{}$500$}}%
      \csname LTb\endcsname
      \put(576,2411){\makebox(0,0)[r]{\strut{}$1000$}}%
      \csname LTb\endcsname
      \put(576,3256){\makebox(0,0)[r]{\strut{}$1500$}}%
      \csname LTb\endcsname
      \put(576,4102){\makebox(0,0)[r]{\strut{}$2000$}}%
      \csname LTb\endcsname
      \put(576,4947){\makebox(0,0)[r]{\strut{}$2500$}}%
      \csname LTb\endcsname
      \put(1365,576){\rotatebox{-45.00}{\makebox(0,0)[l]{\strut{}0}}}%
      \csname LTb\endcsname
      \put(2010,576){\rotatebox{-45.00}{\makebox(0,0)[l]{\strut{}20}}}%
      \csname LTb\endcsname
      \put(2655,576){\rotatebox{-45.00}{\makebox(0,0)[l]{\strut{}40}}}%
      \csname LTb\endcsname
      \put(3300,576){\rotatebox{-45.00}{\makebox(0,0)[l]{\strut{}60}}}%
      \csname LTb\endcsname
      \put(3945,576){\rotatebox{-45.00}{\makebox(0,0)[l]{\strut{}80}}}%
      \csname LTb\endcsname
      \put(4590,576){\rotatebox{-45.00}{\makebox(0,0)[l]{\strut{}100}}}%
    }%
    \gplgaddtomacro\gplfronttext{%
      \csname LTb\endcsname
      \put(-228,2833){\rotatebox{-270.00}{\makebox(0,0){\strut{}Time (s)}}}%
      \put(2977,55){\makebox(0,0){\strut{}Malleability (\%)}}%
    }%
    \gplbacktext
    \put(0,0){\includegraphics[width={283.40bp},height={283.40bp}]{img//HaswellMakespanMeanPartialStatistic}}%
    \gplfronttext
  \end{picture}%
\endgroup

%% file: img/HaswellWaittimeMeanPartialStatistic.tex
\begingroup
\LARGE
  \makeatletter
  \providecommand\color[2][]{%
    \GenericError{(gnuplot) \space\space\space\@spaces}{%
      Package color not loaded in conjunction with
      terminal option `colourtext'%
    }{See the gnuplot documentation for explanation.%
    }{Either use 'blacktext' in gnuplot or load the package
      color.sty in LaTeX.}%
    \renewcommand\color[2][]{}%
  }%
  \providecommand\includegraphics[2][]{%
    \GenericError{(gnuplot) \space\space\space\@spaces}{%
      Package graphicx or graphics not loaded%
    }{See the gnuplot documentation for explanation.%
    }{The gnuplot epslatex terminal needs graphicx.sty or graphics.sty.}%
    \renewcommand\includegraphics[2][]{}%
  }%
  \providecommand\rotatebox[2]{#2}%
  \@ifundefined{ifGPcolor}{%
    \newif\ifGPcolor
    \GPcolortrue
  }{}%
  \@ifundefined{ifGPblacktext}{%
    \newif\ifGPblacktext
    \GPblacktextfalse
  }{}%
  \let\gplgaddtomacro\g@addto@macro
  \gdef\gplbacktext{}%
  \gdef\gplfronttext{}%
  \makeatother
  \ifGPblacktext
    \def\colorrgb#1{}%
    \def\colorgray#1{}%
  \else
    \ifGPcolor
      \def\colorrgb#1{\color[rgb]{#1}}%
      \def\colorgray#1{\color[gray]{#1}}%
      \expandafter\def\csname LTw\endcsname{\color{white}}%
      \expandafter\def\csname LTb\endcsname{\color{black}}%
      \expandafter\def\csname LTa\endcsname{\color{black}}%
      \expandafter\def\csname LT0\endcsname{\color[rgb]{1,0,0}}%
      \expandafter\def\csname LT1\endcsname{\color[rgb]{0,1,0}}%
      \expandafter\def\csname LT2\endcsname{\color[rgb]{0,0,1}}%
      \expandafter\def\csname LT3\endcsname{\color[rgb]{1,0,1}}%
      \expandafter\def\csname LT4\endcsname{\color[rgb]{0,1,1}}%
      \expandafter\def\csname LT5\endcsname{\color[rgb]{1,1,0}}%
      \expandafter\def\csname LT6\endcsname{\color[rgb]{0,0,0}}%
      \expandafter\def\csname LT7\endcsname{\color[rgb]{1,0.3,0}}%
      \expandafter\def\csname LT8\endcsname{\color[rgb]{0.5,0.5,0.5}}%
    \else
      \def\colorrgb#1{\color{black}}%
      \def\colorgray#1{\color[gray]{#1}}%
      \expandafter\def\csname LTw\endcsname{\color{white}}%
      \expandafter\def\csname LTb\endcsname{\color{black}}%
      \expandafter\def\csname LTa\endcsname{\color{black}}%
      \expandafter\def\csname LT0\endcsname{\color{black}}%
      \expandafter\def\csname LT1\endcsname{\color{black}}%
      \expandafter\def\csname LT2\endcsname{\color{black}}%
      \expandafter\def\csname LT3\endcsname{\color{black}}%
      \expandafter\def\csname LT4\endcsname{\color{black}}%
      \expandafter\def\csname LT5\endcsname{\color{black}}%
      \expandafter\def\csname LT6\endcsname{\color{black}}%
      \expandafter\def\csname LT7\endcsname{\color{black}}%
      \expandafter\def\csname LT8\endcsname{\color{black}}%
    \fi
  \fi
    \setlength{\unitlength}{0.0500bp}%
    \ifx\gptboxheight\undefined%
      \newlength{\gptboxheight}%
      \newlength{\gptboxwidth}%
      \newsavebox{\gptboxtext}%
    \fi%
    \setlength{\fboxrule}{0.5pt}%
    \setlength{\fboxsep}{1pt}%
    \definecolor{tbcol}{rgb}{1,1,1}%
\begin{picture}(5668.00,5668.00)%
    \gplgaddtomacro\gplbacktext{%
      \csname LTb\endcsname
      \put(576,720){\makebox(0,0)[r]{\strut{}$0$}}%
      \csname LTb\endcsname
      \put(576,1425){\makebox(0,0)[r]{\strut{}$20$}}%
      \csname LTb\endcsname
      \put(576,2129){\makebox(0,0)[r]{\strut{}$40$}}%
      \csname LTb\endcsname
      \put(576,2834){\makebox(0,0)[r]{\strut{}$60$}}%
      \csname LTb\endcsname
      \put(576,3538){\makebox(0,0)[r]{\strut{}$80$}}%
      \csname LTb\endcsname
      \put(576,4243){\makebox(0,0)[r]{\strut{}$100$}}%
      \csname LTb\endcsname
      \put(576,4947){\makebox(0,0)[r]{\strut{}$120$}}%
      \csname LTb\endcsname
      \put(1365,576){\rotatebox{-45.00}{\makebox(0,0)[l]{\strut{}0}}}%
      \csname LTb\endcsname
      \put(2010,576){\rotatebox{-45.00}{\makebox(0,0)[l]{\strut{}20}}}%
      \csname LTb\endcsname
      \put(2655,576){\rotatebox{-45.00}{\makebox(0,0)[l]{\strut{}40}}}%
      \csname LTb\endcsname
      \put(3300,576){\rotatebox{-45.00}{\makebox(0,0)[l]{\strut{}60}}}%
      \csname LTb\endcsname
      \put(3945,576){\rotatebox{-45.00}{\makebox(0,0)[l]{\strut{}80}}}%
      \csname LTb\endcsname
      \put(4590,576){\rotatebox{-45.00}{\makebox(0,0)[l]{\strut{}100}}}%
    }%
    \gplgaddtomacro\gplfronttext{%
      \csname LTb\endcsname
      \put(-84,2833){\rotatebox{-270.00}{\makebox(0,0){\strut{}Time (s)}}}%
      \put(2977,55){\makebox(0,0){\strut{}Malleability (\%)}}%
    }%
    \gplbacktext
    \put(0,0){\includegraphics[width={283.40bp},height={283.40bp}]{img//HaswellWaittimeMeanPartialStatistic}}%
    \gplfronttext
  \end{picture}%
\endgroup

%% file: img/HaswellNodeutilizationMeanPartialStatistic.tex
\begingroup
\LARGE
  \makeatletter
  \providecommand\color[2][]{%
    \GenericError{(gnuplot) \space\space\space\@spaces}{%
      Package color not loaded in conjunction with
      terminal option `colourtext'%
    }{See the gnuplot documentation for explanation.%
    }{Either use 'blacktext' in gnuplot or load the package
      color.sty in LaTeX.}%
    \renewcommand\color[2][]{}%
  }%
  \providecommand\includegraphics[2][]{%
    \GenericError{(gnuplot) \space\space\space\@spaces}{%
      Package graphicx or graphics not loaded%
    }{See the gnuplot documentation for explanation.%
    }{The gnuplot epslatex terminal needs graphicx.sty or graphics.sty.}%
    \renewcommand\includegraphics[2][]{}%
  }%
  \providecommand\rotatebox[2]{#2}%
  \@ifundefined{ifGPcolor}{%
    \newif\ifGPcolor
    \GPcolortrue
  }{}%
  \@ifundefined{ifGPblacktext}{%
    \newif\ifGPblacktext
    \GPblacktextfalse
  }{}%
  \let\gplgaddtomacro\g@addto@macro
  \gdef\gplbacktext{}%
  \gdef\gplfronttext{}%
  \makeatother
  \ifGPblacktext
    \def\colorrgb#1{}%
    \def\colorgray#1{}%
  \else
    \ifGPcolor
      \def\colorrgb#1{\color[rgb]{#1}}%
      \def\colorgray#1{\color[gray]{#1}}%
      \expandafter\def\csname LTw\endcsname{\color{white}}%
      \expandafter\def\csname LTb\endcsname{\color{black}}%
      \expandafter\def\csname LTa\endcsname{\color{black}}%
      \expandafter\def\csname LT0\endcsname{\color[rgb]{1,0,0}}%
      \expandafter\def\csname LT1\endcsname{\color[rgb]{0,1,0}}%
      \expandafter\def\csname LT2\endcsname{\color[rgb]{0,0,1}}%
      \expandafter\def\csname LT3\endcsname{\color[rgb]{1,0,1}}%
      \expandafter\def\csname LT4\endcsname{\color[rgb]{0,1,1}}%
      \expandafter\def\csname LT5\endcsname{\color[rgb]{1,1,0}}%
      \expandafter\def\csname LT6\endcsname{\color[rgb]{0,0,0}}%
      \expandafter\def\csname LT7\endcsname{\color[rgb]{1,0.3,0}}%
      \expandafter\def\csname LT8\endcsname{\color[rgb]{0.5,0.5,0.5}}%
    \else
      \def\colorrgb#1{\color{black}}%
      \def\colorgray#1{\color[gray]{#1}}%
      \expandafter\def\csname LTw\endcsname{\color{white}}%
      \expandafter\def\csname LTb\endcsname{\color{black}}%
      \expandafter\def\csname LTa\endcsname{\color{black}}%
      \expandafter\def\csname LT0\endcsname{\color{black}}%
      \expandafter\def\csname LT1\endcsname{\color{black}}%
      \expandafter\def\csname LT2\endcsname{\color{black}}%
      \expandafter\def\csname LT3\endcsname{\color{black}}%
      \expandafter\def\csname LT4\endcsname{\color{black}}%
      \expandafter\def\csname LT5\endcsname{\color{black}}%
      \expandafter\def\csname LT6\endcsname{\color{black}}%
      \expandafter\def\csname LT7\endcsname{\color{black}}%
      \expandafter\def\csname LT8\endcsname{\color{black}}%
    \fi
  \fi
    \setlength{\unitlength}{0.0500bp}%
    \ifx\gptboxheight\undefined%
      \newlength{\gptboxheight}%
      \newlength{\gptboxwidth}%
      \newsavebox{\gptboxtext}%
    \fi%
    \setlength{\fboxrule}{0.5pt}%
    \setlength{\fboxsep}{1pt}%
    \definecolor{tbcol}{rgb}{1,1,1}%
\begin{picture}(5668.00,5668.00)%
    \gplgaddtomacro\gplbacktext{%
      \csname LTb\endcsname
      \put(576,720){\makebox(0,0)[r]{\strut{}$60$}}%
      \csname LTb\endcsname
      \put(576,1248){\makebox(0,0)[r]{\strut{}$65$}}%
      \csname LTb\endcsname
      \put(576,1777){\makebox(0,0)[r]{\strut{}$70$}}%
      \csname LTb\endcsname
      \put(576,2305){\makebox(0,0)[r]{\strut{}$75$}}%
      \csname LTb\endcsname
      \put(576,2834){\makebox(0,0)[r]{\strut{}$80$}}%
      \csname LTb\endcsname
      \put(576,3362){\makebox(0,0)[r]{\strut{}$85$}}%
      \csname LTb\endcsname
      \put(576,3890){\makebox(0,0)[r]{\strut{}$90$}}%
      \csname LTb\endcsname
      \put(576,4419){\makebox(0,0)[r]{\strut{}$95$}}%
      \csname LTb\endcsname
      \put(576,4947){\makebox(0,0)[r]{\strut{}$100$}}%
      \csname LTb\endcsname
      \put(1365,576){\rotatebox{-45.00}{\makebox(0,0)[l]{\strut{}0}}}%
      \csname LTb\endcsname
      \put(2010,576){\rotatebox{-45.00}{\makebox(0,0)[l]{\strut{}20}}}%
      \csname LTb\endcsname
      \put(2655,576){\rotatebox{-45.00}{\makebox(0,0)[l]{\strut{}40}}}%
      \csname LTb\endcsname
      \put(3300,576){\rotatebox{-45.00}{\makebox(0,0)[l]{\strut{}60}}}%
      \csname LTb\endcsname
      \put(3945,576){\rotatebox{-45.00}{\makebox(0,0)[l]{\strut{}80}}}%
      \csname LTb\endcsname
      \put(4590,576){\rotatebox{-45.00}{\makebox(0,0)[l]{\strut{}100}}}%
    }%
    \gplgaddtomacro\gplfronttext{%
      \csname LTb\endcsname
      \put(-84,2833){\rotatebox{-270.00}{\makebox(0,0){\strut{}Node Utilization (\%)}}}%
      \put(2977,55){\makebox(0,0){\strut{}Malleability (\%)}}%
    }%
    \gplbacktext
    \put(0,0){\includegraphics[width={283.40bp},height={283.40bp}]{img//HaswellNodeutilizationMeanPartialStatistic}}%
    \gplfronttext
  \end{picture}%
\endgroup

%% file: img/HaswellExpandsJobPartialStatistic.tex
\begingroup
\LARGE
  \makeatletter
  \providecommand\color[2][]{%
    \GenericError{(gnuplot) \space\space\space\@spaces}{%
      Package color not loaded in conjunction with
      terminal option `colourtext'%
    }{See the gnuplot documentation for explanation.%
    }{Either use 'blacktext' in gnuplot or load the package
      color.sty in LaTeX.}%
    \renewcommand\color[2][]{}%
  }%
  \providecommand\includegraphics[2][]{%
    \GenericError{(gnuplot) \space\space\space\@spaces}{%
      Package graphicx or graphics not loaded%
    }{See the gnuplot documentation for explanation.%
    }{The gnuplot epslatex terminal needs graphicx.sty or graphics.sty.}%
    \renewcommand\includegraphics[2][]{}%
  }%
  \providecommand\rotatebox[2]{#2}%
  \@ifundefined{ifGPcolor}{%
    \newif\ifGPcolor
    \GPcolortrue
  }{}%
  \@ifundefined{ifGPblacktext}{%
    \newif\ifGPblacktext
    \GPblacktextfalse
  }{}%
  \let\gplgaddtomacro\g@addto@macro
  \gdef\gplbacktext{}%
  \gdef\gplfronttext{}%
  \makeatother
  \ifGPblacktext
    \def\colorrgb#1{}%
    \def\colorgray#1{}%
  \else
    \ifGPcolor
      \def\colorrgb#1{\color[rgb]{#1}}%
      \def\colorgray#1{\color[gray]{#1}}%
      \expandafter\def\csname LTw\endcsname{\color{white}}%
      \expandafter\def\csname LTb\endcsname{\color{black}}%
      \expandafter\def\csname LTa\endcsname{\color{black}}%
      \expandafter\def\csname LT0\endcsname{\color[rgb]{1,0,0}}%
      \expandafter\def\csname LT1\endcsname{\color[rgb]{0,1,0}}%
      \expandafter\def\csname LT2\endcsname{\color[rgb]{0,0,1}}%
      \expandafter\def\csname LT3\endcsname{\color[rgb]{1,0,1}}%
      \expandafter\def\csname LT4\endcsname{\color[rgb]{0,1,1}}%
      \expandafter\def\csname LT5\endcsname{\color[rgb]{1,1,0}}%
      \expandafter\def\csname LT6\endcsname{\color[rgb]{0,0,0}}%
      \expandafter\def\csname LT7\endcsname{\color[rgb]{1,0.3,0}}%
      \expandafter\def\csname LT8\endcsname{\color[rgb]{0.5,0.5,0.5}}%
    \else
      \def\colorrgb#1{\color{black}}%
      \def\colorgray#1{\color[gray]{#1}}%
      \expandafter\def\csname LTw\endcsname{\color{white}}%
      \expandafter\def\csname LTb\endcsname{\color{black}}%
      \expandafter\def\csname LTa\endcsname{\color{black}}%
      \expandafter\def\csname LT0\endcsname{\color{black}}%
      \expandafter\def\csname LT1\endcsname{\color{black}}%
      \expandafter\def\csname LT2\endcsname{\color{black}}%
      \expandafter\def\csname LT3\endcsname{\color{black}}%
      \expandafter\def\csname LT4\endcsname{\color{black}}%
      \expandafter\def\csname LT5\endcsname{\color{black}}%
      \expandafter\def\csname LT6\endcsname{\color{black}}%
      \expandafter\def\csname LT7\endcsname{\color{black}}%
      \expandafter\def\csname LT8\endcsname{\color{black}}%
    \fi
  \fi
    \setlength{\unitlength}{0.0500bp}%
    \ifx\gptboxheight\undefined%
      \newlength{\gptboxheight}%
      \newlength{\gptboxwidth}%
      \newsavebox{\gptboxtext}%
    \fi%
    \setlength{\fboxrule}{0.5pt}%
    \setlength{\fboxsep}{1pt}%
    \definecolor{tbcol}{rgb}{1,1,1}%
\begin{picture}(5668.00,5668.00)%
    \gplgaddtomacro\gplbacktext{%
      \csname LTb\endcsname
      \put(576,720){\makebox(0,0)[r]{\strut{}$0$}}%
      \csname LTb\endcsname
      \put(576,1143){\makebox(0,0)[r]{\strut{}$2$}}%
      \csname LTb\endcsname
      \put(576,1565){\makebox(0,0)[r]{\strut{}$4$}}%
      \csname LTb\endcsname
      \put(576,1988){\makebox(0,0)[r]{\strut{}$6$}}%
      \csname LTb\endcsname
      \put(576,2411){\makebox(0,0)[r]{\strut{}$8$}}%
      \csname LTb\endcsname
      \put(576,2834){\makebox(0,0)[r]{\strut{}$10$}}%
      \csname LTb\endcsname
      \put(576,3256){\makebox(0,0)[r]{\strut{}$12$}}%
      \csname LTb\endcsname
      \put(576,3679){\makebox(0,0)[r]{\strut{}$14$}}%
      \csname LTb\endcsname
      \put(576,4102){\makebox(0,0)[r]{\strut{}$16$}}%
      \csname LTb\endcsname
      \put(576,4524){\makebox(0,0)[r]{\strut{}$18$}}%
      \csname LTb\endcsname
      \put(576,4947){\makebox(0,0)[r]{\strut{}$20$}}%
      \csname LTb\endcsname
      \put(1365,576){\rotatebox{-45.00}{\makebox(0,0)[l]{\strut{}0}}}%
      \csname LTb\endcsname
      \put(2010,576){\rotatebox{-45.00}{\makebox(0,0)[l]{\strut{}20}}}%
      \csname LTb\endcsname
      \put(2655,576){\rotatebox{-45.00}{\makebox(0,0)[l]{\strut{}40}}}%
      \csname LTb\endcsname
      \put(3300,576){\rotatebox{-45.00}{\makebox(0,0)[l]{\strut{}60}}}%
      \csname LTb\endcsname
      \put(3945,576){\rotatebox{-45.00}{\makebox(0,0)[l]{\strut{}80}}}%
      \csname LTb\endcsname
      \put(4590,576){\rotatebox{-45.00}{\makebox(0,0)[l]{\strut{}100}}}%
    }%
    \gplgaddtomacro\gplfronttext{%
      \csname LTb\endcsname
      \put(60,2833){\rotatebox{-270.00}{\makebox(0,0){\strut{}Events}}}%
      \put(2977,55){\makebox(0,0){\strut{}Malleability (\%)}}%
    }%
    \gplbacktext
    \put(0,0){\includegraphics[width={283.40bp},height={283.40bp}]{img//HaswellExpandsJobPartialStatistic}}%
    \gplfronttext
  \end{picture}%
\endgroup

%% file: img/HaswellShrinksJobPartialStatistic.tex
\begingroup
\LARGE
  \makeatletter
  \providecommand\color[2][]{%
    \GenericError{(gnuplot) \space\space\space\@spaces}{%
      Package color not loaded in conjunction with
      terminal option `colourtext'%
    }{See the gnuplot documentation for explanation.%
    }{Either use 'blacktext' in gnuplot or load the package
      color.sty in LaTeX.}%
    \renewcommand\color[2][]{}%
  }%
  \providecommand\includegraphics[2][]{%
    \GenericError{(gnuplot) \space\space\space\@spaces}{%
      Package graphicx or graphics not loaded%
    }{See the gnuplot documentation for explanation.%
    }{The gnuplot epslatex terminal needs graphicx.sty or graphics.sty.}%
    \renewcommand\includegraphics[2][]{}%
  }%
  \providecommand\rotatebox[2]{#2}%
  \@ifundefined{ifGPcolor}{%
    \newif\ifGPcolor
    \GPcolortrue
  }{}%
  \@ifundefined{ifGPblacktext}{%
    \newif\ifGPblacktext
    \GPblacktextfalse
  }{}%
  \let\gplgaddtomacro\g@addto@macro
  \gdef\gplbacktext{}%
  \gdef\gplfronttext{}%
  \makeatother
  \ifGPblacktext
    \def\colorrgb#1{}%
    \def\colorgray#1{}%
  \else
    \ifGPcolor
      \def\colorrgb#1{\color[rgb]{#1}}%
      \def\colorgray#1{\color[gray]{#1}}%
      \expandafter\def\csname LTw\endcsname{\color{white}}%
      \expandafter\def\csname LTb\endcsname{\color{black}}%
      \expandafter\def\csname LTa\endcsname{\color{black}}%
      \expandafter\def\csname LT0\endcsname{\color[rgb]{1,0,0}}%
      \expandafter\def\csname LT1\endcsname{\color[rgb]{0,1,0}}%
      \expandafter\def\csname LT2\endcsname{\color[rgb]{0,0,1}}%
      \expandafter\def\csname LT3\endcsname{\color[rgb]{1,0,1}}%
      \expandafter\def\csname LT4\endcsname{\color[rgb]{0,1,1}}%
      \expandafter\def\csname LT5\endcsname{\color[rgb]{1,1,0}}%
      \expandafter\def\csname LT6\endcsname{\color[rgb]{0,0,0}}%
      \expandafter\def\csname LT7\endcsname{\color[rgb]{1,0.3,0}}%
      \expandafter\def\csname LT8\endcsname{\color[rgb]{0.5,0.5,0.5}}%
    \else
      \def\colorrgb#1{\color{black}}%
      \def\colorgray#1{\color[gray]{#1}}%
      \expandafter\def\csname LTw\endcsname{\color{white}}%
      \expandafter\def\csname LTb\endcsname{\color{black}}%
      \expandafter\def\csname LTa\endcsname{\color{black}}%
      \expandafter\def\csname LT0\endcsname{\color{black}}%
      \expandafter\def\csname LT1\endcsname{\color{black}}%
      \expandafter\def\csname LT2\endcsname{\color{black}}%
      \expandafter\def\csname LT3\endcsname{\color{black}}%
      \expandafter\def\csname LT4\endcsname{\color{black}}%
      \expandafter\def\csname LT5\endcsname{\color{black}}%
      \expandafter\def\csname LT6\endcsname{\color{black}}%
      \expandafter\def\csname LT7\endcsname{\color{black}}%
      \expandafter\def\csname LT8\endcsname{\color{black}}%
    \fi
  \fi
    \setlength{\unitlength}{0.0500bp}%
    \ifx\gptboxheight\undefined%
      \newlength{\gptboxheight}%
      \newlength{\gptboxwidth}%
      \newsavebox{\gptboxtext}%
    \fi%
    \setlength{\fboxrule}{0.5pt}%
    \setlength{\fboxsep}{1pt}%
    \definecolor{tbcol}{rgb}{1,1,1}%
\begin{picture}(5668.00,5668.00)%
    \gplgaddtomacro\gplbacktext{%
      \csname LTb\endcsname
      \put(576,720){\makebox(0,0)[r]{\strut{}$0$}}%
      \csname LTb\endcsname
      \put(576,1425){\makebox(0,0)[r]{\strut{}$5$}}%
      \csname LTb\endcsname
      \put(576,2129){\makebox(0,0)[r]{\strut{}$10$}}%
      \csname LTb\endcsname
      \put(576,2834){\makebox(0,0)[r]{\strut{}$15$}}%
      \csname LTb\endcsname
      \put(576,3538){\makebox(0,0)[r]{\strut{}$20$}}%
      \csname LTb\endcsname
      \put(576,4243){\makebox(0,0)[r]{\strut{}$25$}}%
      \csname LTb\endcsname
      \put(576,4947){\makebox(0,0)[r]{\strut{}$30$}}%
      \csname LTb\endcsname
      \put(1365,576){\rotatebox{-45.00}{\makebox(0,0)[l]{\strut{}0}}}%
      \csname LTb\endcsname
      \put(2010,576){\rotatebox{-45.00}{\makebox(0,0)[l]{\strut{}20}}}%
      \csname LTb\endcsname
      \put(2655,576){\rotatebox{-45.00}{\makebox(0,0)[l]{\strut{}40}}}%
      \csname LTb\endcsname
      \put(3300,576){\rotatebox{-45.00}{\makebox(0,0)[l]{\strut{}60}}}%
      \csname LTb\endcsname
      \put(3945,576){\rotatebox{-45.00}{\makebox(0,0)[l]{\strut{}80}}}%
      \csname LTb\endcsname
      \put(4590,576){\rotatebox{-45.00}{\makebox(0,0)[l]{\strut{}100}}}%
    }%
    \gplgaddtomacro\gplfronttext{%
      \csname LTb\endcsname
      \put(60,2833){\rotatebox{-270.00}{\makebox(0,0){\strut{}Events}}}%
      \put(2977,55){\makebox(0,0){\strut{}Malleability (\%)}}%
    }%
    \gplbacktext
    \put(0,0){\includegraphics[width={283.40bp},height={283.40bp}]{img//HaswellShrinksJobPartialStatistic}}%
    \gplfronttext
  \end{picture}%
\endgroup

%% file: img/Legend.tex
\begingroup
\LARGE
  \makeatletter
  \providecommand\color[2][]{%
    \GenericError{(gnuplot) \space\space\space\@spaces}{%
      Package color not loaded in conjunction with
      terminal option `colourtext'%
    }{See the gnuplot documentation for explanation.%
    }{Either use 'blacktext' in gnuplot or load the package
      color.sty in LaTeX.}%
    \renewcommand\color[2][]{}%
  }%
  \providecommand\includegraphics[2][]{%
    \GenericError{(gnuplot) \space\space\space\@spaces}{%
      Package graphicx or graphics not loaded%
    }{See the gnuplot documentation for explanation.%
    }{The gnuplot epslatex terminal needs graphicx.sty or graphics.sty.}%
    \renewcommand\includegraphics[2][]{}%
  }%
  \providecommand\rotatebox[2]{#2}%
  \@ifundefined{ifGPcolor}{%
    \newif\ifGPcolor
    \GPcolortrue
  }{}%
  \@ifundefined{ifGPblacktext}{%
    \newif\ifGPblacktext
    \GPblacktextfalse
  }{}%
  \let\gplgaddtomacro\g@addto@macro
  \gdef\gplbacktext{}%
  \gdef\gplfronttext{}%
  \makeatother
  \ifGPblacktext
    \def\colorrgb#1{}%
    \def\colorgray#1{}%
  \else
    \ifGPcolor
      \def\colorrgb#1{\color[rgb]{#1}}%
      \def\colorgray#1{\color[gray]{#1}}%
      \expandafter\def\csname LTw\endcsname{\color{white}}%
      \expandafter\def\csname LTb\endcsname{\color{black}}%
      \expandafter\def\csname LTa\endcsname{\color{black}}%
      \expandafter\def\csname LT0\endcsname{\color[rgb]{1,0,0}}%
      \expandafter\def\csname LT1\endcsname{\color[rgb]{0,1,0}}%
      \expandafter\def\csname LT2\endcsname{\color[rgb]{0,0,1}}%
      \expandafter\def\csname LT3\endcsname{\color[rgb]{1,0,1}}%
      \expandafter\def\csname LT4\endcsname{\color[rgb]{0,1,1}}%
      \expandafter\def\csname LT5\endcsname{\color[rgb]{1,1,0}}%
      \expandafter\def\csname LT6\endcsname{\color[rgb]{0,0,0}}%
      \expandafter\def\csname LT7\endcsname{\color[rgb]{1,0.3,0}}%
      \expandafter\def\csname LT8\endcsname{\color[rgb]{0.5,0.5,0.5}}%
    \else
      \def\colorrgb#1{\color{black}}%
      \def\colorgray#1{\color[gray]{#1}}%
      \expandafter\def\csname LTw\endcsname{\color{white}}%
      \expandafter\def\csname LTb\endcsname{\color{black}}%
      \expandafter\def\csname LTa\endcsname{\color{black}}%
      \expandafter\def\csname LT0\endcsname{\color{black}}%
      \expandafter\def\csname LT1\endcsname{\color{black}}%
      \expandafter\def\csname LT2\endcsname{\color{black}}%
      \expandafter\def\csname LT3\endcsname{\color{black}}%
      \expandafter\def\csname LT4\endcsname{\color{black}}%
      \expandafter\def\csname LT5\endcsname{\color{black}}%
      \expandafter\def\csname LT6\endcsname{\color{black}}%
      \expandafter\def\csname LT7\endcsname{\color{black}}%
      \expandafter\def\csname LT8\endcsname{\color{black}}%
    \fi
  \fi
    \setlength{\unitlength}{0.0500bp}%
    \ifx\gptboxheight\undefined%
      \newlength{\gptboxheight}%
      \newlength{\gptboxwidth}%
      \newsavebox{\gptboxtext}%
    \fi%
    \setlength{\fboxrule}{0.5pt}%
    \setlength{\fboxsep}{1pt}%
    \definecolor{tbcol}{rgb}{1,1,1}%
\begin{picture}(14172.00,1132.00)%
    \gplgaddtomacro\gplbacktext{%
    }%
    \gplgaddtomacro\gplfronttext{%
      \csname LTb\endcsname
      \put(4075,648){\makebox(0,0)[r]{\strut{}\easybackfill}}%
      \csname LTb\endcsname
      \put(4075,288){\makebox(0,0)[r]{\strut{}\prefkeeper}}%
      \csname LTb\endcsname
      \put(8170,648){\makebox(0,0)[r]{\strut{}\pref}}%
      \csname LTb\endcsname
      \put(8170,288){\makebox(0,0)[r]{\strut{}\min}}%
      \csname LTb\endcsname
      \put(12265,648){\makebox(0,0)[r]{\strut{}\avg}}%
    }%
    \gplbacktext
    \put(0,0){\includegraphics[width={708.60bp},height={56.60bp}]{img//Legend}}%
    \gplfronttext
  \end{picture}%
\endgroup

%% file: img/KNLTurnaroundMeanPartialStatistic.tex
\begingroup
\LARGE
  \makeatletter
  \providecommand\color[2][]{%
    \GenericError{(gnuplot) \space\space\space\@spaces}{%
      Package color not loaded in conjunction with
      terminal option `colourtext'%
    }{See the gnuplot documentation for explanation.%
    }{Either use 'blacktext' in gnuplot or load the package
      color.sty in LaTeX.}%
    \renewcommand\color[2][]{}%
  }%
  \providecommand\includegraphics[2][]{%
    \GenericError{(gnuplot) \space\space\space\@spaces}{%
      Package graphicx or graphics not loaded%
    }{See the gnuplot documentation for explanation.%
    }{The gnuplot epslatex terminal needs graphicx.sty or graphics.sty.}%
    \renewcommand\includegraphics[2][]{}%
  }%
  \providecommand\rotatebox[2]{#2}%
  \@ifundefined{ifGPcolor}{%
    \newif\ifGPcolor
    \GPcolortrue
  }{}%
  \@ifundefined{ifGPblacktext}{%
    \newif\ifGPblacktext
    \GPblacktextfalse
  }{}%
  \let\gplgaddtomacro\g@addto@macro
  \gdef\gplbacktext{}%
  \gdef\gplfronttext{}%
  \makeatother
  \ifGPblacktext
    \def\colorrgb#1{}%
    \def\colorgray#1{}%
  \else
    \ifGPcolor
      \def\colorrgb#1{\color[rgb]{#1}}%
      \def\colorgray#1{\color[gray]{#1}}%
      \expandafter\def\csname LTw\endcsname{\color{white}}%
      \expandafter\def\csname LTb\endcsname{\color{black}}%
      \expandafter\def\csname LTa\endcsname{\color{black}}%
      \expandafter\def\csname LT0\endcsname{\color[rgb]{1,0,0}}%
      \expandafter\def\csname LT1\endcsname{\color[rgb]{0,1,0}}%
      \expandafter\def\csname LT2\endcsname{\color[rgb]{0,0,1}}%
      \expandafter\def\csname LT3\endcsname{\color[rgb]{1,0,1}}%
      \expandafter\def\csname LT4\endcsname{\color[rgb]{0,1,1}}%
      \expandafter\def\csname LT5\endcsname{\color[rgb]{1,1,0}}%
      \expandafter\def\csname LT6\endcsname{\color[rgb]{0,0,0}}%
      \expandafter\def\csname LT7\endcsname{\color[rgb]{1,0.3,0}}%
      \expandafter\def\csname LT8\endcsname{\color[rgb]{0.5,0.5,0.5}}%
    \else
      \def\colorrgb#1{\color{black}}%
      \def\colorgray#1{\color[gray]{#1}}%
      \expandafter\def\csname LTw\endcsname{\color{white}}%
      \expandafter\def\csname LTb\endcsname{\color{black}}%
      \expandafter\def\csname LTa\endcsname{\color{black}}%
      \expandafter\def\csname LT0\endcsname{\color{black}}%
      \expandafter\def\csname LT1\endcsname{\color{black}}%
      \expandafter\def\csname LT2\endcsname{\color{black}}%
      \expandafter\def\csname LT3\endcsname{\color{black}}%
      \expandafter\def\csname LT4\endcsname{\color{black}}%
      \expandafter\def\csname LT5\endcsname{\color{black}}%
      \expandafter\def\csname LT6\endcsname{\color{black}}%
      \expandafter\def\csname LT7\endcsname{\color{black}}%
      \expandafter\def\csname LT8\endcsname{\color{black}}%
    \fi
  \fi
    \setlength{\unitlength}{0.0500bp}%
    \ifx\gptboxheight\undefined%
      \newlength{\gptboxheight}%
      \newlength{\gptboxwidth}%
      \newsavebox{\gptboxtext}%
    \fi%
    \setlength{\fboxrule}{0.5pt}%
    \setlength{\fboxsep}{1pt}%
    \definecolor{tbcol}{rgb}{1,1,1}%
\begin{picture}(5668.00,5668.00)%
    \gplgaddtomacro\gplbacktext{%
      \csname LTb\endcsname
      \put(576,720){\makebox(0,0)[r]{\strut{}$0$}}%
      \csname LTb\endcsname
      \put(576,1143){\makebox(0,0)[r]{\strut{}$500$}}%
      \csname LTb\endcsname
      \put(576,1565){\makebox(0,0)[r]{\strut{}$1000$}}%
      \csname LTb\endcsname
      \put(576,1988){\makebox(0,0)[r]{\strut{}$1500$}}%
      \csname LTb\endcsname
      \put(576,2411){\makebox(0,0)[r]{\strut{}$2000$}}%
      \csname LTb\endcsname
      \put(576,2834){\makebox(0,0)[r]{\strut{}$2500$}}%
      \csname LTb\endcsname
      \put(576,3256){\makebox(0,0)[r]{\strut{}$3000$}}%
      \csname LTb\endcsname
      \put(576,3679){\makebox(0,0)[r]{\strut{}$3500$}}%
      \csname LTb\endcsname
      \put(576,4102){\makebox(0,0)[r]{\strut{}$4000$}}%
      \csname LTb\endcsname
      \put(576,4524){\makebox(0,0)[r]{\strut{}$4500$}}%
      \csname LTb\endcsname
      \put(576,4947){\makebox(0,0)[r]{\strut{}$5000$}}%
      \csname LTb\endcsname
      \put(1365,576){\rotatebox{-45.00}{\makebox(0,0)[l]{\strut{}0}}}%
      \csname LTb\endcsname
      \put(2010,576){\rotatebox{-45.00}{\makebox(0,0)[l]{\strut{}20}}}%
      \csname LTb\endcsname
      \put(2655,576){\rotatebox{-45.00}{\makebox(0,0)[l]{\strut{}40}}}%
      \csname LTb\endcsname
      \put(3300,576){\rotatebox{-45.00}{\makebox(0,0)[l]{\strut{}60}}}%
      \csname LTb\endcsname
      \put(3945,576){\rotatebox{-45.00}{\makebox(0,0)[l]{\strut{}80}}}%
      \csname LTb\endcsname
      \put(4590,576){\rotatebox{-45.00}{\makebox(0,0)[l]{\strut{}100}}}%
    }%
    \gplgaddtomacro\gplfronttext{%
      \csname LTb\endcsname
      \put(-228,2833){\rotatebox{-270.00}{\makebox(0,0){\strut{}Time (s)}}}%
      \put(2977,55){\makebox(0,0){\strut{}Malleability (\%)}}%
    }%
    \gplbacktext
    \put(0,0){\includegraphics[width={283.40bp},height={283.40bp}]{img//KNLTurnaroundMeanPartialStatistic}}%
    \gplfronttext
  \end{picture}%
\endgroup

%% file: img/KNLMakespanMeanPartialStatistic.tex
\begingroup
\LARGE
  \makeatletter
  \providecommand\color[2][]{%
    \GenericError{(gnuplot) \space\space\space\@spaces}{%
      Package color not loaded in conjunction with
      terminal option `colourtext'%
    }{See the gnuplot documentation for explanation.%
    }{Either use 'blacktext' in gnuplot or load the package
      color.sty in LaTeX.}%
    \renewcommand\color[2][]{}%
  }%
  \providecommand\includegraphics[2][]{%
    \GenericError{(gnuplot) \space\space\space\@spaces}{%
      Package graphicx or graphics not loaded%
    }{See the gnuplot documentation for explanation.%
    }{The gnuplot epslatex terminal needs graphicx.sty or graphics.sty.}%
    \renewcommand\includegraphics[2][]{}%
  }%
  \providecommand\rotatebox[2]{#2}%
  \@ifundefined{ifGPcolor}{%
    \newif\ifGPcolor
    \GPcolortrue
  }{}%
  \@ifundefined{ifGPblacktext}{%
    \newif\ifGPblacktext
    \GPblacktextfalse
  }{}%
  \let\gplgaddtomacro\g@addto@macro
  \gdef\gplbacktext{}%
  \gdef\gplfronttext{}%
  \makeatother
  \ifGPblacktext
    \def\colorrgb#1{}%
    \def\colorgray#1{}%
  \else
    \ifGPcolor
      \def\colorrgb#1{\color[rgb]{#1}}%
      \def\colorgray#1{\color[gray]{#1}}%
      \expandafter\def\csname LTw\endcsname{\color{white}}%
      \expandafter\def\csname LTb\endcsname{\color{black}}%
      \expandafter\def\csname LTa\endcsname{\color{black}}%
      \expandafter\def\csname LT0\endcsname{\color[rgb]{1,0,0}}%
      \expandafter\def\csname LT1\endcsname{\color[rgb]{0,1,0}}%
      \expandafter\def\csname LT2\endcsname{\color[rgb]{0,0,1}}%
      \expandafter\def\csname LT3\endcsname{\color[rgb]{1,0,1}}%
      \expandafter\def\csname LT4\endcsname{\color[rgb]{0,1,1}}%
      \expandafter\def\csname LT5\endcsname{\color[rgb]{1,1,0}}%
      \expandafter\def\csname LT6\endcsname{\color[rgb]{0,0,0}}%
      \expandafter\def\csname LT7\endcsname{\color[rgb]{1,0.3,0}}%
      \expandafter\def\csname LT8\endcsname{\color[rgb]{0.5,0.5,0.5}}%
    \else
      \def\colorrgb#1{\color{black}}%
      \def\colorgray#1{\color[gray]{#1}}%
      \expandafter\def\csname LTw\endcsname{\color{white}}%
      \expandafter\def\csname LTb\endcsname{\color{black}}%
      \expandafter\def\csname LTa\endcsname{\color{black}}%
      \expandafter\def\csname LT0\endcsname{\color{black}}%
      \expandafter\def\csname LT1\endcsname{\color{black}}%
      \expandafter\def\csname LT2\endcsname{\color{black}}%
      \expandafter\def\csname LT3\endcsname{\color{black}}%
      \expandafter\def\csname LT4\endcsname{\color{black}}%
      \expandafter\def\csname LT5\endcsname{\color{black}}%
      \expandafter\def\csname LT6\endcsname{\color{black}}%
      \expandafter\def\csname LT7\endcsname{\color{black}}%
      \expandafter\def\csname LT8\endcsname{\color{black}}%
    \fi
  \fi
    \setlength{\unitlength}{0.0500bp}%
    \ifx\gptboxheight\undefined%
      \newlength{\gptboxheight}%
      \newlength{\gptboxwidth}%
      \newsavebox{\gptboxtext}%
    \fi%
    \setlength{\fboxrule}{0.5pt}%
    \setlength{\fboxsep}{1pt}%
    \definecolor{tbcol}{rgb}{1,1,1}%
\begin{picture}(5668.00,5668.00)%
    \gplgaddtomacro\gplbacktext{%
      \csname LTb\endcsname
      \put(576,720){\makebox(0,0)[r]{\strut{}$0$}}%
      \csname LTb\endcsname
      \put(576,1190){\makebox(0,0)[r]{\strut{}$500$}}%
      \csname LTb\endcsname
      \put(576,1659){\makebox(0,0)[r]{\strut{}$1000$}}%
      \csname LTb\endcsname
      \put(576,2129){\makebox(0,0)[r]{\strut{}$1500$}}%
      \csname LTb\endcsname
      \put(576,2599){\makebox(0,0)[r]{\strut{}$2000$}}%
      \csname LTb\endcsname
      \put(576,3068){\makebox(0,0)[r]{\strut{}$2500$}}%
      \csname LTb\endcsname
      \put(576,3538){\makebox(0,0)[r]{\strut{}$3000$}}%
      \csname LTb\endcsname
      \put(576,4008){\makebox(0,0)[r]{\strut{}$3500$}}%
      \csname LTb\endcsname
      \put(576,4477){\makebox(0,0)[r]{\strut{}$4000$}}%
      \csname LTb\endcsname
      \put(576,4947){\makebox(0,0)[r]{\strut{}$4500$}}%
      \csname LTb\endcsname
      \put(1365,576){\rotatebox{-45.00}{\makebox(0,0)[l]{\strut{}0}}}%
      \csname LTb\endcsname
      \put(2010,576){\rotatebox{-45.00}{\makebox(0,0)[l]{\strut{}20}}}%
      \csname LTb\endcsname
      \put(2655,576){\rotatebox{-45.00}{\makebox(0,0)[l]{\strut{}40}}}%
      \csname LTb\endcsname
      \put(3300,576){\rotatebox{-45.00}{\makebox(0,0)[l]{\strut{}60}}}%
      \csname LTb\endcsname
      \put(3945,576){\rotatebox{-45.00}{\makebox(0,0)[l]{\strut{}80}}}%
      \csname LTb\endcsname
      \put(4590,576){\rotatebox{-45.00}{\makebox(0,0)[l]{\strut{}100}}}%
    }%
    \gplgaddtomacro\gplfronttext{%
      \csname LTb\endcsname
      \put(-228,2833){\rotatebox{-270.00}{\makebox(0,0){\strut{}Time (s)}}}%
      \put(2977,55){\makebox(0,0){\strut{}Malleability (\%)}}%
    }%
    \gplbacktext
    \put(0,0){\includegraphics[width={283.40bp},height={283.40bp}]{img//KNLMakespanMeanPartialStatistic}}%
    \gplfronttext
  \end{picture}%
\endgroup

%% file: img/KNLWaittimeMeanPartialStatistic.tex
\begingroup
\LARGE
  \makeatletter
  \providecommand\color[2][]{%
    \GenericError{(gnuplot) \space\space\space\@spaces}{%
      Package color not loaded in conjunction with
      terminal option `colourtext'%
    }{See the gnuplot documentation for explanation.%
    }{Either use 'blacktext' in gnuplot or load the package
      color.sty in LaTeX.}%
    \renewcommand\color[2][]{}%
  }%
  \providecommand\includegraphics[2][]{%
    \GenericError{(gnuplot) \space\space\space\@spaces}{%
      Package graphicx or graphics not loaded%
    }{See the gnuplot documentation for explanation.%
    }{The gnuplot epslatex terminal needs graphicx.sty or graphics.sty.}%
    \renewcommand\includegraphics[2][]{}%
  }%
  \providecommand\rotatebox[2]{#2}%
  \@ifundefined{ifGPcolor}{%
    \newif\ifGPcolor
    \GPcolortrue
  }{}%
  \@ifundefined{ifGPblacktext}{%
    \newif\ifGPblacktext
    \GPblacktextfalse
  }{}%
  \let\gplgaddtomacro\g@addto@macro
  \gdef\gplbacktext{}%
  \gdef\gplfronttext{}%
  \makeatother
  \ifGPblacktext
    \def\colorrgb#1{}%
    \def\colorgray#1{}%
  \else
    \ifGPcolor
      \def\colorrgb#1{\color[rgb]{#1}}%
      \def\colorgray#1{\color[gray]{#1}}%
      \expandafter\def\csname LTw\endcsname{\color{white}}%
      \expandafter\def\csname LTb\endcsname{\color{black}}%
      \expandafter\def\csname LTa\endcsname{\color{black}}%
      \expandafter\def\csname LT0\endcsname{\color[rgb]{1,0,0}}%
      \expandafter\def\csname LT1\endcsname{\color[rgb]{0,1,0}}%
      \expandafter\def\csname LT2\endcsname{\color[rgb]{0,0,1}}%
      \expandafter\def\csname LT3\endcsname{\color[rgb]{1,0,1}}%
      \expandafter\def\csname LT4\endcsname{\color[rgb]{0,1,1}}%
      \expandafter\def\csname LT5\endcsname{\color[rgb]{1,1,0}}%
      \expandafter\def\csname LT6\endcsname{\color[rgb]{0,0,0}}%
      \expandafter\def\csname LT7\endcsname{\color[rgb]{1,0.3,0}}%
      \expandafter\def\csname LT8\endcsname{\color[rgb]{0.5,0.5,0.5}}%
    \else
      \def\colorrgb#1{\color{black}}%
      \def\colorgray#1{\color[gray]{#1}}%
      \expandafter\def\csname LTw\endcsname{\color{white}}%
      \expandafter\def\csname LTb\endcsname{\color{black}}%
      \expandafter\def\csname LTa\endcsname{\color{black}}%
      \expandafter\def\csname LT0\endcsname{\color{black}}%
      \expandafter\def\csname LT1\endcsname{\color{black}}%
      \expandafter\def\csname LT2\endcsname{\color{black}}%
      \expandafter\def\csname LT3\endcsname{\color{black}}%
      \expandafter\def\csname LT4\endcsname{\color{black}}%
      \expandafter\def\csname LT5\endcsname{\color{black}}%
      \expandafter\def\csname LT6\endcsname{\color{black}}%
      \expandafter\def\csname LT7\endcsname{\color{black}}%
      \expandafter\def\csname LT8\endcsname{\color{black}}%
    \fi
  \fi
    \setlength{\unitlength}{0.0500bp}%
    \ifx\gptboxheight\undefined%
      \newlength{\gptboxheight}%
      \newlength{\gptboxwidth}%
      \newsavebox{\gptboxtext}%
    \fi%
    \setlength{\fboxrule}{0.5pt}%
    \setlength{\fboxsep}{1pt}%
    \definecolor{tbcol}{rgb}{1,1,1}%
\begin{picture}(5668.00,5668.00)%
    \gplgaddtomacro\gplbacktext{%
      \csname LTb\endcsname
      \put(576,720){\makebox(0,0)[r]{\strut{}$0$}}%
      \csname LTb\endcsname
      \put(576,1425){\makebox(0,0)[r]{\strut{}$50$}}%
      \csname LTb\endcsname
      \put(576,2129){\makebox(0,0)[r]{\strut{}$100$}}%
      \csname LTb\endcsname
      \put(576,2834){\makebox(0,0)[r]{\strut{}$150$}}%
      \csname LTb\endcsname
      \put(576,3538){\makebox(0,0)[r]{\strut{}$200$}}%
      \csname LTb\endcsname
      \put(576,4243){\makebox(0,0)[r]{\strut{}$250$}}%
      \csname LTb\endcsname
      \put(576,4947){\makebox(0,0)[r]{\strut{}$300$}}%
      \csname LTb\endcsname
      \put(1365,576){\rotatebox{-45.00}{\makebox(0,0)[l]{\strut{}0}}}%
      \csname LTb\endcsname
      \put(2010,576){\rotatebox{-45.00}{\makebox(0,0)[l]{\strut{}20}}}%
      \csname LTb\endcsname
      \put(2655,576){\rotatebox{-45.00}{\makebox(0,0)[l]{\strut{}40}}}%
      \csname LTb\endcsname
      \put(3300,576){\rotatebox{-45.00}{\makebox(0,0)[l]{\strut{}60}}}%
      \csname LTb\endcsname
      \put(3945,576){\rotatebox{-45.00}{\makebox(0,0)[l]{\strut{}80}}}%
      \csname LTb\endcsname
      \put(4590,576){\rotatebox{-45.00}{\makebox(0,0)[l]{\strut{}100}}}%
    }%
    \gplgaddtomacro\gplfronttext{%
      \csname LTb\endcsname
      \put(-84,2833){\rotatebox{-270.00}{\makebox(0,0){\strut{}Time (s)}}}%
      \put(2977,55){\makebox(0,0){\strut{}Malleability (\%)}}%
    }%
    \gplbacktext
    \put(0,0){\includegraphics[width={283.40bp},height={283.40bp}]{img//KNLWaittimeMeanPartialStatistic}}%
    \gplfronttext
  \end{picture}%
\endgroup

%% file: img/KNLNodeutilizationMeanPartialStatistic.tex
\begingroup
\LARGE
  \makeatletter
  \providecommand\color[2][]{%
    \GenericError{(gnuplot) \space\space\space\@spaces}{%
      Package color not loaded in conjunction with
      terminal option `colourtext'%
    }{See the gnuplot documentation for explanation.%
    }{Either use 'blacktext' in gnuplot or load the package
      color.sty in LaTeX.}%
    \renewcommand\color[2][]{}%
  }%
  \providecommand\includegraphics[2][]{%
    \GenericError{(gnuplot) \space\space\space\@spaces}{%
      Package graphicx or graphics not loaded%
    }{See the gnuplot documentation for explanation.%
    }{The gnuplot epslatex terminal needs graphicx.sty or graphics.sty.}%
    \renewcommand\includegraphics[2][]{}%
  }%
  \providecommand\rotatebox[2]{#2}%
  \@ifundefined{ifGPcolor}{%
    \newif\ifGPcolor
    \GPcolortrue
  }{}%
  \@ifundefined{ifGPblacktext}{%
    \newif\ifGPblacktext
    \GPblacktextfalse
  }{}%
  \let\gplgaddtomacro\g@addto@macro
  \gdef\gplbacktext{}%
  \gdef\gplfronttext{}%
  \makeatother
  \ifGPblacktext
    \def\colorrgb#1{}%
    \def\colorgray#1{}%
  \else
    \ifGPcolor
      \def\colorrgb#1{\color[rgb]{#1}}%
      \def\colorgray#1{\color[gray]{#1}}%
      \expandafter\def\csname LTw\endcsname{\color{white}}%
      \expandafter\def\csname LTb\endcsname{\color{black}}%
      \expandafter\def\csname LTa\endcsname{\color{black}}%
      \expandafter\def\csname LT0\endcsname{\color[rgb]{1,0,0}}%
      \expandafter\def\csname LT1\endcsname{\color[rgb]{0,1,0}}%
      \expandafter\def\csname LT2\endcsname{\color[rgb]{0,0,1}}%
      \expandafter\def\csname LT3\endcsname{\color[rgb]{1,0,1}}%
      \expandafter\def\csname LT4\endcsname{\color[rgb]{0,1,1}}%
      \expandafter\def\csname LT5\endcsname{\color[rgb]{1,1,0}}%
      \expandafter\def\csname LT6\endcsname{\color[rgb]{0,0,0}}%
      \expandafter\def\csname LT7\endcsname{\color[rgb]{1,0.3,0}}%
      \expandafter\def\csname LT8\endcsname{\color[rgb]{0.5,0.5,0.5}}%
    \else
      \def\colorrgb#1{\color{black}}%
      \def\colorgray#1{\color[gray]{#1}}%
      \expandafter\def\csname LTw\endcsname{\color{white}}%
      \expandafter\def\csname LTb\endcsname{\color{black}}%
      \expandafter\def\csname LTa\endcsname{\color{black}}%
      \expandafter\def\csname LT0\endcsname{\color{black}}%
      \expandafter\def\csname LT1\endcsname{\color{black}}%
      \expandafter\def\csname LT2\endcsname{\color{black}}%
      \expandafter\def\csname LT3\endcsname{\color{black}}%
      \expandafter\def\csname LT4\endcsname{\color{black}}%
      \expandafter\def\csname LT5\endcsname{\color{black}}%
      \expandafter\def\csname LT6\endcsname{\color{black}}%
      \expandafter\def\csname LT7\endcsname{\color{black}}%
      \expandafter\def\csname LT8\endcsname{\color{black}}%
    \fi
  \fi
    \setlength{\unitlength}{0.0500bp}%
    \ifx\gptboxheight\undefined%
      \newlength{\gptboxheight}%
      \newlength{\gptboxwidth}%
      \newsavebox{\gptboxtext}%
    \fi%
    \setlength{\fboxrule}{0.5pt}%
    \setlength{\fboxsep}{1pt}%
    \definecolor{tbcol}{rgb}{1,1,1}%
\begin{picture}(5668.00,5668.00)%
    \gplgaddtomacro\gplbacktext{%
      \csname LTb\endcsname
      \put(576,720){\makebox(0,0)[r]{\strut{}$65$}}%
      \csname LTb\endcsname
      \put(576,1324){\makebox(0,0)[r]{\strut{}$70$}}%
      \csname LTb\endcsname
      \put(576,1928){\makebox(0,0)[r]{\strut{}$75$}}%
      \csname LTb\endcsname
      \put(576,2532){\makebox(0,0)[r]{\strut{}$80$}}%
      \csname LTb\endcsname
      \put(576,3135){\makebox(0,0)[r]{\strut{}$85$}}%
      \csname LTb\endcsname
      \put(576,3739){\makebox(0,0)[r]{\strut{}$90$}}%
      \csname LTb\endcsname
      \put(576,4343){\makebox(0,0)[r]{\strut{}$95$}}%
      \csname LTb\endcsname
      \put(576,4947){\makebox(0,0)[r]{\strut{}$100$}}%
      \csname LTb\endcsname
      \put(1365,576){\rotatebox{-45.00}{\makebox(0,0)[l]{\strut{}0}}}%
      \csname LTb\endcsname
      \put(2010,576){\rotatebox{-45.00}{\makebox(0,0)[l]{\strut{}20}}}%
      \csname LTb\endcsname
      \put(2655,576){\rotatebox{-45.00}{\makebox(0,0)[l]{\strut{}40}}}%
      \csname LTb\endcsname
      \put(3300,576){\rotatebox{-45.00}{\makebox(0,0)[l]{\strut{}60}}}%
      \csname LTb\endcsname
      \put(3945,576){\rotatebox{-45.00}{\makebox(0,0)[l]{\strut{}80}}}%
      \csname LTb\endcsname
      \put(4590,576){\rotatebox{-45.00}{\makebox(0,0)[l]{\strut{}100}}}%
    }%
    \gplgaddtomacro\gplfronttext{%
      \csname LTb\endcsname
      \put(-84,2833){\rotatebox{-270.00}{\makebox(0,0){\strut{}Node Utilization (\%)}}}%
      \put(2977,55){\makebox(0,0){\strut{}Malleability (\%)}}%
    }%
    \gplbacktext
    \put(0,0){\includegraphics[width={283.40bp},height={283.40bp}]{img//KNLNodeutilizationMeanPartialStatistic}}%
    \gplfronttext
  \end{picture}%
\endgroup

%% file: img/KNLExpandsJobPartialStatistic.tex
\begingroup
\LARGE
  \makeatletter
  \providecommand\color[2][]{%
    \GenericError{(gnuplot) \space\space\space\@spaces}{%
      Package color not loaded in conjunction with
      terminal option `colourtext'%
    }{See the gnuplot documentation for explanation.%
    }{Either use 'blacktext' in gnuplot or load the package
      color.sty in LaTeX.}%
    \renewcommand\color[2][]{}%
  }%
  \providecommand\includegraphics[2][]{%
    \GenericError{(gnuplot) \space\space\space\@spaces}{%
      Package graphicx or graphics not loaded%
    }{See the gnuplot documentation for explanation.%
    }{The gnuplot epslatex terminal needs graphicx.sty or graphics.sty.}%
    \renewcommand\includegraphics[2][]{}%
  }%
  \providecommand\rotatebox[2]{#2}%
  \@ifundefined{ifGPcolor}{%
    \newif\ifGPcolor
    \GPcolortrue
  }{}%
  \@ifundefined{ifGPblacktext}{%
    \newif\ifGPblacktext
    \GPblacktextfalse
  }{}%
  \let\gplgaddtomacro\g@addto@macro
  \gdef\gplbacktext{}%
  \gdef\gplfronttext{}%
  \makeatother
  \ifGPblacktext
    \def\colorrgb#1{}%
    \def\colorgray#1{}%
  \else
    \ifGPcolor
      \def\colorrgb#1{\color[rgb]{#1}}%
      \def\colorgray#1{\color[gray]{#1}}%
      \expandafter\def\csname LTw\endcsname{\color{white}}%
      \expandafter\def\csname LTb\endcsname{\color{black}}%
      \expandafter\def\csname LTa\endcsname{\color{black}}%
      \expandafter\def\csname LT0\endcsname{\color[rgb]{1,0,0}}%
      \expandafter\def\csname LT1\endcsname{\color[rgb]{0,1,0}}%
      \expandafter\def\csname LT2\endcsname{\color[rgb]{0,0,1}}%
      \expandafter\def\csname LT3\endcsname{\color[rgb]{1,0,1}}%
      \expandafter\def\csname LT4\endcsname{\color[rgb]{0,1,1}}%
      \expandafter\def\csname LT5\endcsname{\color[rgb]{1,1,0}}%
      \expandafter\def\csname LT6\endcsname{\color[rgb]{0,0,0}}%
      \expandafter\def\csname LT7\endcsname{\color[rgb]{1,0.3,0}}%
      \expandafter\def\csname LT8\endcsname{\color[rgb]{0.5,0.5,0.5}}%
    \else
      \def\colorrgb#1{\color{black}}%
      \def\colorgray#1{\color[gray]{#1}}%
      \expandafter\def\csname LTw\endcsname{\color{white}}%
      \expandafter\def\csname LTb\endcsname{\color{black}}%
      \expandafter\def\csname LTa\endcsname{\color{black}}%
      \expandafter\def\csname LT0\endcsname{\color{black}}%
      \expandafter\def\csname LT1\endcsname{\color{black}}%
      \expandafter\def\csname LT2\endcsname{\color{black}}%
      \expandafter\def\csname LT3\endcsname{\color{black}}%
      \expandafter\def\csname LT4\endcsname{\color{black}}%
      \expandafter\def\csname LT5\endcsname{\color{black}}%
      \expandafter\def\csname LT6\endcsname{\color{black}}%
      \expandafter\def\csname LT7\endcsname{\color{black}}%
      \expandafter\def\csname LT8\endcsname{\color{black}}%
    \fi
  \fi
    \setlength{\unitlength}{0.0500bp}%
    \ifx\gptboxheight\undefined%
      \newlength{\gptboxheight}%
      \newlength{\gptboxwidth}%
      \newsavebox{\gptboxtext}%
    \fi%
    \setlength{\fboxrule}{0.5pt}%
    \setlength{\fboxsep}{1pt}%
    \definecolor{tbcol}{rgb}{1,1,1}%
\begin{picture}(5668.00,5668.00)%
    \gplgaddtomacro\gplbacktext{%
      \csname LTb\endcsname
      \put(576,720){\makebox(0,0)[r]{\strut{}$0$}}%
      \csname LTb\endcsname
      \put(576,1190){\makebox(0,0)[r]{\strut{}$2$}}%
      \csname LTb\endcsname
      \put(576,1659){\makebox(0,0)[r]{\strut{}$4$}}%
      \csname LTb\endcsname
      \put(576,2129){\makebox(0,0)[r]{\strut{}$6$}}%
      \csname LTb\endcsname
      \put(576,2599){\makebox(0,0)[r]{\strut{}$8$}}%
      \csname LTb\endcsname
      \put(576,3068){\makebox(0,0)[r]{\strut{}$10$}}%
      \csname LTb\endcsname
      \put(576,3538){\makebox(0,0)[r]{\strut{}$12$}}%
      \csname LTb\endcsname
      \put(576,4008){\makebox(0,0)[r]{\strut{}$14$}}%
      \csname LTb\endcsname
      \put(576,4477){\makebox(0,0)[r]{\strut{}$16$}}%
      \csname LTb\endcsname
      \put(576,4947){\makebox(0,0)[r]{\strut{}$18$}}%
      \csname LTb\endcsname
      \put(1365,576){\rotatebox{-45.00}{\makebox(0,0)[l]{\strut{}0}}}%
      \csname LTb\endcsname
      \put(2010,576){\rotatebox{-45.00}{\makebox(0,0)[l]{\strut{}20}}}%
      \csname LTb\endcsname
      \put(2655,576){\rotatebox{-45.00}{\makebox(0,0)[l]{\strut{}40}}}%
      \csname LTb\endcsname
      \put(3300,576){\rotatebox{-45.00}{\makebox(0,0)[l]{\strut{}60}}}%
      \csname LTb\endcsname
      \put(3945,576){\rotatebox{-45.00}{\makebox(0,0)[l]{\strut{}80}}}%
      \csname LTb\endcsname
      \put(4590,576){\rotatebox{-45.00}{\makebox(0,0)[l]{\strut{}100}}}%
    }%
    \gplgaddtomacro\gplfronttext{%
      \csname LTb\endcsname
      \put(60,2833){\rotatebox{-270.00}{\makebox(0,0){\strut{}Events}}}%
      \put(2977,55){\makebox(0,0){\strut{}Malleability (\%)}}%
    }%
    \gplbacktext
    \put(0,0){\includegraphics[width={283.40bp},height={283.40bp}]{img//KNLExpandsJobPartialStatistic}}%
    \gplfronttext
  \end{picture}%
\endgroup

%% file: img/KNLShrinksJobPartialStatistic.tex
\begingroup
\LARGE
  \makeatletter
  \providecommand\color[2][]{%
    \GenericError{(gnuplot) \space\space\space\@spaces}{%
      Package color not loaded in conjunction with
      terminal option `colourtext'%
    }{See the gnuplot documentation for explanation.%
    }{Either use 'blacktext' in gnuplot or load the package
      color.sty in LaTeX.}%
    \renewcommand\color[2][]{}%
  }%
  \providecommand\includegraphics[2][]{%
    \GenericError{(gnuplot) \space\space\space\@spaces}{%
      Package graphicx or graphics not loaded%
    }{See the gnuplot documentation for explanation.%
    }{The gnuplot epslatex terminal needs graphicx.sty or graphics.sty.}%
    \renewcommand\includegraphics[2][]{}%
  }%
  \providecommand\rotatebox[2]{#2}%
  \@ifundefined{ifGPcolor}{%
    \newif\ifGPcolor
    \GPcolortrue
  }{}%
  \@ifundefined{ifGPblacktext}{%
    \newif\ifGPblacktext
    \GPblacktextfalse
  }{}%
  \let\gplgaddtomacro\g@addto@macro
  \gdef\gplbacktext{}%
  \gdef\gplfronttext{}%
  \makeatother
  \ifGPblacktext
    \def\colorrgb#1{}%
    \def\colorgray#1{}%
  \else
    \ifGPcolor
      \def\colorrgb#1{\color[rgb]{#1}}%
      \def\colorgray#1{\color[gray]{#1}}%
      \expandafter\def\csname LTw\endcsname{\color{white}}%
      \expandafter\def\csname LTb\endcsname{\color{black}}%
      \expandafter\def\csname LTa\endcsname{\color{black}}%
      \expandafter\def\csname LT0\endcsname{\color[rgb]{1,0,0}}%
      \expandafter\def\csname LT1\endcsname{\color[rgb]{0,1,0}}%
      \expandafter\def\csname LT2\endcsname{\color[rgb]{0,0,1}}%
      \expandafter\def\csname LT3\endcsname{\color[rgb]{1,0,1}}%
      \expandafter\def\csname LT4\endcsname{\color[rgb]{0,1,1}}%
      \expandafter\def\csname LT5\endcsname{\color[rgb]{1,1,0}}%
      \expandafter\def\csname LT6\endcsname{\color[rgb]{0,0,0}}%
      \expandafter\def\csname LT7\endcsname{\color[rgb]{1,0.3,0}}%
      \expandafter\def\csname LT8\endcsname{\color[rgb]{0.5,0.5,0.5}}%
    \else
      \def\colorrgb#1{\color{black}}%
      \def\colorgray#1{\color[gray]{#1}}%
      \expandafter\def\csname LTw\endcsname{\color{white}}%
      \expandafter\def\csname LTb\endcsname{\color{black}}%
      \expandafter\def\csname LTa\endcsname{\color{black}}%
      \expandafter\def\csname LT0\endcsname{\color{black}}%
      \expandafter\def\csname LT1\endcsname{\color{black}}%
      \expandafter\def\csname LT2\endcsname{\color{black}}%
      \expandafter\def\csname LT3\endcsname{\color{black}}%
      \expandafter\def\csname LT4\endcsname{\color{black}}%
      \expandafter\def\csname LT5\endcsname{\color{black}}%
      \expandafter\def\csname LT6\endcsname{\color{black}}%
      \expandafter\def\csname LT7\endcsname{\color{black}}%
      \expandafter\def\csname LT8\endcsname{\color{black}}%
    \fi
  \fi
    \setlength{\unitlength}{0.0500bp}%
    \ifx\gptboxheight\undefined%
      \newlength{\gptboxheight}%
      \newlength{\gptboxwidth}%
      \newsavebox{\gptboxtext}%
    \fi%
    \setlength{\fboxrule}{0.5pt}%
    \setlength{\fboxsep}{1pt}%
    \definecolor{tbcol}{rgb}{1,1,1}%
\begin{picture}(5668.00,5668.00)%
    \gplgaddtomacro\gplbacktext{%
      \csname LTb\endcsname
      \put(576,720){\makebox(0,0)[r]{\strut{}$0$}}%
      \csname LTb\endcsname
      \put(576,1425){\makebox(0,0)[r]{\strut{}$10$}}%
      \csname LTb\endcsname
      \put(576,2129){\makebox(0,0)[r]{\strut{}$20$}}%
      \csname LTb\endcsname
      \put(576,2834){\makebox(0,0)[r]{\strut{}$30$}}%
      \csname LTb\endcsname
      \put(576,3538){\makebox(0,0)[r]{\strut{}$40$}}%
      \csname LTb\endcsname
      \put(576,4243){\makebox(0,0)[r]{\strut{}$50$}}%
      \csname LTb\endcsname
      \put(576,4947){\makebox(0,0)[r]{\strut{}$60$}}%
      \csname LTb\endcsname
      \put(1365,576){\rotatebox{-45.00}{\makebox(0,0)[l]{\strut{}0}}}%
      \csname LTb\endcsname
      \put(2010,576){\rotatebox{-45.00}{\makebox(0,0)[l]{\strut{}20}}}%
      \csname LTb\endcsname
      \put(2655,576){\rotatebox{-45.00}{\makebox(0,0)[l]{\strut{}40}}}%
      \csname LTb\endcsname
      \put(3300,576){\rotatebox{-45.00}{\makebox(0,0)[l]{\strut{}60}}}%
      \csname LTb\endcsname
      \put(3945,576){\rotatebox{-45.00}{\makebox(0,0)[l]{\strut{}80}}}%
      \csname LTb\endcsname
      \put(4590,576){\rotatebox{-45.00}{\makebox(0,0)[l]{\strut{}100}}}%
    }%
    \gplgaddtomacro\gplfronttext{%
      \csname LTb\endcsname
      \put(60,2833){\rotatebox{-270.00}{\makebox(0,0){\strut{}Events}}}%
      \put(2977,55){\makebox(0,0){\strut{}Malleability (\%)}}%
    }%
    \gplbacktext
    \put(0,0){\includegraphics[width={283.40bp},height={283.40bp}]{img//KNLShrinksJobPartialStatistic}}%
    \gplfronttext
  \end{picture}%
\endgroup

%% file: 03eagle-theta-figs-2.tex
\begin{figure}
    \centering
    \vspace*{-10pt}

    \begin{minipage}{\linewidth}
        \centering

        \begin{subfigure}{.32\linewidth}
            \centering
            \resizebox{\linewidth}{!}{\input{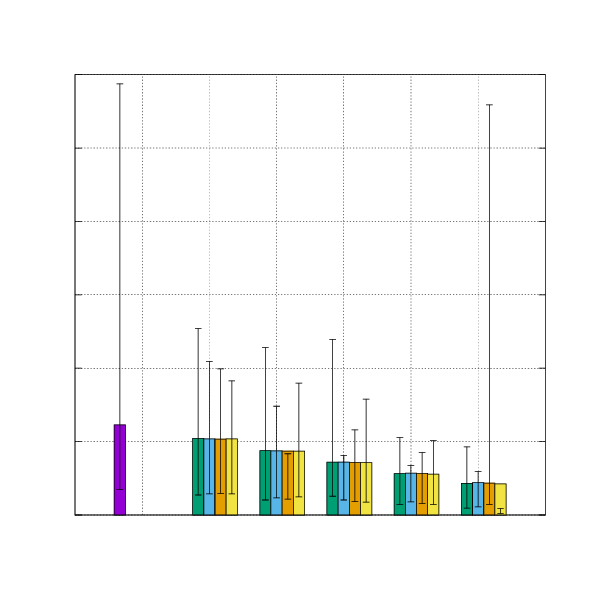}}
            \caption{Turnaround Time}
            \label{fig:eagle-statistics-turnaround}
        \end{subfigure}
        \hfill
        \begin{subfigure}{.32\linewidth}
            \centering
            \resizebox{\linewidth}{!}{\input{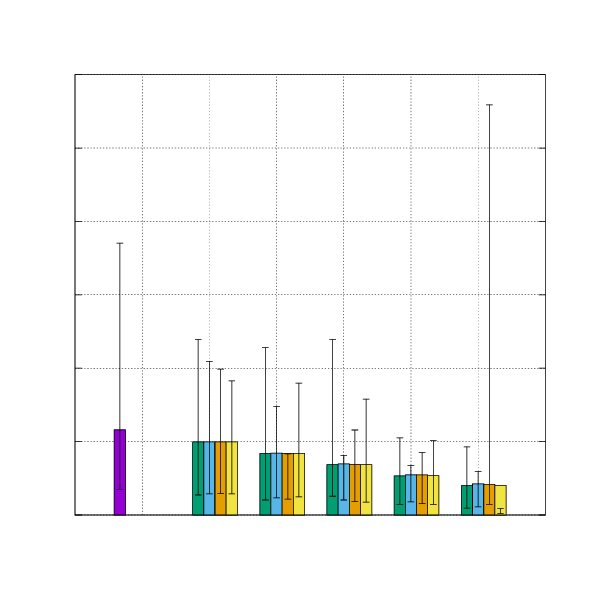}}
            \caption{Makespan}
            \label{fig:eagle-statistics-makespan}
        \end{subfigure}
        \hfill
        \begin{subfigure}{.32\linewidth}
            \centering
            \resizebox{\linewidth}{!}{\input{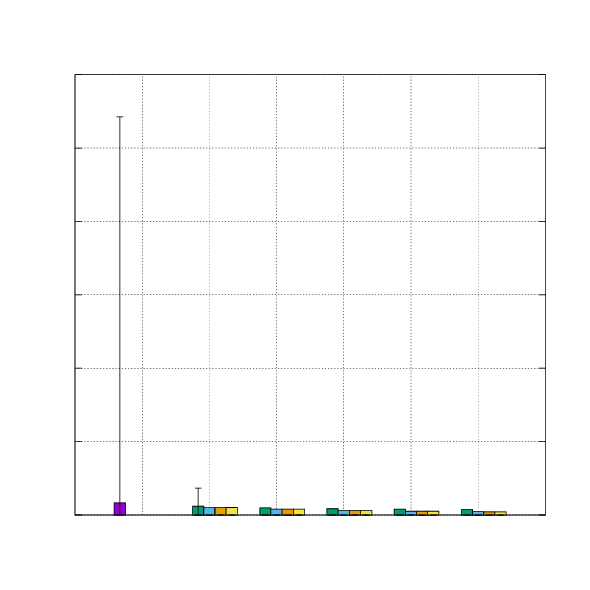}}
            \caption{Wait Time}
            \label{fig:eagle-statistics-waittime}
        \end{subfigure}

        \vspace{1pt}
        \begin{subfigure}{.32\linewidth}
            \centering
            \resizebox{\linewidth}{!}{\input{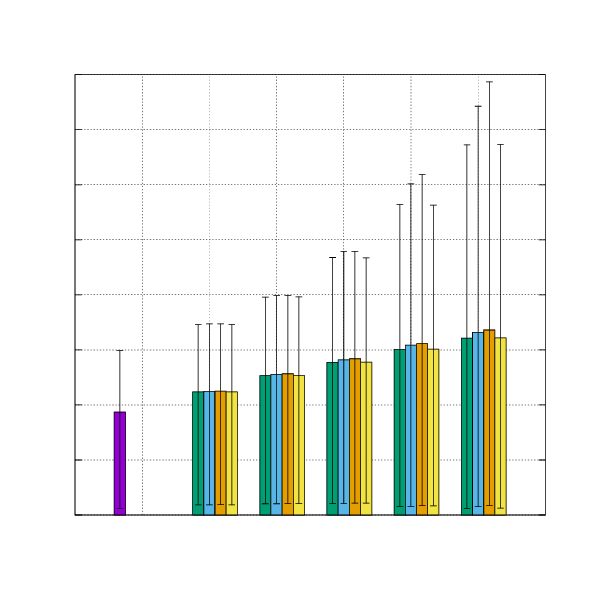}}
            \caption{Node Utilization}
            \label{fig:eagle-statistics-nodeutil}
        \end{subfigure}
        \hfill
        \begin{subfigure}{.32\linewidth}
            \centering
            \resizebox{\linewidth}{!}{\input{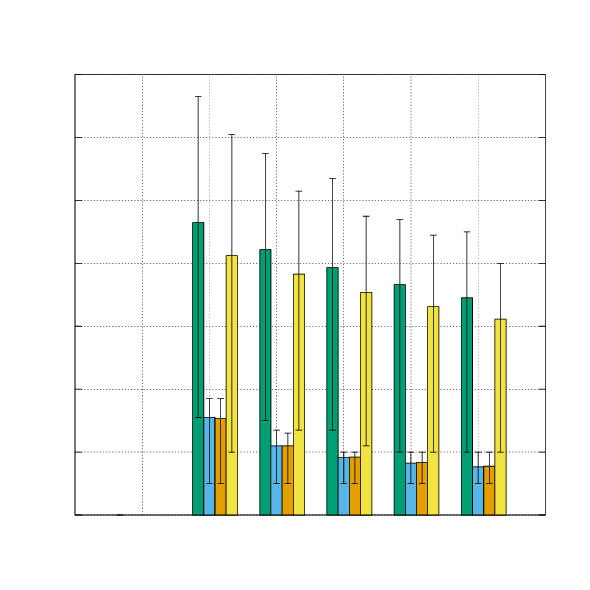}}
            \caption{Expands per Job}
            \label{fig:eagle-expands}
        \end{subfigure}
        \hfill
        \begin{subfigure}{.32\linewidth}
            \centering
            \resizebox{\linewidth}{!}{\input{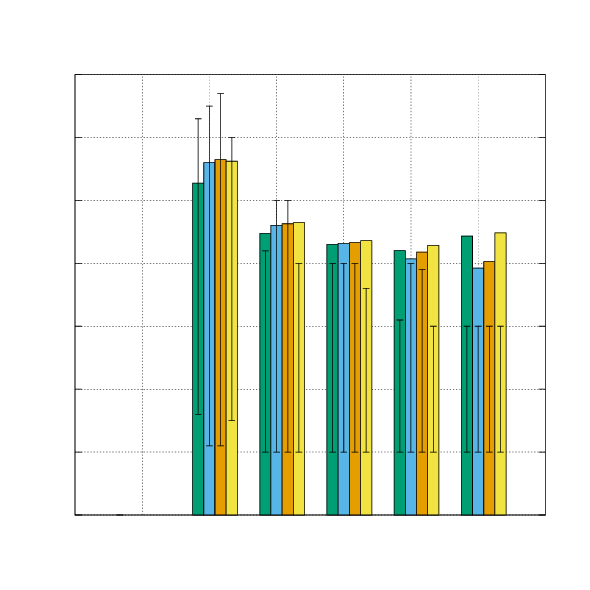}}
            \caption{Shrinks per Job}
            \label{fig:eagle-shrinks}
        \end{subfigure}
        \vspace{0pt}
        \resizebox{0.99\linewidth}{!}{\input{img/Legend}}
        \vspace*{-6pt}
        \caption{\eagle: Simulation results with IQR error bars}
        \label{fig:eagle-all-metrics}
    \end{minipage}

    \vspace{0pt}

    \begin{minipage}{\linewidth}
        \centering

        \begin{subfigure}[t]{.32\linewidth}
            \centering
            \resizebox{\linewidth}{!}{\input{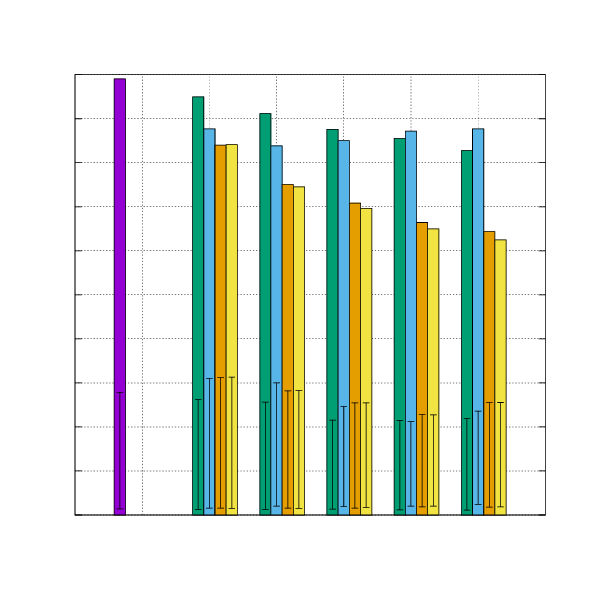}}
            \caption{Turnaround Time}
            \label{fig:theta-statistics-turnaround}
        \end{subfigure}
        \hfill
        \begin{subfigure}[t]{.32\linewidth}
            \centering
            \resizebox{\linewidth}{!}{\input{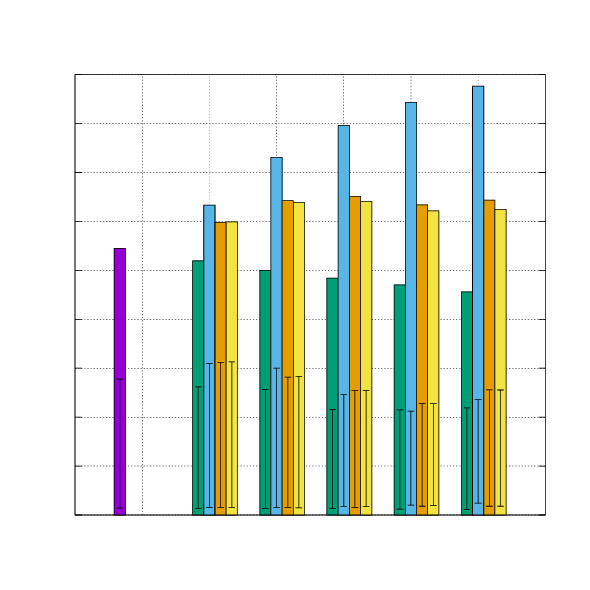}}
            \caption{Makespan}
            \label{fig:theta-statistics-makespan}
        \end{subfigure}
        \hfill
        \begin{subfigure}[t]{.32\linewidth}
            \centering
            \resizebox{\linewidth}{!}{\input{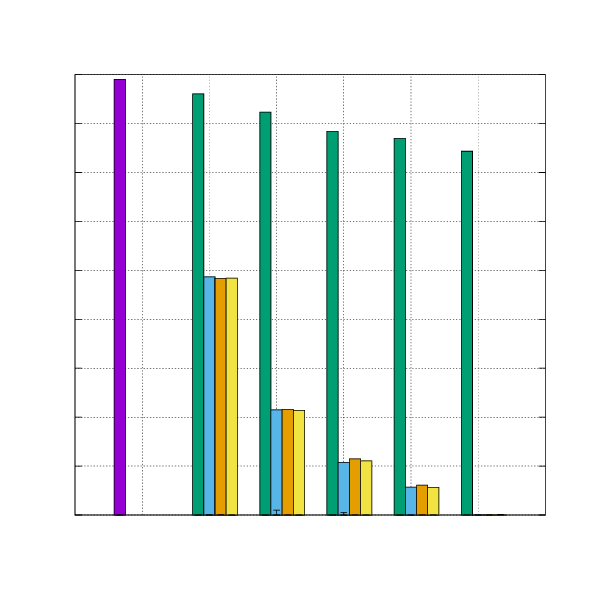}}
            \caption{Wait Time}
            \label{fig:theta-statistics-waittime}
        \end{subfigure}

        \vspace{-10pt}
        \begin{subfigure}{.32\linewidth}
            \centering
            \resizebox{\linewidth}{!}{\input{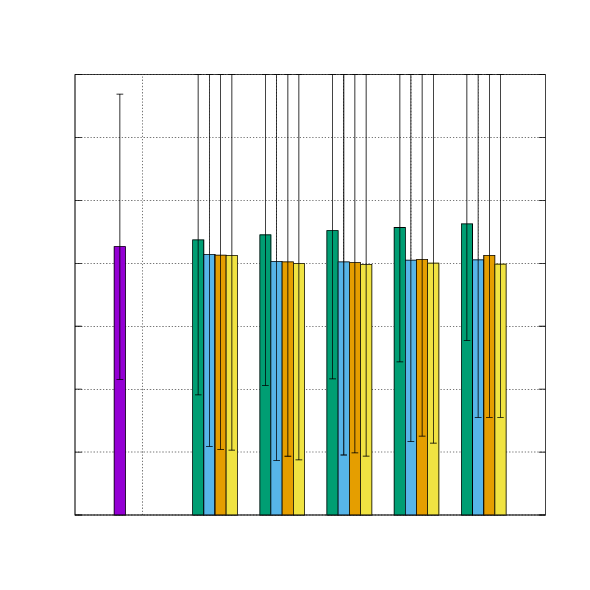}}
            \caption{Node Utilization}
            \label{fig:theta-statistics-nodeutil}
        \end{subfigure}
        \hfill
        \begin{subfigure}{.32\linewidth}
            \centering
            \resizebox{\linewidth}{!}{\input{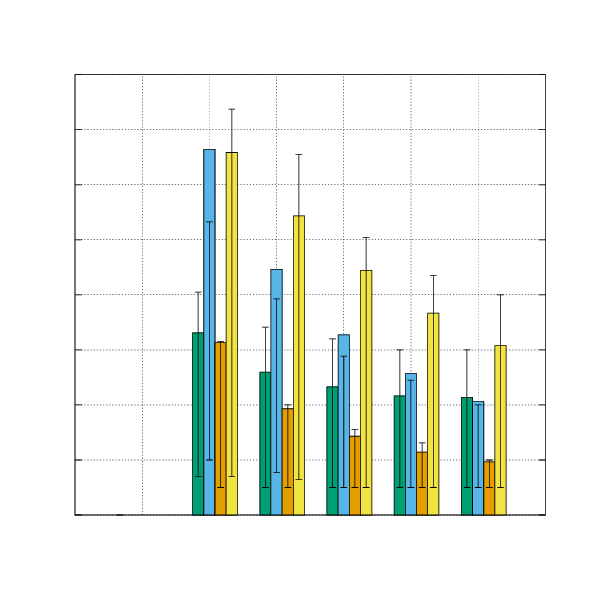}}
            \caption{Expands per Job}
            \label{fig:theta-expands}
        \end{subfigure}
        \hfill
        \begin{subfigure}{.32\linewidth}
            \centering
            \resizebox{\linewidth}{!}{\input{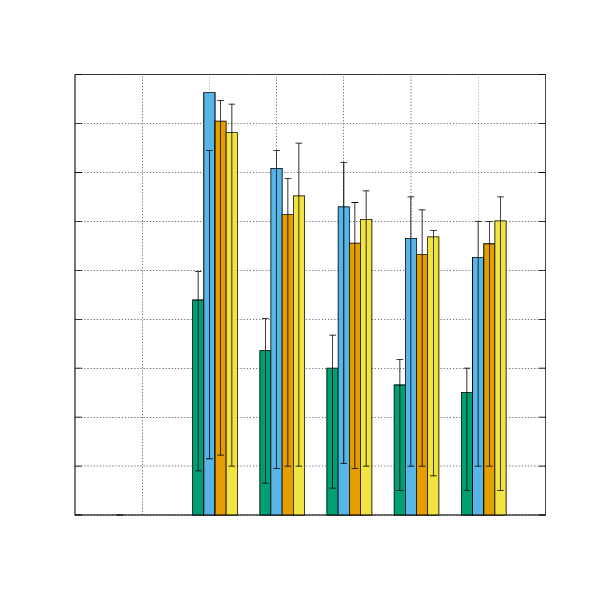}}
            \caption{Shrinks per Job}
            \label{fig:theta-shrinks}
        \end{subfigure}
        \vspace{0pt}
        \resizebox{0.99\linewidth}{!}{\input{img/Legend}}
        \vspace*{-6pt}
        \caption{\theta: Simulation results with IQR error bars}
        \label{fig:theta-all-metrics}
    \end{minipage}
\end{figure}

%% file: img/EagleTurnaroundMeanPartialStatistic.tex
\begingroup
\LARGE
  \makeatletter
  \providecommand\color[2][]{%
    \GenericError{(gnuplot) \space\space\space\@spaces}{%
      Package color not loaded in conjunction with
      terminal option `colourtext'%
    }{See the gnuplot documentation for explanation.%
    }{Either use 'blacktext' in gnuplot or load the package
      color.sty in LaTeX.}%
    \renewcommand\color[2][]{}%
  }%
  \providecommand\includegraphics[2][]{%
    \GenericError{(gnuplot) \space\space\space\@spaces}{%
      Package graphicx or graphics not loaded%
    }{See the gnuplot documentation for explanation.%
    }{The gnuplot epslatex terminal needs graphicx.sty or graphics.sty.}%
    \renewcommand\includegraphics[2][]{}%
  }%
  \providecommand\rotatebox[2]{#2}%
  \@ifundefined{ifGPcolor}{%
    \newif\ifGPcolor
    \GPcolortrue
  }{}%
  \@ifundefined{ifGPblacktext}{%
    \newif\ifGPblacktext
    \GPblacktextfalse
  }{}%
  \let\gplgaddtomacro\g@addto@macro
  \gdef\gplbacktext{}%
  \gdef\gplfronttext{}%
  \makeatother
  \ifGPblacktext
    \def\colorrgb#1{}%
    \def\colorgray#1{}%
  \else
    \ifGPcolor
      \def\colorrgb#1{\color[rgb]{#1}}%
      \def\colorgray#1{\color[gray]{#1}}%
      \expandafter\def\csname LTw\endcsname{\color{white}}%
      \expandafter\def\csname LTb\endcsname{\color{black}}%
      \expandafter\def\csname LTa\endcsname{\color{black}}%
      \expandafter\def\csname LT0\endcsname{\color[rgb]{1,0,0}}%
      \expandafter\def\csname LT1\endcsname{\color[rgb]{0,1,0}}%
      \expandafter\def\csname LT2\endcsname{\color[rgb]{0,0,1}}%
      \expandafter\def\csname LT3\endcsname{\color[rgb]{1,0,1}}%
      \expandafter\def\csname LT4\endcsname{\color[rgb]{0,1,1}}%
      \expandafter\def\csname LT5\endcsname{\color[rgb]{1,1,0}}%
      \expandafter\def\csname LT6\endcsname{\color[rgb]{0,0,0}}%
      \expandafter\def\csname LT7\endcsname{\color[rgb]{1,0.3,0}}%
      \expandafter\def\csname LT8\endcsname{\color[rgb]{0.5,0.5,0.5}}%
    \else
      \def\colorrgb#1{\color{black}}%
      \def\colorgray#1{\color[gray]{#1}}%
      \expandafter\def\csname LTw\endcsname{\color{white}}%
      \expandafter\def\csname LTb\endcsname{\color{black}}%
      \expandafter\def\csname LTa\endcsname{\color{black}}%
      \expandafter\def\csname LT0\endcsname{\color{black}}%
      \expandafter\def\csname LT1\endcsname{\color{black}}%
      \expandafter\def\csname LT2\endcsname{\color{black}}%
      \expandafter\def\csname LT3\endcsname{\color{black}}%
      \expandafter\def\csname LT4\endcsname{\color{black}}%
      \expandafter\def\csname LT5\endcsname{\color{black}}%
      \expandafter\def\csname LT6\endcsname{\color{black}}%
      \expandafter\def\csname LT7\endcsname{\color{black}}%
      \expandafter\def\csname LT8\endcsname{\color{black}}%
    \fi
  \fi
    \setlength{\unitlength}{0.0500bp}%
    \ifx\gptboxheight\undefined%
      \newlength{\gptboxheight}%
      \newlength{\gptboxwidth}%
      \newsavebox{\gptboxtext}%
    \fi%
    \setlength{\fboxrule}{0.5pt}%
    \setlength{\fboxsep}{1pt}%
    \definecolor{tbcol}{rgb}{1,1,1}%
\begin{picture}(5668.00,5668.00)%
    \gplgaddtomacro\gplbacktext{%
      \csname LTb\endcsname
      \put(576,720){\makebox(0,0)[r]{\strut{}$0$}}%
      \csname LTb\endcsname
      \put(576,1425){\makebox(0,0)[r]{\strut{}$5000$}}%
      \csname LTb\endcsname
      \put(576,2129){\makebox(0,0)[r]{\strut{}$10000$}}%
      \csname LTb\endcsname
      \put(576,2834){\makebox(0,0)[r]{\strut{}$15000$}}%
      \csname LTb\endcsname
      \put(576,3538){\makebox(0,0)[r]{\strut{}$20000$}}%
      \csname LTb\endcsname
      \put(576,4243){\makebox(0,0)[r]{\strut{}$25000$}}%
      \csname LTb\endcsname
      \put(576,4947){\makebox(0,0)[r]{\strut{}$30000$}}%
      \csname LTb\endcsname
      \put(1365,576){\rotatebox{-45.00}{\makebox(0,0)[l]{\strut{}0}}}%
      \csname LTb\endcsname
      \put(2010,576){\rotatebox{-45.00}{\makebox(0,0)[l]{\strut{}20}}}%
      \csname LTb\endcsname
      \put(2655,576){\rotatebox{-45.00}{\makebox(0,0)[l]{\strut{}40}}}%
      \csname LTb\endcsname
      \put(3300,576){\rotatebox{-45.00}{\makebox(0,0)[l]{\strut{}60}}}%
      \csname LTb\endcsname
      \put(3945,576){\rotatebox{-45.00}{\makebox(0,0)[l]{\strut{}80}}}%
      \csname LTb\endcsname
      \put(4590,576){\rotatebox{-45.00}{\makebox(0,0)[l]{\strut{}100}}}%
    }%
    \gplgaddtomacro\gplfronttext{%
      \csname LTb\endcsname
      \put(-372,2833){\rotatebox{-270.00}{\makebox(0,0){\strut{}Time (s)}}}%
      \put(2977,55){\makebox(0,0){\strut{}Malleability (\%)}}%
    }%
    \gplbacktext
    \put(0,0){\includegraphics[width={283.40bp},height={283.40bp}]{img//EagleTurnaroundMeanPartialStatistic}}%
    \gplfronttext
  \end{picture}%
\endgroup

%% file: img/EagleMakespanMeanPartialStatistic.tex
\begingroup
\LARGE
  \makeatletter
  \providecommand\color[2][]{%
    \GenericError{(gnuplot) \space\space\space\@spaces}{%
      Package color not loaded in conjunction with
      terminal option `colourtext'%
    }{See the gnuplot documentation for explanation.%
    }{Either use 'blacktext' in gnuplot or load the package
      color.sty in LaTeX.}%
    \renewcommand\color[2][]{}%
  }%
  \providecommand\includegraphics[2][]{%
    \GenericError{(gnuplot) \space\space\space\@spaces}{%
      Package graphicx or graphics not loaded%
    }{See the gnuplot documentation for explanation.%
    }{The gnuplot epslatex terminal needs graphicx.sty or graphics.sty.}%
    \renewcommand\includegraphics[2][]{}%
  }%
  \providecommand\rotatebox[2]{#2}%
  \@ifundefined{ifGPcolor}{%
    \newif\ifGPcolor
    \GPcolortrue
  }{}%
  \@ifundefined{ifGPblacktext}{%
    \newif\ifGPblacktext
    \GPblacktextfalse
  }{}%
  \let\gplgaddtomacro\g@addto@macro
  \gdef\gplbacktext{}%
  \gdef\gplfronttext{}%
  \makeatother
  \ifGPblacktext
    \def\colorrgb#1{}%
    \def\colorgray#1{}%
  \else
    \ifGPcolor
      \def\colorrgb#1{\color[rgb]{#1}}%
      \def\colorgray#1{\color[gray]{#1}}%
      \expandafter\def\csname LTw\endcsname{\color{white}}%
      \expandafter\def\csname LTb\endcsname{\color{black}}%
      \expandafter\def\csname LTa\endcsname{\color{black}}%
      \expandafter\def\csname LT0\endcsname{\color[rgb]{1,0,0}}%
      \expandafter\def\csname LT1\endcsname{\color[rgb]{0,1,0}}%
      \expandafter\def\csname LT2\endcsname{\color[rgb]{0,0,1}}%
      \expandafter\def\csname LT3\endcsname{\color[rgb]{1,0,1}}%
      \expandafter\def\csname LT4\endcsname{\color[rgb]{0,1,1}}%
      \expandafter\def\csname LT5\endcsname{\color[rgb]{1,1,0}}%
      \expandafter\def\csname LT6\endcsname{\color[rgb]{0,0,0}}%
      \expandafter\def\csname LT7\endcsname{\color[rgb]{1,0.3,0}}%
      \expandafter\def\csname LT8\endcsname{\color[rgb]{0.5,0.5,0.5}}%
    \else
      \def\colorrgb#1{\color{black}}%
      \def\colorgray#1{\color[gray]{#1}}%
      \expandafter\def\csname LTw\endcsname{\color{white}}%
      \expandafter\def\csname LTb\endcsname{\color{black}}%
      \expandafter\def\csname LTa\endcsname{\color{black}}%
      \expandafter\def\csname LT0\endcsname{\color{black}}%
      \expandafter\def\csname LT1\endcsname{\color{black}}%
      \expandafter\def\csname LT2\endcsname{\color{black}}%
      \expandafter\def\csname LT3\endcsname{\color{black}}%
      \expandafter\def\csname LT4\endcsname{\color{black}}%
      \expandafter\def\csname LT5\endcsname{\color{black}}%
      \expandafter\def\csname LT6\endcsname{\color{black}}%
      \expandafter\def\csname LT7\endcsname{\color{black}}%
      \expandafter\def\csname LT8\endcsname{\color{black}}%
    \fi
  \fi
    \setlength{\unitlength}{0.0500bp}%
    \ifx\gptboxheight\undefined%
      \newlength{\gptboxheight}%
      \newlength{\gptboxwidth}%
      \newsavebox{\gptboxtext}%
    \fi%
    \setlength{\fboxrule}{0.5pt}%
    \setlength{\fboxsep}{1pt}%
    \definecolor{tbcol}{rgb}{1,1,1}%
\begin{picture}(5668.00,5668.00)%
    \gplgaddtomacro\gplbacktext{%
      \csname LTb\endcsname
      \put(576,720){\makebox(0,0)[r]{\strut{}$0$}}%
      \csname LTb\endcsname
      \put(576,1425){\makebox(0,0)[r]{\strut{}$5000$}}%
      \csname LTb\endcsname
      \put(576,2129){\makebox(0,0)[r]{\strut{}$10000$}}%
      \csname LTb\endcsname
      \put(576,2834){\makebox(0,0)[r]{\strut{}$15000$}}%
      \csname LTb\endcsname
      \put(576,3538){\makebox(0,0)[r]{\strut{}$20000$}}%
      \csname LTb\endcsname
      \put(576,4243){\makebox(0,0)[r]{\strut{}$25000$}}%
      \csname LTb\endcsname
      \put(576,4947){\makebox(0,0)[r]{\strut{}$30000$}}%
      \csname LTb\endcsname
      \put(1365,576){\rotatebox{-45.00}{\makebox(0,0)[l]{\strut{}0}}}%
      \csname LTb\endcsname
      \put(2010,576){\rotatebox{-45.00}{\makebox(0,0)[l]{\strut{}20}}}%
      \csname LTb\endcsname
      \put(2655,576){\rotatebox{-45.00}{\makebox(0,0)[l]{\strut{}40}}}%
      \csname LTb\endcsname
      \put(3300,576){\rotatebox{-45.00}{\makebox(0,0)[l]{\strut{}60}}}%
      \csname LTb\endcsname
      \put(3945,576){\rotatebox{-45.00}{\makebox(0,0)[l]{\strut{}80}}}%
      \csname LTb\endcsname
      \put(4590,576){\rotatebox{-45.00}{\makebox(0,0)[l]{\strut{}100}}}%
    }%
    \gplgaddtomacro\gplfronttext{%
      \csname LTb\endcsname
      \put(-372,2833){\rotatebox{-270.00}{\makebox(0,0){\strut{}Time (s)}}}%
      \put(2977,55){\makebox(0,0){\strut{}Malleability (\%)}}%
    }%
    \gplbacktext
    \put(0,0){\includegraphics[width={283.40bp},height={283.40bp}]{img//EagleMakespanMeanPartialStatistic}}%
    \gplfronttext
  \end{picture}%
\endgroup

%% file: img/EagleWaittimeMeanPartialStatistic.tex
\begingroup
\LARGE
  \makeatletter
  \providecommand\color[2][]{%
    \GenericError{(gnuplot) \space\space\space\@spaces}{%
      Package color not loaded in conjunction with
      terminal option `colourtext'%
    }{See the gnuplot documentation for explanation.%
    }{Either use 'blacktext' in gnuplot or load the package
      color.sty in LaTeX.}%
    \renewcommand\color[2][]{}%
  }%
  \providecommand\includegraphics[2][]{%
    \GenericError{(gnuplot) \space\space\space\@spaces}{%
      Package graphicx or graphics not loaded%
    }{See the gnuplot documentation for explanation.%
    }{The gnuplot epslatex terminal needs graphicx.sty or graphics.sty.}%
    \renewcommand\includegraphics[2][]{}%
  }%
  \providecommand\rotatebox[2]{#2}%
  \@ifundefined{ifGPcolor}{%
    \newif\ifGPcolor
    \GPcolortrue
  }{}%
  \@ifundefined{ifGPblacktext}{%
    \newif\ifGPblacktext
    \GPblacktextfalse
  }{}%
  \let\gplgaddtomacro\g@addto@macro
  \gdef\gplbacktext{}%
  \gdef\gplfronttext{}%
  \makeatother
  \ifGPblacktext
    \def\colorrgb#1{}%
    \def\colorgray#1{}%
  \else
    \ifGPcolor
      \def\colorrgb#1{\color[rgb]{#1}}%
      \def\colorgray#1{\color[gray]{#1}}%
      \expandafter\def\csname LTw\endcsname{\color{white}}%
      \expandafter\def\csname LTb\endcsname{\color{black}}%
      \expandafter\def\csname LTa\endcsname{\color{black}}%
      \expandafter\def\csname LT0\endcsname{\color[rgb]{1,0,0}}%
      \expandafter\def\csname LT1\endcsname{\color[rgb]{0,1,0}}%
      \expandafter\def\csname LT2\endcsname{\color[rgb]{0,0,1}}%
      \expandafter\def\csname LT3\endcsname{\color[rgb]{1,0,1}}%
      \expandafter\def\csname LT4\endcsname{\color[rgb]{0,1,1}}%
      \expandafter\def\csname LT5\endcsname{\color[rgb]{1,1,0}}%
      \expandafter\def\csname LT6\endcsname{\color[rgb]{0,0,0}}%
      \expandafter\def\csname LT7\endcsname{\color[rgb]{1,0.3,0}}%
      \expandafter\def\csname LT8\endcsname{\color[rgb]{0.5,0.5,0.5}}%
    \else
      \def\colorrgb#1{\color{black}}%
      \def\colorgray#1{\color[gray]{#1}}%
      \expandafter\def\csname LTw\endcsname{\color{white}}%
      \expandafter\def\csname LTb\endcsname{\color{black}}%
      \expandafter\def\csname LTa\endcsname{\color{black}}%
      \expandafter\def\csname LT0\endcsname{\color{black}}%
      \expandafter\def\csname LT1\endcsname{\color{black}}%
      \expandafter\def\csname LT2\endcsname{\color{black}}%
      \expandafter\def\csname LT3\endcsname{\color{black}}%
      \expandafter\def\csname LT4\endcsname{\color{black}}%
      \expandafter\def\csname LT5\endcsname{\color{black}}%
      \expandafter\def\csname LT6\endcsname{\color{black}}%
      \expandafter\def\csname LT7\endcsname{\color{black}}%
      \expandafter\def\csname LT8\endcsname{\color{black}}%
    \fi
  \fi
    \setlength{\unitlength}{0.0500bp}%
    \ifx\gptboxheight\undefined%
      \newlength{\gptboxheight}%
      \newlength{\gptboxwidth}%
      \newsavebox{\gptboxtext}%
    \fi%
    \setlength{\fboxrule}{0.5pt}%
    \setlength{\fboxsep}{1pt}%
    \definecolor{tbcol}{rgb}{1,1,1}%
\begin{picture}(5668.00,5668.00)%
    \gplgaddtomacro\gplbacktext{%
      \csname LTb\endcsname
      \put(576,720){\makebox(0,0)[r]{\strut{}$0$}}%
      \csname LTb\endcsname
      \put(576,1425){\makebox(0,0)[r]{\strut{}$2000$}}%
      \csname LTb\endcsname
      \put(576,2129){\makebox(0,0)[r]{\strut{}$4000$}}%
      \csname LTb\endcsname
      \put(576,2834){\makebox(0,0)[r]{\strut{}$6000$}}%
      \csname LTb\endcsname
      \put(576,3538){\makebox(0,0)[r]{\strut{}$8000$}}%
      \csname LTb\endcsname
      \put(576,4243){\makebox(0,0)[r]{\strut{}$10000$}}%
      \csname LTb\endcsname
      \put(576,4947){\makebox(0,0)[r]{\strut{}$12000$}}%
      \csname LTb\endcsname
      \put(1365,576){\rotatebox{-45.00}{\makebox(0,0)[l]{\strut{}0}}}%
      \csname LTb\endcsname
      \put(2010,576){\rotatebox{-45.00}{\makebox(0,0)[l]{\strut{}20}}}%
      \csname LTb\endcsname
      \put(2655,576){\rotatebox{-45.00}{\makebox(0,0)[l]{\strut{}40}}}%
      \csname LTb\endcsname
      \put(3300,576){\rotatebox{-45.00}{\makebox(0,0)[l]{\strut{}60}}}%
      \csname LTb\endcsname
      \put(3945,576){\rotatebox{-45.00}{\makebox(0,0)[l]{\strut{}80}}}%
      \csname LTb\endcsname
      \put(4590,576){\rotatebox{-45.00}{\makebox(0,0)[l]{\strut{}100}}}%
    }%
    \gplgaddtomacro\gplfronttext{%
      \csname LTb\endcsname
      \put(-372,2833){\rotatebox{-270.00}{\makebox(0,0){\strut{}Time (s)}}}%
      \put(2977,55){\makebox(0,0){\strut{}Malleability (\%)}}%
    }%
    \gplbacktext
    \put(0,0){\includegraphics[width={283.40bp},height={283.40bp}]{img//EagleWaittimeMeanPartialStatistic}}%
    \gplfronttext
  \end{picture}%
\endgroup

%% file: img/EagleNodeutilizationMeanPartialStatistic.tex
\begingroup
\LARGE
  \makeatletter
  \providecommand\color[2][]{%
    \GenericError{(gnuplot) \space\space\space\@spaces}{%
      Package color not loaded in conjunction with
      terminal option `colourtext'%
    }{See the gnuplot documentation for explanation.%
    }{Either use 'blacktext' in gnuplot or load the package
      color.sty in LaTeX.}%
    \renewcommand\color[2][]{}%
  }%
  \providecommand\includegraphics[2][]{%
    \GenericError{(gnuplot) \space\space\space\@spaces}{%
      Package graphicx or graphics not loaded%
    }{See the gnuplot documentation for explanation.%
    }{The gnuplot epslatex terminal needs graphicx.sty or graphics.sty.}%
    \renewcommand\includegraphics[2][]{}%
  }%
  \providecommand\rotatebox[2]{#2}%
  \@ifundefined{ifGPcolor}{%
    \newif\ifGPcolor
    \GPcolortrue
  }{}%
  \@ifundefined{ifGPblacktext}{%
    \newif\ifGPblacktext
    \GPblacktextfalse
  }{}%
  \let\gplgaddtomacro\g@addto@macro
  \gdef\gplbacktext{}%
  \gdef\gplfronttext{}%
  \makeatother
  \ifGPblacktext
    \def\colorrgb#1{}%
    \def\colorgray#1{}%
  \else
    \ifGPcolor
      \def\colorrgb#1{\color[rgb]{#1}}%
      \def\colorgray#1{\color[gray]{#1}}%
      \expandafter\def\csname LTw\endcsname{\color{white}}%
      \expandafter\def\csname LTb\endcsname{\color{black}}%
      \expandafter\def\csname LTa\endcsname{\color{black}}%
      \expandafter\def\csname LT0\endcsname{\color[rgb]{1,0,0}}%
      \expandafter\def\csname LT1\endcsname{\color[rgb]{0,1,0}}%
      \expandafter\def\csname LT2\endcsname{\color[rgb]{0,0,1}}%
      \expandafter\def\csname LT3\endcsname{\color[rgb]{1,0,1}}%
      \expandafter\def\csname LT4\endcsname{\color[rgb]{0,1,1}}%
      \expandafter\def\csname LT5\endcsname{\color[rgb]{1,1,0}}%
      \expandafter\def\csname LT6\endcsname{\color[rgb]{0,0,0}}%
      \expandafter\def\csname LT7\endcsname{\color[rgb]{1,0.3,0}}%
      \expandafter\def\csname LT8\endcsname{\color[rgb]{0.5,0.5,0.5}}%
    \else
      \def\colorrgb#1{\color{black}}%
      \def\colorgray#1{\color[gray]{#1}}%
      \expandafter\def\csname LTw\endcsname{\color{white}}%
      \expandafter\def\csname LTb\endcsname{\color{black}}%
      \expandafter\def\csname LTa\endcsname{\color{black}}%
      \expandafter\def\csname LT0\endcsname{\color{black}}%
      \expandafter\def\csname LT1\endcsname{\color{black}}%
      \expandafter\def\csname LT2\endcsname{\color{black}}%
      \expandafter\def\csname LT3\endcsname{\color{black}}%
      \expandafter\def\csname LT4\endcsname{\color{black}}%
      \expandafter\def\csname LT5\endcsname{\color{black}}%
      \expandafter\def\csname LT6\endcsname{\color{black}}%
      \expandafter\def\csname LT7\endcsname{\color{black}}%
      \expandafter\def\csname LT8\endcsname{\color{black}}%
    \fi
  \fi
    \setlength{\unitlength}{0.0500bp}%
    \ifx\gptboxheight\undefined%
      \newlength{\gptboxheight}%
      \newlength{\gptboxwidth}%
      \newsavebox{\gptboxtext}%
    \fi%
    \setlength{\fboxrule}{0.5pt}%
    \setlength{\fboxsep}{1pt}%
    \definecolor{tbcol}{rgb}{1,1,1}%
\begin{picture}(5668.00,5668.00)%
    \gplgaddtomacro\gplbacktext{%
      \csname LTb\endcsname
      \put(576,720){\makebox(0,0)[r]{\strut{}$10$}}%
      \csname LTb\endcsname
      \put(576,1248){\makebox(0,0)[r]{\strut{}$20$}}%
      \csname LTb\endcsname
      \put(576,1777){\makebox(0,0)[r]{\strut{}$30$}}%
      \csname LTb\endcsname
      \put(576,2305){\makebox(0,0)[r]{\strut{}$40$}}%
      \csname LTb\endcsname
      \put(576,2834){\makebox(0,0)[r]{\strut{}$50$}}%
      \csname LTb\endcsname
      \put(576,3362){\makebox(0,0)[r]{\strut{}$60$}}%
      \csname LTb\endcsname
      \put(576,3890){\makebox(0,0)[r]{\strut{}$70$}}%
      \csname LTb\endcsname
      \put(576,4419){\makebox(0,0)[r]{\strut{}$80$}}%
      \csname LTb\endcsname
      \put(576,4947){\makebox(0,0)[r]{\strut{}$90$}}%
      \csname LTb\endcsname
      \put(1365,576){\rotatebox{-45.00}{\makebox(0,0)[l]{\strut{}0}}}%
      \csname LTb\endcsname
      \put(2010,576){\rotatebox{-45.00}{\makebox(0,0)[l]{\strut{}20}}}%
      \csname LTb\endcsname
      \put(2655,576){\rotatebox{-45.00}{\makebox(0,0)[l]{\strut{}40}}}%
      \csname LTb\endcsname
      \put(3300,576){\rotatebox{-45.00}{\makebox(0,0)[l]{\strut{}60}}}%
      \csname LTb\endcsname
      \put(3945,576){\rotatebox{-45.00}{\makebox(0,0)[l]{\strut{}80}}}%
      \csname LTb\endcsname
      \put(4590,576){\rotatebox{-45.00}{\makebox(0,0)[l]{\strut{}100}}}%
    }%
    \gplgaddtomacro\gplfronttext{%
      \csname LTb\endcsname
      \put(60,2833){\rotatebox{-270.00}{\makebox(0,0){\strut{}Node Utilization (\%)}}}%
      \put(2977,55){\makebox(0,0){\strut{}Malleability (\%)}}%
    }%
    \gplbacktext
    \put(0,0){\includegraphics[width={283.40bp},height={283.40bp}]{img//EagleNodeutilizationMeanPartialStatistic}}%
    \gplfronttext
  \end{picture}%
\endgroup

%% file: img/EagleExpandsJobPartialStatistic.tex
\begingroup
\LARGE
  \makeatletter
  \providecommand\color[2][]{%
    \GenericError{(gnuplot) \space\space\space\@spaces}{%
      Package color not loaded in conjunction with
      terminal option `colourtext'%
    }{See the gnuplot documentation for explanation.%
    }{Either use 'blacktext' in gnuplot or load the package
      color.sty in LaTeX.}%
    \renewcommand\color[2][]{}%
  }%
  \providecommand\includegraphics[2][]{%
    \GenericError{(gnuplot) \space\space\space\@spaces}{%
      Package graphicx or graphics not loaded%
    }{See the gnuplot documentation for explanation.%
    }{The gnuplot epslatex terminal needs graphicx.sty or graphics.sty.}%
    \renewcommand\includegraphics[2][]{}%
  }%
  \providecommand\rotatebox[2]{#2}%
  \@ifundefined{ifGPcolor}{%
    \newif\ifGPcolor
    \GPcolortrue
  }{}%
  \@ifundefined{ifGPblacktext}{%
    \newif\ifGPblacktext
    \GPblacktextfalse
  }{}%
  \let\gplgaddtomacro\g@addto@macro
  \gdef\gplbacktext{}%
  \gdef\gplfronttext{}%
  \makeatother
  \ifGPblacktext
    \def\colorrgb#1{}%
    \def\colorgray#1{}%
  \else
    \ifGPcolor
      \def\colorrgb#1{\color[rgb]{#1}}%
      \def\colorgray#1{\color[gray]{#1}}%
      \expandafter\def\csname LTw\endcsname{\color{white}}%
      \expandafter\def\csname LTb\endcsname{\color{black}}%
      \expandafter\def\csname LTa\endcsname{\color{black}}%
      \expandafter\def\csname LT0\endcsname{\color[rgb]{1,0,0}}%
      \expandafter\def\csname LT1\endcsname{\color[rgb]{0,1,0}}%
      \expandafter\def\csname LT2\endcsname{\color[rgb]{0,0,1}}%
      \expandafter\def\csname LT3\endcsname{\color[rgb]{1,0,1}}%
      \expandafter\def\csname LT4\endcsname{\color[rgb]{0,1,1}}%
      \expandafter\def\csname LT5\endcsname{\color[rgb]{1,1,0}}%
      \expandafter\def\csname LT6\endcsname{\color[rgb]{0,0,0}}%
      \expandafter\def\csname LT7\endcsname{\color[rgb]{1,0.3,0}}%
      \expandafter\def\csname LT8\endcsname{\color[rgb]{0.5,0.5,0.5}}%
    \else
      \def\colorrgb#1{\color{black}}%
      \def\colorgray#1{\color[gray]{#1}}%
      \expandafter\def\csname LTw\endcsname{\color{white}}%
      \expandafter\def\csname LTb\endcsname{\color{black}}%
      \expandafter\def\csname LTa\endcsname{\color{black}}%
      \expandafter\def\csname LT0\endcsname{\color{black}}%
      \expandafter\def\csname LT1\endcsname{\color{black}}%
      \expandafter\def\csname LT2\endcsname{\color{black}}%
      \expandafter\def\csname LT3\endcsname{\color{black}}%
      \expandafter\def\csname LT4\endcsname{\color{black}}%
      \expandafter\def\csname LT5\endcsname{\color{black}}%
      \expandafter\def\csname LT6\endcsname{\color{black}}%
      \expandafter\def\csname LT7\endcsname{\color{black}}%
      \expandafter\def\csname LT8\endcsname{\color{black}}%
    \fi
  \fi
    \setlength{\unitlength}{0.0500bp}%
    \ifx\gptboxheight\undefined%
      \newlength{\gptboxheight}%
      \newlength{\gptboxwidth}%
      \newsavebox{\gptboxtext}%
    \fi%
    \setlength{\fboxrule}{0.5pt}%
    \setlength{\fboxsep}{1pt}%
    \definecolor{tbcol}{rgb}{1,1,1}%
\begin{picture}(5668.00,5668.00)%
    \gplgaddtomacro\gplbacktext{%
      \csname LTb\endcsname
      \put(576,720){\makebox(0,0)[r]{\strut{}$0$}}%
      \csname LTb\endcsname
      \put(576,1324){\makebox(0,0)[r]{\strut{}$2$}}%
      \csname LTb\endcsname
      \put(576,1928){\makebox(0,0)[r]{\strut{}$4$}}%
      \csname LTb\endcsname
      \put(576,2532){\makebox(0,0)[r]{\strut{}$6$}}%
      \csname LTb\endcsname
      \put(576,3135){\makebox(0,0)[r]{\strut{}$8$}}%
      \csname LTb\endcsname
      \put(576,3739){\makebox(0,0)[r]{\strut{}$10$}}%
      \csname LTb\endcsname
      \put(576,4343){\makebox(0,0)[r]{\strut{}$12$}}%
      \csname LTb\endcsname
      \put(576,4947){\makebox(0,0)[r]{\strut{}$14$}}%
      \csname LTb\endcsname
      \put(1365,576){\rotatebox{-45.00}{\makebox(0,0)[l]{\strut{}0}}}%
      \csname LTb\endcsname
      \put(2010,576){\rotatebox{-45.00}{\makebox(0,0)[l]{\strut{}20}}}%
      \csname LTb\endcsname
      \put(2655,576){\rotatebox{-45.00}{\makebox(0,0)[l]{\strut{}40}}}%
      \csname LTb\endcsname
      \put(3300,576){\rotatebox{-45.00}{\makebox(0,0)[l]{\strut{}60}}}%
      \csname LTb\endcsname
      \put(3945,576){\rotatebox{-45.00}{\makebox(0,0)[l]{\strut{}80}}}%
      \csname LTb\endcsname
      \put(4590,576){\rotatebox{-45.00}{\makebox(0,0)[l]{\strut{}100}}}%
    }%
    \gplgaddtomacro\gplfronttext{%
      \csname LTb\endcsname
      \put(60,2833){\rotatebox{-270.00}{\makebox(0,0){\strut{}Events}}}%
      \put(2977,55){\makebox(0,0){\strut{}Malleability (\%)}}%
    }%
    \gplbacktext
    \put(0,0){\includegraphics[width={283.40bp},height={283.40bp}]{img//EagleExpandsJobPartialStatistic}}%
    \gplfronttext
  \end{picture}%
\endgroup

%% file: img/EagleShrinksJobPartialStatistic.tex
\begingroup
\LARGE
  \makeatletter
  \providecommand\color[2][]{%
    \GenericError{(gnuplot) \space\space\space\@spaces}{%
      Package color not loaded in conjunction with
      terminal option `colourtext'%
    }{See the gnuplot documentation for explanation.%
    }{Either use 'blacktext' in gnuplot or load the package
      color.sty in LaTeX.}%
    \renewcommand\color[2][]{}%
  }%
  \providecommand\includegraphics[2][]{%
    \GenericError{(gnuplot) \space\space\space\@spaces}{%
      Package graphicx or graphics not loaded%
    }{See the gnuplot documentation for explanation.%
    }{The gnuplot epslatex terminal needs graphicx.sty or graphics.sty.}%
    \renewcommand\includegraphics[2][]{}%
  }%
  \providecommand\rotatebox[2]{#2}%
  \@ifundefined{ifGPcolor}{%
    \newif\ifGPcolor
    \GPcolortrue
  }{}%
  \@ifundefined{ifGPblacktext}{%
    \newif\ifGPblacktext
    \GPblacktextfalse
  }{}%
  \let\gplgaddtomacro\g@addto@macro
  \gdef\gplbacktext{}%
  \gdef\gplfronttext{}%
  \makeatother
  \ifGPblacktext
    \def\colorrgb#1{}%
    \def\colorgray#1{}%
  \else
    \ifGPcolor
      \def\colorrgb#1{\color[rgb]{#1}}%
      \def\colorgray#1{\color[gray]{#1}}%
      \expandafter\def\csname LTw\endcsname{\color{white}}%
      \expandafter\def\csname LTb\endcsname{\color{black}}%
      \expandafter\def\csname LTa\endcsname{\color{black}}%
      \expandafter\def\csname LT0\endcsname{\color[rgb]{1,0,0}}%
      \expandafter\def\csname LT1\endcsname{\color[rgb]{0,1,0}}%
      \expandafter\def\csname LT2\endcsname{\color[rgb]{0,0,1}}%
      \expandafter\def\csname LT3\endcsname{\color[rgb]{1,0,1}}%
      \expandafter\def\csname LT4\endcsname{\color[rgb]{0,1,1}}%
      \expandafter\def\csname LT5\endcsname{\color[rgb]{1,1,0}}%
      \expandafter\def\csname LT6\endcsname{\color[rgb]{0,0,0}}%
      \expandafter\def\csname LT7\endcsname{\color[rgb]{1,0.3,0}}%
      \expandafter\def\csname LT8\endcsname{\color[rgb]{0.5,0.5,0.5}}%
    \else
      \def\colorrgb#1{\color{black}}%
      \def\colorgray#1{\color[gray]{#1}}%
      \expandafter\def\csname LTw\endcsname{\color{white}}%
      \expandafter\def\csname LTb\endcsname{\color{black}}%
      \expandafter\def\csname LTa\endcsname{\color{black}}%
      \expandafter\def\csname LT0\endcsname{\color{black}}%
      \expandafter\def\csname LT1\endcsname{\color{black}}%
      \expandafter\def\csname LT2\endcsname{\color{black}}%
      \expandafter\def\csname LT3\endcsname{\color{black}}%
      \expandafter\def\csname LT4\endcsname{\color{black}}%
      \expandafter\def\csname LT5\endcsname{\color{black}}%
      \expandafter\def\csname LT6\endcsname{\color{black}}%
      \expandafter\def\csname LT7\endcsname{\color{black}}%
      \expandafter\def\csname LT8\endcsname{\color{black}}%
    \fi
  \fi
    \setlength{\unitlength}{0.0500bp}%
    \ifx\gptboxheight\undefined%
      \newlength{\gptboxheight}%
      \newlength{\gptboxwidth}%
      \newsavebox{\gptboxtext}%
    \fi%
    \setlength{\fboxrule}{0.5pt}%
    \setlength{\fboxsep}{1pt}%
    \definecolor{tbcol}{rgb}{1,1,1}%
\begin{picture}(5668.00,5668.00)%
    \gplgaddtomacro\gplbacktext{%
      \csname LTb\endcsname
      \put(576,720){\makebox(0,0)[r]{\strut{}$0$}}%
      \csname LTb\endcsname
      \put(576,1324){\makebox(0,0)[r]{\strut{}$1$}}%
      \csname LTb\endcsname
      \put(576,1928){\makebox(0,0)[r]{\strut{}$2$}}%
      \csname LTb\endcsname
      \put(576,2532){\makebox(0,0)[r]{\strut{}$3$}}%
      \csname LTb\endcsname
      \put(576,3135){\makebox(0,0)[r]{\strut{}$4$}}%
      \csname LTb\endcsname
      \put(576,3739){\makebox(0,0)[r]{\strut{}$5$}}%
      \csname LTb\endcsname
      \put(576,4343){\makebox(0,0)[r]{\strut{}$6$}}%
      \csname LTb\endcsname
      \put(576,4947){\makebox(0,0)[r]{\strut{}$7$}}%
      \csname LTb\endcsname
      \put(1365,576){\rotatebox{-45.00}{\makebox(0,0)[l]{\strut{}0}}}%
      \csname LTb\endcsname
      \put(2010,576){\rotatebox{-45.00}{\makebox(0,0)[l]{\strut{}20}}}%
      \csname LTb\endcsname
      \put(2655,576){\rotatebox{-45.00}{\makebox(0,0)[l]{\strut{}40}}}%
      \csname LTb\endcsname
      \put(3300,576){\rotatebox{-45.00}{\makebox(0,0)[l]{\strut{}60}}}%
      \csname LTb\endcsname
      \put(3945,576){\rotatebox{-45.00}{\makebox(0,0)[l]{\strut{}80}}}%
      \csname LTb\endcsname
      \put(4590,576){\rotatebox{-45.00}{\makebox(0,0)[l]{\strut{}100}}}%
    }%
    \gplgaddtomacro\gplfronttext{%
      \csname LTb\endcsname
      \put(204,2833){\rotatebox{-270.00}{\makebox(0,0){\strut{}Events}}}%
      \put(2977,55){\makebox(0,0){\strut{}Malleability (\%)}}%
    }%
    \gplbacktext
    \put(0,0){\includegraphics[width={283.40bp},height={283.40bp}]{img//EagleShrinksJobPartialStatistic}}%
    \gplfronttext
  \end{picture}%
\endgroup

%% file: img/ThetaTurnaroundMeanPartialStatistic.tex
\begingroup
\LARGE
  \makeatletter
  \providecommand\color[2][]{%
    \GenericError{(gnuplot) \space\space\space\@spaces}{%
      Package color not loaded in conjunction with
      terminal option `colourtext'%
    }{See the gnuplot documentation for explanation.%
    }{Either use 'blacktext' in gnuplot or load the package
      color.sty in LaTeX.}%
    \renewcommand\color[2][]{}%
  }%
  \providecommand\includegraphics[2][]{%
    \GenericError{(gnuplot) \space\space\space\@spaces}{%
      Package graphicx or graphics not loaded%
    }{See the gnuplot documentation for explanation.%
    }{The gnuplot epslatex terminal needs graphicx.sty or graphics.sty.}%
    \renewcommand\includegraphics[2][]{}%
  }%
  \providecommand\rotatebox[2]{#2}%
  \@ifundefined{ifGPcolor}{%
    \newif\ifGPcolor
    \GPcolortrue
  }{}%
  \@ifundefined{ifGPblacktext}{%
    \newif\ifGPblacktext
    \GPblacktextfalse
  }{}%
  \let\gplgaddtomacro\g@addto@macro
  \gdef\gplbacktext{}%
  \gdef\gplfronttext{}%
  \makeatother
  \ifGPblacktext
    \def\colorrgb#1{}%
    \def\colorgray#1{}%
  \else
    \ifGPcolor
      \def\colorrgb#1{\color[rgb]{#1}}%
      \def\colorgray#1{\color[gray]{#1}}%
      \expandafter\def\csname LTw\endcsname{\color{white}}%
      \expandafter\def\csname LTb\endcsname{\color{black}}%
      \expandafter\def\csname LTa\endcsname{\color{black}}%
      \expandafter\def\csname LT0\endcsname{\color[rgb]{1,0,0}}%
      \expandafter\def\csname LT1\endcsname{\color[rgb]{0,1,0}}%
      \expandafter\def\csname LT2\endcsname{\color[rgb]{0,0,1}}%
      \expandafter\def\csname LT3\endcsname{\color[rgb]{1,0,1}}%
      \expandafter\def\csname LT4\endcsname{\color[rgb]{0,1,1}}%
      \expandafter\def\csname LT5\endcsname{\color[rgb]{1,1,0}}%
      \expandafter\def\csname LT6\endcsname{\color[rgb]{0,0,0}}%
      \expandafter\def\csname LT7\endcsname{\color[rgb]{1,0.3,0}}%
      \expandafter\def\csname LT8\endcsname{\color[rgb]{0.5,0.5,0.5}}%
    \else
      \def\colorrgb#1{\color{black}}%
      \def\colorgray#1{\color[gray]{#1}}%
      \expandafter\def\csname LTw\endcsname{\color{white}}%
      \expandafter\def\csname LTb\endcsname{\color{black}}%
      \expandafter\def\csname LTa\endcsname{\color{black}}%
      \expandafter\def\csname LT0\endcsname{\color{black}}%
      \expandafter\def\csname LT1\endcsname{\color{black}}%
      \expandafter\def\csname LT2\endcsname{\color{black}}%
      \expandafter\def\csname LT3\endcsname{\color{black}}%
      \expandafter\def\csname LT4\endcsname{\color{black}}%
      \expandafter\def\csname LT5\endcsname{\color{black}}%
      \expandafter\def\csname LT6\endcsname{\color{black}}%
      \expandafter\def\csname LT7\endcsname{\color{black}}%
      \expandafter\def\csname LT8\endcsname{\color{black}}%
    \fi
  \fi
    \setlength{\unitlength}{0.0500bp}%
    \ifx\gptboxheight\undefined%
      \newlength{\gptboxheight}%
      \newlength{\gptboxwidth}%
      \newsavebox{\gptboxtext}%
    \fi%
    \setlength{\fboxrule}{0.5pt}%
    \setlength{\fboxsep}{1pt}%
    \definecolor{tbcol}{rgb}{1,1,1}%
\begin{picture}(5668.00,5668.00)%
    \gplgaddtomacro\gplbacktext{%
      \csname LTb\endcsname
      \put(576,720){\makebox(0,0)[r]{\strut{}$0$}}%
      \csname LTb\endcsname
      \put(576,1143){\makebox(0,0)[r]{\strut{}$1000$}}%
      \csname LTb\endcsname
      \put(576,1565){\makebox(0,0)[r]{\strut{}$2000$}}%
      \csname LTb\endcsname
      \put(576,1988){\makebox(0,0)[r]{\strut{}$3000$}}%
      \csname LTb\endcsname
      \put(576,2411){\makebox(0,0)[r]{\strut{}$4000$}}%
      \csname LTb\endcsname
      \put(576,2834){\makebox(0,0)[r]{\strut{}$5000$}}%
      \csname LTb\endcsname
      \put(576,3256){\makebox(0,0)[r]{\strut{}$6000$}}%
      \csname LTb\endcsname
      \put(576,3679){\makebox(0,0)[r]{\strut{}$7000$}}%
      \csname LTb\endcsname
      \put(576,4102){\makebox(0,0)[r]{\strut{}$8000$}}%
      \csname LTb\endcsname
      \put(576,4524){\makebox(0,0)[r]{\strut{}$9000$}}%
      \csname LTb\endcsname
      \put(576,4947){\makebox(0,0)[r]{\strut{}$10000$}}%
      \csname LTb\endcsname
      \put(1365,576){\rotatebox{-45.00}{\makebox(0,0)[l]{\strut{}0}}}%
      \csname LTb\endcsname
      \put(2010,576){\rotatebox{-45.00}{\makebox(0,0)[l]{\strut{}20}}}%
      \csname LTb\endcsname
      \put(2655,576){\rotatebox{-45.00}{\makebox(0,0)[l]{\strut{}40}}}%
      \csname LTb\endcsname
      \put(3300,576){\rotatebox{-45.00}{\makebox(0,0)[l]{\strut{}60}}}%
      \csname LTb\endcsname
      \put(3945,576){\rotatebox{-45.00}{\makebox(0,0)[l]{\strut{}80}}}%
      \csname LTb\endcsname
      \put(4590,576){\rotatebox{-45.00}{\makebox(0,0)[l]{\strut{}100}}}%
    }%
    \gplgaddtomacro\gplfronttext{%
      \csname LTb\endcsname
      \put(-372,2833){\rotatebox{-270.00}{\makebox(0,0){\strut{}Time (s)}}}%
      \put(2977,55){\makebox(0,0){\strut{}Malleability (\%)}}%
    }%
    \gplbacktext
    \put(0,0){\includegraphics[width={283.40bp},height={283.40bp}]{img//ThetaTurnaroundMeanPartialStatistic}}%
    \gplfronttext
  \end{picture}%
\endgroup

%% file: img/ThetaMakespanMeanPartialStatistic.tex
\begingroup
\LARGE
  \makeatletter
  \providecommand\color[2][]{%
    \GenericError{(gnuplot) \space\space\space\@spaces}{%
      Package color not loaded in conjunction with
      terminal option `colourtext'%
    }{See the gnuplot documentation for explanation.%
    }{Either use 'blacktext' in gnuplot or load the package
      color.sty in LaTeX.}%
    \renewcommand\color[2][]{}%
  }%
  \providecommand\includegraphics[2][]{%
    \GenericError{(gnuplot) \space\space\space\@spaces}{%
      Package graphicx or graphics not loaded%
    }{See the gnuplot documentation for explanation.%
    }{The gnuplot epslatex terminal needs graphicx.sty or graphics.sty.}%
    \renewcommand\includegraphics[2][]{}%
  }%
  \providecommand\rotatebox[2]{#2}%
  \@ifundefined{ifGPcolor}{%
    \newif\ifGPcolor
    \GPcolortrue
  }{}%
  \@ifundefined{ifGPblacktext}{%
    \newif\ifGPblacktext
    \GPblacktextfalse
  }{}%
  \let\gplgaddtomacro\g@addto@macro
  \gdef\gplbacktext{}%
  \gdef\gplfronttext{}%
  \makeatother
  \ifGPblacktext
    \def\colorrgb#1{}%
    \def\colorgray#1{}%
  \else
    \ifGPcolor
      \def\colorrgb#1{\color[rgb]{#1}}%
      \def\colorgray#1{\color[gray]{#1}}%
      \expandafter\def\csname LTw\endcsname{\color{white}}%
      \expandafter\def\csname LTb\endcsname{\color{black}}%
      \expandafter\def\csname LTa\endcsname{\color{black}}%
      \expandafter\def\csname LT0\endcsname{\color[rgb]{1,0,0}}%
      \expandafter\def\csname LT1\endcsname{\color[rgb]{0,1,0}}%
      \expandafter\def\csname LT2\endcsname{\color[rgb]{0,0,1}}%
      \expandafter\def\csname LT3\endcsname{\color[rgb]{1,0,1}}%
      \expandafter\def\csname LT4\endcsname{\color[rgb]{0,1,1}}%
      \expandafter\def\csname LT5\endcsname{\color[rgb]{1,1,0}}%
      \expandafter\def\csname LT6\endcsname{\color[rgb]{0,0,0}}%
      \expandafter\def\csname LT7\endcsname{\color[rgb]{1,0.3,0}}%
      \expandafter\def\csname LT8\endcsname{\color[rgb]{0.5,0.5,0.5}}%
    \else
      \def\colorrgb#1{\color{black}}%
      \def\colorgray#1{\color[gray]{#1}}%
      \expandafter\def\csname LTw\endcsname{\color{white}}%
      \expandafter\def\csname LTb\endcsname{\color{black}}%
      \expandafter\def\csname LTa\endcsname{\color{black}}%
      \expandafter\def\csname LT0\endcsname{\color{black}}%
      \expandafter\def\csname LT1\endcsname{\color{black}}%
      \expandafter\def\csname LT2\endcsname{\color{black}}%
      \expandafter\def\csname LT3\endcsname{\color{black}}%
      \expandafter\def\csname LT4\endcsname{\color{black}}%
      \expandafter\def\csname LT5\endcsname{\color{black}}%
      \expandafter\def\csname LT6\endcsname{\color{black}}%
      \expandafter\def\csname LT7\endcsname{\color{black}}%
      \expandafter\def\csname LT8\endcsname{\color{black}}%
    \fi
  \fi
    \setlength{\unitlength}{0.0500bp}%
    \ifx\gptboxheight\undefined%
      \newlength{\gptboxheight}%
      \newlength{\gptboxwidth}%
      \newsavebox{\gptboxtext}%
    \fi%
    \setlength{\fboxrule}{0.5pt}%
    \setlength{\fboxsep}{1pt}%
    \definecolor{tbcol}{rgb}{1,1,1}%
\begin{picture}(5668.00,5668.00)%
    \gplgaddtomacro\gplbacktext{%
      \csname LTb\endcsname
      \put(576,720){\makebox(0,0)[r]{\strut{}$0$}}%
      \csname LTb\endcsname
      \put(576,1190){\makebox(0,0)[r]{\strut{}$1000$}}%
      \csname LTb\endcsname
      \put(576,1659){\makebox(0,0)[r]{\strut{}$2000$}}%
      \csname LTb\endcsname
      \put(576,2129){\makebox(0,0)[r]{\strut{}$3000$}}%
      \csname LTb\endcsname
      \put(576,2599){\makebox(0,0)[r]{\strut{}$4000$}}%
      \csname LTb\endcsname
      \put(576,3068){\makebox(0,0)[r]{\strut{}$5000$}}%
      \csname LTb\endcsname
      \put(576,3538){\makebox(0,0)[r]{\strut{}$6000$}}%
      \csname LTb\endcsname
      \put(576,4008){\makebox(0,0)[r]{\strut{}$7000$}}%
      \csname LTb\endcsname
      \put(576,4477){\makebox(0,0)[r]{\strut{}$8000$}}%
      \csname LTb\endcsname
      \put(576,4947){\makebox(0,0)[r]{\strut{}$9000$}}%
      \csname LTb\endcsname
      \put(1365,576){\rotatebox{-45.00}{\makebox(0,0)[l]{\strut{}0}}}%
      \csname LTb\endcsname
      \put(2010,576){\rotatebox{-45.00}{\makebox(0,0)[l]{\strut{}20}}}%
      \csname LTb\endcsname
      \put(2655,576){\rotatebox{-45.00}{\makebox(0,0)[l]{\strut{}40}}}%
      \csname LTb\endcsname
      \put(3300,576){\rotatebox{-45.00}{\makebox(0,0)[l]{\strut{}60}}}%
      \csname LTb\endcsname
      \put(3945,576){\rotatebox{-45.00}{\makebox(0,0)[l]{\strut{}80}}}%
      \csname LTb\endcsname
      \put(4590,576){\rotatebox{-45.00}{\makebox(0,0)[l]{\strut{}100}}}%
    }%
    \gplgaddtomacro\gplfronttext{%
      \csname LTb\endcsname
      \put(-228,2833){\rotatebox{-270.00}{\makebox(0,0){\strut{}Time (s)}}}%
      \put(2977,55){\makebox(0,0){\strut{}Malleability (\%)}}%
    }%
    \gplbacktext
    \put(0,0){\includegraphics[width={283.40bp},height={283.40bp}]{img//ThetaMakespanMeanPartialStatistic}}%
    \gplfronttext
  \end{picture}%
\endgroup

%% file: img/ThetaWaittimeMeanPartialStatistic.tex
\begingroup
\LARGE
  \makeatletter
  \providecommand\color[2][]{%
    \GenericError{(gnuplot) \space\space\space\@spaces}{%
      Package color not loaded in conjunction with
      terminal option `colourtext'%
    }{See the gnuplot documentation for explanation.%
    }{Either use 'blacktext' in gnuplot or load the package
      color.sty in LaTeX.}%
    \renewcommand\color[2][]{}%
  }%
  \providecommand\includegraphics[2][]{%
    \GenericError{(gnuplot) \space\space\space\@spaces}{%
      Package graphicx or graphics not loaded%
    }{See the gnuplot documentation for explanation.%
    }{The gnuplot epslatex terminal needs graphicx.sty or graphics.sty.}%
    \renewcommand\includegraphics[2][]{}%
  }%
  \providecommand\rotatebox[2]{#2}%
  \@ifundefined{ifGPcolor}{%
    \newif\ifGPcolor
    \GPcolortrue
  }{}%
  \@ifundefined{ifGPblacktext}{%
    \newif\ifGPblacktext
    \GPblacktextfalse
  }{}%
  \let\gplgaddtomacro\g@addto@macro
  \gdef\gplbacktext{}%
  \gdef\gplfronttext{}%
  \makeatother
  \ifGPblacktext
    \def\colorrgb#1{}%
    \def\colorgray#1{}%
  \else
    \ifGPcolor
      \def\colorrgb#1{\color[rgb]{#1}}%
      \def\colorgray#1{\color[gray]{#1}}%
      \expandafter\def\csname LTw\endcsname{\color{white}}%
      \expandafter\def\csname LTb\endcsname{\color{black}}%
      \expandafter\def\csname LTa\endcsname{\color{black}}%
      \expandafter\def\csname LT0\endcsname{\color[rgb]{1,0,0}}%
      \expandafter\def\csname LT1\endcsname{\color[rgb]{0,1,0}}%
      \expandafter\def\csname LT2\endcsname{\color[rgb]{0,0,1}}%
      \expandafter\def\csname LT3\endcsname{\color[rgb]{1,0,1}}%
      \expandafter\def\csname LT4\endcsname{\color[rgb]{0,1,1}}%
      \expandafter\def\csname LT5\endcsname{\color[rgb]{1,1,0}}%
      \expandafter\def\csname LT6\endcsname{\color[rgb]{0,0,0}}%
      \expandafter\def\csname LT7\endcsname{\color[rgb]{1,0.3,0}}%
      \expandafter\def\csname LT8\endcsname{\color[rgb]{0.5,0.5,0.5}}%
    \else
      \def\colorrgb#1{\color{black}}%
      \def\colorgray#1{\color[gray]{#1}}%
      \expandafter\def\csname LTw\endcsname{\color{white}}%
      \expandafter\def\csname LTb\endcsname{\color{black}}%
      \expandafter\def\csname LTa\endcsname{\color{black}}%
      \expandafter\def\csname LT0\endcsname{\color{black}}%
      \expandafter\def\csname LT1\endcsname{\color{black}}%
      \expandafter\def\csname LT2\endcsname{\color{black}}%
      \expandafter\def\csname LT3\endcsname{\color{black}}%
      \expandafter\def\csname LT4\endcsname{\color{black}}%
      \expandafter\def\csname LT5\endcsname{\color{black}}%
      \expandafter\def\csname LT6\endcsname{\color{black}}%
      \expandafter\def\csname LT7\endcsname{\color{black}}%
      \expandafter\def\csname LT8\endcsname{\color{black}}%
    \fi
  \fi
    \setlength{\unitlength}{0.0500bp}%
    \ifx\gptboxheight\undefined%
      \newlength{\gptboxheight}%
      \newlength{\gptboxwidth}%
      \newsavebox{\gptboxtext}%
    \fi%
    \setlength{\fboxrule}{0.5pt}%
    \setlength{\fboxsep}{1pt}%
    \definecolor{tbcol}{rgb}{1,1,1}%
\begin{picture}(5668.00,5668.00)%
    \gplgaddtomacro\gplbacktext{%
      \csname LTb\endcsname
      \put(576,720){\makebox(0,0)[r]{\strut{}$0$}}%
      \csname LTb\endcsname
      \put(576,1190){\makebox(0,0)[r]{\strut{}$500$}}%
      \csname LTb\endcsname
      \put(576,1659){\makebox(0,0)[r]{\strut{}$1000$}}%
      \csname LTb\endcsname
      \put(576,2129){\makebox(0,0)[r]{\strut{}$1500$}}%
      \csname LTb\endcsname
      \put(576,2599){\makebox(0,0)[r]{\strut{}$2000$}}%
      \csname LTb\endcsname
      \put(576,3068){\makebox(0,0)[r]{\strut{}$2500$}}%
      \csname LTb\endcsname
      \put(576,3538){\makebox(0,0)[r]{\strut{}$3000$}}%
      \csname LTb\endcsname
      \put(576,4008){\makebox(0,0)[r]{\strut{}$3500$}}%
      \csname LTb\endcsname
      \put(576,4477){\makebox(0,0)[r]{\strut{}$4000$}}%
      \csname LTb\endcsname
      \put(576,4947){\makebox(0,0)[r]{\strut{}$4500$}}%
      \csname LTb\endcsname
      \put(1365,576){\rotatebox{-45.00}{\makebox(0,0)[l]{\strut{}0}}}%
      \csname LTb\endcsname
      \put(2010,576){\rotatebox{-45.00}{\makebox(0,0)[l]{\strut{}20}}}%
      \csname LTb\endcsname
      \put(2655,576){\rotatebox{-45.00}{\makebox(0,0)[l]{\strut{}40}}}%
      \csname LTb\endcsname
      \put(3300,576){\rotatebox{-45.00}{\makebox(0,0)[l]{\strut{}60}}}%
      \csname LTb\endcsname
      \put(3945,576){\rotatebox{-45.00}{\makebox(0,0)[l]{\strut{}80}}}%
      \csname LTb\endcsname
      \put(4590,576){\rotatebox{-45.00}{\makebox(0,0)[l]{\strut{}100}}}%
    }%
    \gplgaddtomacro\gplfronttext{%
      \csname LTb\endcsname
      \put(-228,2833){\rotatebox{-270.00}{\makebox(0,0){\strut{}Time (s)}}}%
      \put(2977,55){\makebox(0,0){\strut{}Malleability (\%)}}%
    }%
    \gplbacktext
    \put(0,0){\includegraphics[width={283.40bp},height={283.40bp}]{img//ThetaWaittimeMeanPartialStatistic}}%
    \gplfronttext
  \end{picture}%
\endgroup

%% file: img/ThetaNodeutilizationMeanPartialStatistic.tex
\begingroup
\LARGE
  \makeatletter
  \providecommand\color[2][]{%
    \GenericError{(gnuplot) \space\space\space\@spaces}{%
      Package color not loaded in conjunction with
      terminal option `colourtext'%
    }{See the gnuplot documentation for explanation.%
    }{Either use 'blacktext' in gnuplot or load the package
      color.sty in LaTeX.}%
    \renewcommand\color[2][]{}%
  }%
  \providecommand\includegraphics[2][]{%
    \GenericError{(gnuplot) \space\space\space\@spaces}{%
      Package graphicx or graphics not loaded%
    }{See the gnuplot documentation for explanation.%
    }{The gnuplot epslatex terminal needs graphicx.sty or graphics.sty.}%
    \renewcommand\includegraphics[2][]{}%
  }%
  \providecommand\rotatebox[2]{#2}%
  \@ifundefined{ifGPcolor}{%
    \newif\ifGPcolor
    \GPcolortrue
  }{}%
  \@ifundefined{ifGPblacktext}{%
    \newif\ifGPblacktext
    \GPblacktextfalse
  }{}%
  \let\gplgaddtomacro\g@addto@macro
  \gdef\gplbacktext{}%
  \gdef\gplfronttext{}%
  \makeatother
  \ifGPblacktext
    \def\colorrgb#1{}%
    \def\colorgray#1{}%
  \else
    \ifGPcolor
      \def\colorrgb#1{\color[rgb]{#1}}%
      \def\colorgray#1{\color[gray]{#1}}%
      \expandafter\def\csname LTw\endcsname{\color{white}}%
      \expandafter\def\csname LTb\endcsname{\color{black}}%
      \expandafter\def\csname LTa\endcsname{\color{black}}%
      \expandafter\def\csname LT0\endcsname{\color[rgb]{1,0,0}}%
      \expandafter\def\csname LT1\endcsname{\color[rgb]{0,1,0}}%
      \expandafter\def\csname LT2\endcsname{\color[rgb]{0,0,1}}%
      \expandafter\def\csname LT3\endcsname{\color[rgb]{1,0,1}}%
      \expandafter\def\csname LT4\endcsname{\color[rgb]{0,1,1}}%
      \expandafter\def\csname LT5\endcsname{\color[rgb]{1,1,0}}%
      \expandafter\def\csname LT6\endcsname{\color[rgb]{0,0,0}}%
      \expandafter\def\csname LT7\endcsname{\color[rgb]{1,0.3,0}}%
      \expandafter\def\csname LT8\endcsname{\color[rgb]{0.5,0.5,0.5}}%
    \else
      \def\colorrgb#1{\color{black}}%
      \def\colorgray#1{\color[gray]{#1}}%
      \expandafter\def\csname LTw\endcsname{\color{white}}%
      \expandafter\def\csname LTb\endcsname{\color{black}}%
      \expandafter\def\csname LTa\endcsname{\color{black}}%
      \expandafter\def\csname LT0\endcsname{\color{black}}%
      \expandafter\def\csname LT1\endcsname{\color{black}}%
      \expandafter\def\csname LT2\endcsname{\color{black}}%
      \expandafter\def\csname LT3\endcsname{\color{black}}%
      \expandafter\def\csname LT4\endcsname{\color{black}}%
      \expandafter\def\csname LT5\endcsname{\color{black}}%
      \expandafter\def\csname LT6\endcsname{\color{black}}%
      \expandafter\def\csname LT7\endcsname{\color{black}}%
      \expandafter\def\csname LT8\endcsname{\color{black}}%
    \fi
  \fi
    \setlength{\unitlength}{0.0500bp}%
    \ifx\gptboxheight\undefined%
      \newlength{\gptboxheight}%
      \newlength{\gptboxwidth}%
      \newsavebox{\gptboxtext}%
    \fi%
    \setlength{\fboxrule}{0.5pt}%
    \setlength{\fboxsep}{1pt}%
    \definecolor{tbcol}{rgb}{1,1,1}%
\begin{picture}(5668.00,5668.00)%
    \gplgaddtomacro\gplbacktext{%
      \csname LTb\endcsname
      \put(576,720){\makebox(0,0)[r]{\strut{}$30$}}%
      \csname LTb\endcsname
      \put(576,1324){\makebox(0,0)[r]{\strut{}$40$}}%
      \csname LTb\endcsname
      \put(576,1928){\makebox(0,0)[r]{\strut{}$50$}}%
      \csname LTb\endcsname
      \put(576,2532){\makebox(0,0)[r]{\strut{}$60$}}%
      \csname LTb\endcsname
      \put(576,3135){\makebox(0,0)[r]{\strut{}$70$}}%
      \csname LTb\endcsname
      \put(576,3739){\makebox(0,0)[r]{\strut{}$80$}}%
      \csname LTb\endcsname
      \put(576,4343){\makebox(0,0)[r]{\strut{}$90$}}%
      \csname LTb\endcsname
      \put(576,4947){\makebox(0,0)[r]{\strut{}$100$}}%
      \csname LTb\endcsname
      \put(1365,576){\rotatebox{-45.00}{\makebox(0,0)[l]{\strut{}0}}}%
      \csname LTb\endcsname
      \put(2010,576){\rotatebox{-45.00}{\makebox(0,0)[l]{\strut{}20}}}%
      \csname LTb\endcsname
      \put(2655,576){\rotatebox{-45.00}{\makebox(0,0)[l]{\strut{}40}}}%
      \csname LTb\endcsname
      \put(3300,576){\rotatebox{-45.00}{\makebox(0,0)[l]{\strut{}60}}}%
      \csname LTb\endcsname
      \put(3945,576){\rotatebox{-45.00}{\makebox(0,0)[l]{\strut{}80}}}%
      \csname LTb\endcsname
      \put(4590,576){\rotatebox{-45.00}{\makebox(0,0)[l]{\strut{}100}}}%
    }%
    \gplgaddtomacro\gplfronttext{%
      \csname LTb\endcsname
      \put(-84,2833){\rotatebox{-270.00}{\makebox(0,0){\strut{}Node Utilization (\%)}}}%
      \put(2977,55){\makebox(0,0){\strut{}Malleability (\%)}}%
    }%
    \gplbacktext
    \put(0,0){\includegraphics[width={283.40bp},height={283.40bp}]{img//ThetaNodeutilizationMeanPartialStatistic}}%
    \gplfronttext
  \end{picture}%
\endgroup

%% file: img/ThetaExpandsJobPartialStatistic.tex
\begingroup
\LARGE
  \makeatletter
  \providecommand\color[2][]{%
    \GenericError{(gnuplot) \space\space\space\@spaces}{%
      Package color not loaded in conjunction with
      terminal option `colourtext'%
    }{See the gnuplot documentation for explanation.%
    }{Either use 'blacktext' in gnuplot or load the package
      color.sty in LaTeX.}%
    \renewcommand\color[2][]{}%
  }%
  \providecommand\includegraphics[2][]{%
    \GenericError{(gnuplot) \space\space\space\@spaces}{%
      Package graphicx or graphics not loaded%
    }{See the gnuplot documentation for explanation.%
    }{The gnuplot epslatex terminal needs graphicx.sty or graphics.sty.}%
    \renewcommand\includegraphics[2][]{}%
  }%
  \providecommand\rotatebox[2]{#2}%
  \@ifundefined{ifGPcolor}{%
    \newif\ifGPcolor
    \GPcolortrue
  }{}%
  \@ifundefined{ifGPblacktext}{%
    \newif\ifGPblacktext
    \GPblacktextfalse
  }{}%
  \let\gplgaddtomacro\g@addto@macro
  \gdef\gplbacktext{}%
  \gdef\gplfronttext{}%
  \makeatother
  \ifGPblacktext
    \def\colorrgb#1{}%
    \def\colorgray#1{}%
  \else
    \ifGPcolor
      \def\colorrgb#1{\color[rgb]{#1}}%
      \def\colorgray#1{\color[gray]{#1}}%
      \expandafter\def\csname LTw\endcsname{\color{white}}%
      \expandafter\def\csname LTb\endcsname{\color{black}}%
      \expandafter\def\csname LTa\endcsname{\color{black}}%
      \expandafter\def\csname LT0\endcsname{\color[rgb]{1,0,0}}%
      \expandafter\def\csname LT1\endcsname{\color[rgb]{0,1,0}}%
      \expandafter\def\csname LT2\endcsname{\color[rgb]{0,0,1}}%
      \expandafter\def\csname LT3\endcsname{\color[rgb]{1,0,1}}%
      \expandafter\def\csname LT4\endcsname{\color[rgb]{0,1,1}}%
      \expandafter\def\csname LT5\endcsname{\color[rgb]{1,1,0}}%
      \expandafter\def\csname LT6\endcsname{\color[rgb]{0,0,0}}%
      \expandafter\def\csname LT7\endcsname{\color[rgb]{1,0.3,0}}%
      \expandafter\def\csname LT8\endcsname{\color[rgb]{0.5,0.5,0.5}}%
    \else
      \def\colorrgb#1{\color{black}}%
      \def\colorgray#1{\color[gray]{#1}}%
      \expandafter\def\csname LTw\endcsname{\color{white}}%
      \expandafter\def\csname LTb\endcsname{\color{black}}%
      \expandafter\def\csname LTa\endcsname{\color{black}}%
      \expandafter\def\csname LT0\endcsname{\color{black}}%
      \expandafter\def\csname LT1\endcsname{\color{black}}%
      \expandafter\def\csname LT2\endcsname{\color{black}}%
      \expandafter\def\csname LT3\endcsname{\color{black}}%
      \expandafter\def\csname LT4\endcsname{\color{black}}%
      \expandafter\def\csname LT5\endcsname{\color{black}}%
      \expandafter\def\csname LT6\endcsname{\color{black}}%
      \expandafter\def\csname LT7\endcsname{\color{black}}%
      \expandafter\def\csname LT8\endcsname{\color{black}}%
    \fi
  \fi
    \setlength{\unitlength}{0.0500bp}%
    \ifx\gptboxheight\undefined%
      \newlength{\gptboxheight}%
      \newlength{\gptboxwidth}%
      \newsavebox{\gptboxtext}%
    \fi%
    \setlength{\fboxrule}{0.5pt}%
    \setlength{\fboxsep}{1pt}%
    \definecolor{tbcol}{rgb}{1,1,1}%
\begin{picture}(5668.00,5668.00)%
    \gplgaddtomacro\gplbacktext{%
      \csname LTb\endcsname
      \put(576,720){\makebox(0,0)[r]{\strut{}$0$}}%
      \csname LTb\endcsname
      \put(576,1248){\makebox(0,0)[r]{\strut{}$2$}}%
      \csname LTb\endcsname
      \put(576,1777){\makebox(0,0)[r]{\strut{}$4$}}%
      \csname LTb\endcsname
      \put(576,2305){\makebox(0,0)[r]{\strut{}$6$}}%
      \csname LTb\endcsname
      \put(576,2834){\makebox(0,0)[r]{\strut{}$8$}}%
      \csname LTb\endcsname
      \put(576,3362){\makebox(0,0)[r]{\strut{}$10$}}%
      \csname LTb\endcsname
      \put(576,3890){\makebox(0,0)[r]{\strut{}$12$}}%
      \csname LTb\endcsname
      \put(576,4419){\makebox(0,0)[r]{\strut{}$14$}}%
      \csname LTb\endcsname
      \put(576,4947){\makebox(0,0)[r]{\strut{}$16$}}%
      \csname LTb\endcsname
      \put(1365,576){\rotatebox{-45.00}{\makebox(0,0)[l]{\strut{}0}}}%
      \csname LTb\endcsname
      \put(2010,576){\rotatebox{-45.00}{\makebox(0,0)[l]{\strut{}20}}}%
      \csname LTb\endcsname
      \put(2655,576){\rotatebox{-45.00}{\makebox(0,0)[l]{\strut{}40}}}%
      \csname LTb\endcsname
      \put(3300,576){\rotatebox{-45.00}{\makebox(0,0)[l]{\strut{}60}}}%
      \csname LTb\endcsname
      \put(3945,576){\rotatebox{-45.00}{\makebox(0,0)[l]{\strut{}80}}}%
      \csname LTb\endcsname
      \put(4590,576){\rotatebox{-45.00}{\makebox(0,0)[l]{\strut{}100}}}%
    }%
    \gplgaddtomacro\gplfronttext{%
      \csname LTb\endcsname
      \put(60,2833){\rotatebox{-270.00}{\makebox(0,0){\strut{}Events}}}%
      \put(2977,55){\makebox(0,0){\strut{}Malleability (\%)}}%
    }%
    \gplbacktext
    \put(0,0){\includegraphics[width={283.40bp},height={283.40bp}]{img//ThetaExpandsJobPartialStatistic}}%
    \gplfronttext
  \end{picture}%
\endgroup

%% file: img/ThetaShrinksJobPartialStatistic.tex
\begingroup
\LARGE
  \makeatletter
  \providecommand\color[2][]{%
    \GenericError{(gnuplot) \space\space\space\@spaces}{%
      Package color not loaded in conjunction with
      terminal option `colourtext'%
    }{See the gnuplot documentation for explanation.%
    }{Either use 'blacktext' in gnuplot or load the package
      color.sty in LaTeX.}%
    \renewcommand\color[2][]{}%
  }%
  \providecommand\includegraphics[2][]{%
    \GenericError{(gnuplot) \space\space\space\@spaces}{%
      Package graphicx or graphics not loaded%
    }{See the gnuplot documentation for explanation.%
    }{The gnuplot epslatex terminal needs graphicx.sty or graphics.sty.}%
    \renewcommand\includegraphics[2][]{}%
  }%
  \providecommand\rotatebox[2]{#2}%
  \@ifundefined{ifGPcolor}{%
    \newif\ifGPcolor
    \GPcolortrue
  }{}%
  \@ifundefined{ifGPblacktext}{%
    \newif\ifGPblacktext
    \GPblacktextfalse
  }{}%
  \let\gplgaddtomacro\g@addto@macro
  \gdef\gplbacktext{}%
  \gdef\gplfronttext{}%
  \makeatother
  \ifGPblacktext
    \def\colorrgb#1{}%
    \def\colorgray#1{}%
  \else
    \ifGPcolor
      \def\colorrgb#1{\color[rgb]{#1}}%
      \def\colorgray#1{\color[gray]{#1}}%
      \expandafter\def\csname LTw\endcsname{\color{white}}%
      \expandafter\def\csname LTb\endcsname{\color{black}}%
      \expandafter\def\csname LTa\endcsname{\color{black}}%
      \expandafter\def\csname LT0\endcsname{\color[rgb]{1,0,0}}%
      \expandafter\def\csname LT1\endcsname{\color[rgb]{0,1,0}}%
      \expandafter\def\csname LT2\endcsname{\color[rgb]{0,0,1}}%
      \expandafter\def\csname LT3\endcsname{\color[rgb]{1,0,1}}%
      \expandafter\def\csname LT4\endcsname{\color[rgb]{0,1,1}}%
      \expandafter\def\csname LT5\endcsname{\color[rgb]{1,1,0}}%
      \expandafter\def\csname LT6\endcsname{\color[rgb]{0,0,0}}%
      \expandafter\def\csname LT7\endcsname{\color[rgb]{1,0.3,0}}%
      \expandafter\def\csname LT8\endcsname{\color[rgb]{0.5,0.5,0.5}}%
    \else
      \def\colorrgb#1{\color{black}}%
      \def\colorgray#1{\color[gray]{#1}}%
      \expandafter\def\csname LTw\endcsname{\color{white}}%
      \expandafter\def\csname LTb\endcsname{\color{black}}%
      \expandafter\def\csname LTa\endcsname{\color{black}}%
      \expandafter\def\csname LT0\endcsname{\color{black}}%
      \expandafter\def\csname LT1\endcsname{\color{black}}%
      \expandafter\def\csname LT2\endcsname{\color{black}}%
      \expandafter\def\csname LT3\endcsname{\color{black}}%
      \expandafter\def\csname LT4\endcsname{\color{black}}%
      \expandafter\def\csname LT5\endcsname{\color{black}}%
      \expandafter\def\csname LT6\endcsname{\color{black}}%
      \expandafter\def\csname LT7\endcsname{\color{black}}%
      \expandafter\def\csname LT8\endcsname{\color{black}}%
    \fi
  \fi
    \setlength{\unitlength}{0.0500bp}%
    \ifx\gptboxheight\undefined%
      \newlength{\gptboxheight}%
      \newlength{\gptboxwidth}%
      \newsavebox{\gptboxtext}%
    \fi%
    \setlength{\fboxrule}{0.5pt}%
    \setlength{\fboxsep}{1pt}%
    \definecolor{tbcol}{rgb}{1,1,1}%
\begin{picture}(5668.00,5668.00)%
    \gplgaddtomacro\gplbacktext{%
      \csname LTb\endcsname
      \put(576,720){\makebox(0,0)[r]{\strut{}$0$}}%
      \csname LTb\endcsname
      \put(576,1190){\makebox(0,0)[r]{\strut{}$2$}}%
      \csname LTb\endcsname
      \put(576,1659){\makebox(0,0)[r]{\strut{}$4$}}%
      \csname LTb\endcsname
      \put(576,2129){\makebox(0,0)[r]{\strut{}$6$}}%
      \csname LTb\endcsname
      \put(576,2599){\makebox(0,0)[r]{\strut{}$8$}}%
      \csname LTb\endcsname
      \put(576,3068){\makebox(0,0)[r]{\strut{}$10$}}%
      \csname LTb\endcsname
      \put(576,3538){\makebox(0,0)[r]{\strut{}$12$}}%
      \csname LTb\endcsname
      \put(576,4008){\makebox(0,0)[r]{\strut{}$14$}}%
      \csname LTb\endcsname
      \put(576,4477){\makebox(0,0)[r]{\strut{}$16$}}%
      \csname LTb\endcsname
      \put(576,4947){\makebox(0,0)[r]{\strut{}$18$}}%
      \csname LTb\endcsname
      \put(1365,576){\rotatebox{-45.00}{\makebox(0,0)[l]{\strut{}0}}}%
      \csname LTb\endcsname
      \put(2010,576){\rotatebox{-45.00}{\makebox(0,0)[l]{\strut{}20}}}%
      \csname LTb\endcsname
      \put(2655,576){\rotatebox{-45.00}{\makebox(0,0)[l]{\strut{}40}}}%
      \csname LTb\endcsname
      \put(3300,576){\rotatebox{-45.00}{\makebox(0,0)[l]{\strut{}60}}}%
      \csname LTb\endcsname
      \put(3945,576){\rotatebox{-45.00}{\makebox(0,0)[l]{\strut{}80}}}%
      \csname LTb\endcsname
      \put(4590,576){\rotatebox{-45.00}{\makebox(0,0)[l]{\strut{}100}}}%
    }%
    \gplgaddtomacro\gplfronttext{%
      \csname LTb\endcsname
      \put(60,2833){\rotatebox{-270.00}{\makebox(0,0){\strut{}Events}}}%
      \put(2977,55){\makebox(0,0){\strut{}Malleability (\%)}}%
    }%
    \gplbacktext
    \put(0,0){\includegraphics[width={283.40bp},height={283.40bp}]{img//ThetaShrinksJobPartialStatistic}}%
    \gplfronttext
  \end{picture}%
\endgroup

%% file: 04discussion.tex
\section{Limitations}\label{sec:05-Discussion}
Simulating the impact of malleability in real HPC clusters remains inherently challenging due to the complexity of modeling workloads, system architectures, and scheduling behavior.
While our simulations reveal clear trends, several limitations should be addressed in future work to further enhance accuracy.

First, more comprehensive and longer-duration workload traces would improve simulation fidelity and support stronger generalizations.
Our current simulations do not account for data movement or I/O, which is a notable omission, as such phases can significantly limit scheduling flexibility.

As GPUs are now widespread in modern HPC architectures, supporting heterogeneous workloads that use both CPUs and GPUs is a critical next step.
This study focuses solely on CPU-based workloads.

Additionally, the transformation of rigid into malleable jobs was based on a heuristic model.
Developing a systematic, empirically grounded method for this conversion would increase the realism and generalizability of simulation results.

The M100 dataset~\cite{M100} from the Marconi100 supercomputer provides hundreds of performance and sensor data, offering strong potential for future enhancements such as phase-based workload modeling and improved performance profiling.

Furthermore, schedulers could target consistent node utilization levels to enable selective node shutdowns for energy savings, maintenance, or workload-aware scaling.
This aligns with the growing focus on sustainability and system availability in HPC infrastructure planning.

%% file: 05relatedwork.tex
\section{Related Work}\label{sec:06-Related-Work}
Numerous studies have demonstrated the benefits of job malleability in HPC, e.g.~\cite{Taylan2024,ElasticJobsIntelSC23,ScheduleElasticIPDPS22,JonasMallScheduling23}.
Their findings align roughly with our results, particularly that even modest proportions of malleable jobs yield significant improvements.

In the following, we compare our work to that of Eberius~\textsl{et\,al.}~\cite{ElasticJobsIntelSC23}, who also use simulations to evaluate the impact of malleable jobs on the \cori workloads.
However, the methodological approach differs in several respects.
Notably, the rigid node utilization differs significantly, especially for the \haswell workload.
We merged split jobs and excluded shared-node jobs (see Section~\ref{subsec:03-workload-preparation}), which Eberius~\textsl{et\,al.} appear not to do.
After merging, the original \num{119781} jobs were reduced to \num{110555}, while Eberius~\textsl{et\,al.} reported \num{118818}.
Their rigid simulations consistently demonstrate higher node utilization, frequently approaching 100\%, likely due to unfiltered shared-node jobs.

Our approach differs in several additional aspects.
We explicitly exclude warm-up and drain-down periods to avoid analytical artifacts and employ variable tick rates (\num{1}\,s for \haswell, \num{10}\,s for \knl) to balance simulation runtime and temporal resolution.
We support fine-grained expand/shrink operations, while their malleable schedulers rely on doubling/halving the number of nodes, either conservatively (one change per tick) or aggressively (unlimited).

Despite methodological differences, the \knl results show consistency across both studies, with similar trends and utilization gains (35\% at 100\% malleability).
This convergence---despite differing assumptions and implementations---underscores the robustness and practical value of malleability.

%% file: 06conclusions.tex
\section{Conclusion}\label{sec:07-Conclusions}
This work evaluated the impact of malleable jobs using real-world workload traces from \haswell, \knl, \eagle, and \theta.
Each workload was preprocessed and simulated in both rigid and malleable configurations using ElastiSim.
In addition to the rigid EASY-Backfill job scheduling strategy, we evaluated four malleable strategies, including a novel preference-nodes--preserving strategy.

Across all experiments, malleability consistently yielded substantial performance gains.
With the best-performing scheduler for each supercomputer, job turnaround times improved by 37--67\%, job makespan by 16--65\%, job wait times by 73--99\%, and node utilization by 5--52\%.
Although the extent of improvement varied with workload characteristics and scheduling strategy, significant benefits were already evident at low malleability ratios (e.g., 20\%).

Overall, this work demonstrates the potential of malleability to enhance both resource efficiency and user experience.